\documentclass[aip,pof,amsmath,amssymb,reprint]{revtex4-1}
\usepackage{graphicx}
\usepackage{bm}
\usepackage[utf8]{inputenc}
\usepackage[T1]{fontenc}
\usepackage{mathptmx}
\usepackage{booktabs}
\usepackage{CJKutf8}
\usepackage{amsmath}
\usepackage{hyperref}
\usepackage[capitalise]{cleveref}

\begin{document}

\begin{CJK*}{UTF8}{gbsn}
\title{Physics-Guided Generative Surrogates for Parametric Rarefied Flows
with Neural-Field Auto-Decoders: A Pipeline-Level Study of Flow Matching and
Diffusion}

\author{Yiming Qi (岐亦铭)}
\affiliation{
	Centre for Interdisciplinary Research in Fluids,
	Institute of Mechanics, Chinese Academy of Sciences, Beijing 100190, China
}
\author{Guan Zhang (张冠)}
\affiliation{
	Centre for Interdisciplinary Research in Fluids,
	Institute of Mechanics, Chinese Academy of Sciences, Beijing 100190, China
}
\author{Xu Wang (汪旭)}
\affiliation{
	Centre for Interdisciplinary Research in Fluids,
	Institute of Mechanics, Chinese Academy of Sciences, Beijing 100190, China
}
\author{Yonghao Zhang (张勇豪)}
\affiliation{
	Centre for Interdisciplinary Research in Fluids,
	Institute of Mechanics, Chinese Academy of Sciences, Beijing 100190, China
}
\author{Tianbai Xiao (肖天白)}
\email{txiao@imech.ac.cn}
\affiliation{
	Centre for Interdisciplinary Research in Fluids,
	Institute of Mechanics, Chinese Academy of Sciences, Beijing 100190, China
}
\keywords{Rarefied flows, generative modeling, neural fields, physics constraints}

\date{\today}

\begin{abstract}
We present a conditional latent generative framework for parametric rarefied
flows that separates neural-field representation, latent transport, and frozen
physics adaptation. Neural-field auto-decoders compress discrete-velocity
cavity solutions and direct simulation Monte Carlo cylinder solutions into
shared coordinate decoders. Train-only principal-component charts support
conditional flow matching (FM) and diffusion without a deterministic
condition-to-latent backbone, and structured low-rank adapters correct
selected decoder outputs while the upstream pipeline remains frozen. On two
steady benchmarks, the frozen pipelines interpolate out-of-sample conditions
with cavity kinetic relative $L_1$ errors at the $10^{-5}$ level and cylinder
per-field area-weighted RMSEs of 0.038 (density), 0.041 (temperature), and
below $0.01$ (velocities). For the cavity, physics adaptation reduces the
matched-grid Bhatnagar--Gross--Krook diagnostic by 28.65\% while preserving
field accuracy; for the cylinder, the analytic wall map enforces
no-penetration exactly and, jointly with the learned FM adapter, reduces the
inlet violation to 0.277 and the global mass-balance ratio to 0.963 of the
frozen values with negligible field-error change. A five-seed controlled
comparison with deterministic condition-to-chart multilayer perceptrons shows
that, although the generative pipelines do not surpass the compact MLP in
point accuracy on these single-valued steady problems, the results validate
sampling-based conditional transport on the shared representation as an
effective steady surrogate, with a natural route to multivalued or stochastic
solution families.
\end{abstract}

\maketitle
\end{CJK*}

\section{Introduction}\label{Introduction}

Rarefied gas flows arise in high-altitude and hypersonic aerodynamics,
atmospheric reentry, and micro- and nanoscale transport systems. Departure from
local thermodynamic equilibrium is commonly characterized by the Knudsen number
$\mathrm{Kn}=\lambda/L$, where $\lambda$ is the molecular mean free path and
$L$ is a characteristic length. As $\mathrm{Kn}$ increases through the slip,
transition, and free-molecular regimes, the constitutive closure underlying the
Navier--Stokes equations becomes progressively inadequate, and kinetic
descriptions that resolve molecular velocity distributions become necessary.
Established approaches include the direct simulation Monte
Carlo method (DSMC) \citep{Bird1994}, the discrete velocity method (DVM)
\citep{Mieussens2000,XiaoTB2021}, high-order lattice Boltzmann formulations
\citep{ChenHD2006,ZhangRY2006,ShiYY2021}, and multiscale gas-kinetic schemes
\citep{XuK2010,XuK2014,GuoZL2021,XinZY2025}. These methods provide access to
strongly nonequilibrium states but remain expensive in many-query studies:
deterministic kinetic solvers discretize the coupled physical--velocity phase
space, whereas DSMC requires extensive particle sampling and often local mesh
refinement. Repeated simulations over $\mathrm{Kn}$, Mach number, and other
physical or geometric controls therefore yield large, heterogeneous datasets
for which storage, reconstruction, and rapid condition-to-field prediction are
separate computational challenges.

This computational burden has motivated data-driven surrogates at several levels 
of the simulation hierarchy. In continuum fluid mechanics, machine learning has been
used for field reconstruction from sparse observations
\citep{Callaham2019,Erichson2020,Dubois2022} and for learned components that
accelerate conventional solvers \citep{Kochkov2021}, within a broader movement
toward data-assisted physical modeling \citep{Buzzicotti2023,Vinuesa2026}.
For many-query partial differential equations, the deep operator network
(DeepONet), the Fourier neural operator, and conditional neural fields
approximate mappings from input
functions or physical parameters to solution fields
\citep{LuL2021,LiZY2021,Hagnberger2024}. In kinetic and rarefied-gas modeling,
related efforts have targeted reduced Bhatnagar--Gross--Krook (BGK)
representations \citep{ZhangP2024},
learned moment closures \citep{HuangJT2022,HuangJT2026,Christlieb2025},
structure-preserving collision operators \citep{LeeJY2024,LeeJY2026}, and
field-level surrogates for microscale flows \citep{Roohi2026b,Roohi2026c}.
Existing studies show that machine learning can approximate selected solver 
components and flow observables. The present task, however, combines three 
distinct requirements: compressing each high-dimensional flow solution, 
predicting unseen physical conditions from a limited number of parameter cases, 
and retaining the governing and boundary constraints that can be evaluated from 
the available variables. Here, data scarcity occurs in the parameter space rather 
than necessarily in the spatial domain: high-fidelity solutions are typically 
available at only a limited number of sampled parameter combinations, whereas 
each sampled condition can be associated with multiple spatial fields and, when 
the kinetic state is represented directly, high-dimensional discrete 
velocity-distribution functions \citep{Roohi2026,WangX2026}. Accurate 
reconstruction of the observed cases therefore does not guarantee accurate 
interpolation between them. This difficulty is particularly pronounced for 
nonequilibrium targets such as stress, heat flux, and velocity-distribution 
functions, whose errors are more sensitive to the flow regime than those of 
conservative macroscopic variables \citep{WangX2026}.

One approach to handling high-dimensional flow data is to represent each
solution using a small number of latent variables. Closely related linear
methods such as principal component analysis (PCA) \citep{JolliffeCadima2016}
and proper orthogonal decomposition (POD) \citep{Berkooz1993,LongY2024}
construct these
variables from dominant global modes and provide a clear hierarchy for 
truncating the representation. Their efficiency may nevertheless decrease 
when the flow solutions vary along a strongly nonlinear manifold. An early
connection between modal reduction and machine learning was established by
Milano and Koumoutsakos, who interpreted POD as a particular linear
neural-network representation and extended it using a nonlinear neural network
for the reconstruction and prediction of near-wall turbulent flows
\citep{Milano2002}. This viewpoint motivates modern autoencoders and
variational autoencoders (VAEs), which learn nonlinear mappings between physical
fields and low-dimensional latent variables \citep{Kingma2014}. However,
convolutional implementations generally assume a fixed grid and a fixed
arrangement of output channels. Neural fields take a different approach by
representing a field as a continuous function of its coordinates
\citep{Sitzmann2020}. Fourier features help these models capture small-scale
spatial variations \citep{Tancik2020}, while auto-decoders assign an optimized
latent code to each observed field without requiring a separate encoder
\citep{Park2019}. This coordinate-based formulation can be applied to fields
stored on different spatial discretizations
\citep{YinY2023,Serrano2023,Hagnberger2024}. However, latent codes learned
jointly with a decoder are not unique: they can be rotated or rescaled without
changing the reconstructed fields, and distances between latent codes do not
necessarily measure differences between the decoded flow solutions
\citep{Arvanitidis2018}. Consequently, accurate reconstruction alone does not
guarantee that the learned latent space is well organized for predicting
solutions at unseen physical conditions.

A second line of research concerns the prediction of solution representations
at previously unseen physical conditions. Deterministic operator-learning and
conditional-field models approximate a single-valued map from governing inputs
to solution fields \citep{LuL2021,LiZY2021,Hagnberger2024}. Conditional
generative models instead provide a sample-based formulation by transforming
samples from a tractable reference distribution into solution states under a
specified physical condition. Unlike a deterministic regressor that returns a
single point estimate for each condition, this construction retains a sampling
dimension and can represent conditional variability or multiple plausible
solution states when such variability is resolved by the training data. This
capability should not, however, be interpreted as physical uncertainty when
only one converged reference solution is available at each condition; in that
setting, the generated spread primarily reflects the learned transport and
sampling procedure. In diffusion models, data are progressively perturbed
according to a prescribed stochastic process, and generation is performed by
numerically solving learned reverse-time dynamics conditioned on the physical
parameters \citep{SohlDickstein2015,Ho2020,Song2021}. Their performance can be
sensitive to the noise schedule, prediction parameterization, conditioning
mechanism, and numerical sampler. Flow matching (FM) instead learns a time-dependent
conditional vector field along a prescribed probability path and generates
samples by integrating an ordinary differential equation (ODE)
\citep{Lipman2023,LiuXC2023,Tong2024}. It replaces reverse-time denoising with
ODE transport and may reduce the number of model evaluations, although its
accuracy remains sensitive to the source--target coupling, path geometry, and
integration error. Conditional diffusion and flow-based models have recently
been investigated for parametric equations, physical forecasting, field
reconstruction, and continuum-flow prediction
\citep{GaoH2025,Shysheya2024,Parikh2026,Ramos2026,Armegioiu2025}. Existing
comparisons are nevertheless often confounded by differences in field
representation, conditioning architecture, training data, and evaluation
protocol. Moreover, for deterministic parametric problems with only one
reference state at each sampled condition, it remains unclear whether
conditional generative transport improves prediction beyond a deterministic
conditional regressor or primarily provides an alternative mechanism for
navigating the learned solution manifold.

A third line of research seeks to incorporate physical information into
learned solution models. Physics-informed neural networks introduce governing
equations, initial or boundary conditions, and other physical relations into
the training objective through differentiable residual terms
\citep{Raissi2019}. Weak formulations can improve robustness when strong-form
derivatives are unreliable or when the target fields contain sharp spatial
variations \citep{deRyck2024}. For diffusion models, physical information has
also been introduced through additional training losses, conditional guidance,
or corrections applied during sampling
\citep{ShuDL2023,Jacobsen2024,Bastek2025}. Such methods can reduce selected
equation or boundary residuals, but a lower residual does not by itself
establish that every generated field satisfies the complete governing system.
The result depends on whether all variables required by a constraint are
available, whether the corresponding derivatives can be evaluated accurately,
and whether residuals with different scales are normalized appropriately.
Moreover, increasing the weight of a physical penalty can improve the targeted
diagnostic while moving predictions away from the data-supported solution
manifold, whereas a weak penalty may have little measurable effect.
Parameter-efficient adaptation provides a general mechanism for limiting such
changes: a pretrained backbone is frozen and only a comparatively small
residual module is optimized for the downstream objective
\citep{Houlsby2019,Hu2022}. These techniques were developed for controlled
model adaptation rather than for enforcing physical laws, and their use alone
therefore provides no guarantee of conservation or kinetic consistency. The
unresolved problem is how to impose only those physical relations that can be
evaluated from the available outputs, restrict the capacity of the resulting
correction, and improve the targeted physical diagnostics without materially
degrading the predictive accuracy of the pretrained generative surrogate.

Taken together, these studies leave a coupled
representation--generation--consistency problem. Existing approaches commonly
address field compression, conditional prediction, and physical regularization
as separate tasks, so an improvement at one stage does not necessarily persist
through the complete parameter-to-field pipeline. In particular, it remains
unclear whether a coordinate-based decoder can provide a common compact
representation for heterogeneous rarefied-flow outputs, whether diffusion and
flow matching can accurately transport that representation across sparsely
sampled physical conditions under a common evaluation protocol, and whether
selected physical diagnostics can subsequently be improved without retraining
or materially perturbing the data-supported generative backbone. Resolving
these questions requires reconstruction, conditional-generation, and
physics-correction errors to be evaluated separately, because the final field
error may otherwise conceal the stage at which predictive accuracy is lost.

To address these questions, we develop a conditional latent generative
framework and evaluate it on two steady two-dimensional rarefied-flow
benchmarks: DVM-resolved lid-driven cavity flow parameterized by
$\mathrm{Kn}$ and DSMC-resolved cylinder flow parameterized by
$(\mathrm{Kn},\mathrm{Ma})$ \citep{WangX2026}. The central contribution is a
unified \textit{neural-field representation--conditional latent
generation--frozen physics adaptation} framework, together with an evaluation
that separates the evidence for each component. First, coordinate-based neural-field
auto-decoders are trained without supplying the physical conditions to the
representation stage, and principal-component charts fitted exclusively to
the training lookup codes provide compact coordinates for subsequent
conditional transport. Second, conditional flow matching and diffusion act
directly on these latent charts and generate condition-dependent terminal
latents from reference or source latent samples. Their frozen pipelines are
compared across a one-dimensional condition coordinate and a sparse
two-dimensional parameter geometry, using common data partitions and
decoded-field metrics within each benchmark. A direct condition-to-chart MLP
is additionally decoded by the same field representation to test whether the
sampling-based formulation improves steady point prediction. Third,
generator-specific
structured low-rank adapters introduce capacity-limited physical corrections
while leaving both the selected generator and the neural-field decoder frozen.
For the cavity flow, parallel adapter branches modify the four macroscopic
outputs $(\log\rho,u_x,u_y,\log T)$, the $G/B$ velocity-basis coefficients, and
the corresponding nonequilibrium bias terms, while preserving the decoder's
positive and bounded distribution parameterization. For the cylinder flow,
the adapter corrects the standardized density, streamwise-velocity,
transverse-velocity, and temperature outputs and includes a wall-localized
normal-velocity branch together with a no-penetration wall map. All learned
adapter output projections are initialized to zero, but the FM and diffusion
branches follow the same freezing principle rather than an identical
optimization protocol: their ranks, latent exposure, sampling density, and
checkpoint criteria are selected separately. Finally, the evaluation
distinguishes auto-decoder reconstruction error, conditional-generation error,
physics-correction error, interpolation, and out-of-support extrapolation. The
resulting evidence supports accurate interpolation over covered parameter
ranges and shows that frozen structured adaptation can improve explicitly
computable physical diagnostics with limited changes in field accuracy. The
evaluation is designed to answer, rather than assume, the question of where
sampling-based transport adds value: it includes a deterministic
condition-to-chart MLP baseline under the same frozen decoder, five-seed
repetitions of every generator, and frozen physics-adaptation audits, so that
the regime of applicability of each formulation is established under common
data partitions and decoded-field metrics.

This paper is organized as follows. \Cref{Sec2} defines the two rarefied-flow
problems, describes the DVM and DSMC datasets, and establishes the common
prediction and evaluation protocol. \Cref{Sec3} presents the neural-field
auto-decoders, train-only latent charts, conditional FM and diffusion models,
and structured physics adaptation. \Cref{Sec4} reports the controlled
representation, generator, sampler, and physics-adapter ablations and fixes the
models used thereafter. \Cref{Sec5} evaluates representative interpolation and
extrapolation cases for the cavity, assesses additional cylinder conditions,
benchmarks the frozen generators against direct latent MLPs, and discusses the
cross-benchmark behavior. Finally, \cref{Conclusions}
summarizes the principal findings, limitations, and directions for further
work.

\section{Problem setup and data generation}\label{Sec2}

\subsection{Lid-driven cavity flow and DVM data}\label{sec:cavity-data}

\subsubsection{Physical configuration and kinetic model}

We consider the steady flow of a monatomic gas in the square cavity
$\Omega=[0,L_0]\times[0,L_0]$, where $L_0$ is the cavity side length. 
The dataset is expressed in nondimensional units with $L_0=1$, reference 
density $\rho_0=1$, wall temperature $T_w=1$, and gas constant $R=0.5$, 
so that the reference most probable speed is $\sqrt{2RT_w}=1$. 
The upper wall moves in the positive $x$ direction at $u_w=0.15$, while 
the remaining walls are stationary. All four walls are isothermal and use 
fully diffuse Maxwell reflection; the density of the incoming wall Maxwellian 
is chosen to give zero net normal mass flux. The flow is two-dimensional in 
physical space: there is no dependence on a third spatial coordinate and 
the third macroscopic velocity component is zero. Molecular translation 
nevertheless remains three-dimensional for the monatomic gas, which 
therefore has specific-heat ratio $\gamma=5/3$. The cavity Knudsen number is 
$\mathrm{Kn}=\lambda_0/L_0$, where $\lambda_0$ is the reference mean free path.

Let $f(t,x,y,\xi,\eta,\zeta)$ denote the particle velocity-distribution
function, where $(t,x,y)$ are the time and physical coordinates and
$(\xi,\eta)$ are the particle-velocity components associated with transport
in the two-dimensional physical plane. The remaining component $\zeta$ is
retained in $f$ because the particle velocity is three-dimensional, but it
produces no third-direction spatial transport and is not explicitly
discretized. The reference solutions are
governed by the BGK model \citep{BGK},
\begin{equation}
 \frac{\partial f}{\partial t}
 +\xi\frac{\partial f}{\partial x}
 +\eta\frac{\partial f}{\partial y}
 =\frac{f_0-f}{\tau},
 \label{eq:cavity-bgk}
\end{equation}
where $\tau$ is the relaxation time and the local three-dimensional
Maxwellian $f_0$ is
\begin{equation}
 f_0(t,x,y,\xi,\eta,\zeta)
 =\frac{\rho}{(2\pi RT)^{3/2}}
 \exp \left[-\frac{(\xi-u)^2+(\eta-v)^2+\zeta^2}{2RT}\right].
 \label{eq:cavity-maxwellian}
\end{equation}
The quantities $\rho$, $u$, $v$, and $T$ in
\cref{eq:cavity-maxwellian} are, respectively, the local mass density, the two
in-plane macroscopic velocity components, and the translational temperature 
obtained from moments of $f$. The saved fields are
steady solutions of \cref{eq:cavity-bgk}. For the parameters used by the
solver, the viscosity model and relaxation time reduce to
\begin{equation}
 \mu=\mu_0\left(\frac{T}{T_w}\right)^{\omega},
 \quad
 \mu_0=\frac{5\sqrt{\pi}}{16} \mathrm{Kn},
 \quad
 \tau=\frac{\mu}{p}
 =\frac{\mu_0(T/T_w)^{\omega}}{\rho RT},
 \quad \omega=0.72,
 \label{eq:cavity-relaxation}
\end{equation}
where $\mu$ is the local dynamic viscosity, $\mu_0$ is its reference value,
$p=\rho RT$ is the local pressure, and $\omega$ is the viscosity exponent.
The value $\omega=0.72$ is inherited from the benchmark DVM configuration
\citep{WangX2026}, rather than introduced here as a new gas-model calibration.
Consequently, $\mathrm{Kn}$ is the only physical condition varied across the
cavity cases.

\subsubsection{Reduced distributions and macroscopic moments}

Because the physical problem has only two spatial directions, the DVM
explicitly discretizes only the in-plane particle velocities $(\xi,\eta)$.
The contribution of the remaining particle-velocity component $\zeta$ is
retained by integrating it into two reduced distributions, for which we use
the paper notation
\begin{equation}
 G(t,x,y,\xi,\eta)=\int f d\zeta,
 \quad
 B(t,x,y,\xi,\eta)=\int \zeta^2 f d\zeta.
 \label{eq:cavity-reduced-def}
\end{equation}
The corresponding equilibrium distributions are
\begin{equation}
 G_0=\frac{\rho}{2\pi RT}
 \exp \left[-\frac{(\xi-u)^2+(\eta-v)^2}{2RT}\right],
 \quad B_0=RTG_0,
 \label{eq:cavity-reduced-equilibrium}
\end{equation}
and integration of \cref{eq:cavity-bgk} over the hidden velocity gives
\begin{equation}
 \frac{\partial}{\partial t}
 \begin{bmatrix}G\\B\end{bmatrix}
 +\xi\frac{\partial}{\partial x}
 \begin{bmatrix}G\\B\end{bmatrix}
 +\eta\frac{\partial}{\partial y}
 \begin{bmatrix}G\\B\end{bmatrix}
 =\frac{1}{\tau}
 \left(
 \begin{bmatrix}G_0\\B_0\end{bmatrix}
 -\begin{bmatrix}G\\B\end{bmatrix}
 \right).
 \label{eq:cavity-reduced-bgk}
\end{equation}

The conserved moments are recovered by integration over the explicit
two-dimensional velocity plane:
\begin{subequations}
\begin{equation}
	\rho=\iint G d\xi d\eta,
\end{equation}
\begin{equation}
	\rho u=\iint \xi G d\xi d\eta,
 \quad \rho v=\iint \eta G d\xi d\eta,
\end{equation}
\begin{equation}
	\rho E=\frac{1}{2}\iint
 \left[(\xi^2+\eta^2)G+B\right]d\xi d\eta,
\end{equation}
 \label{eq:cavity-conservative-moments}
\end{subequations}
where $E=(u^2+v^2+3RT)/2$ is the total energy. With the in-plane peculiar
velocity components $\bm{c}=(c_x,c_y)=(\xi-u,\eta-v)$, the pressure tensor and heat flux are
\begin{equation}
	P_{ij}=\iint c_i c_jG d\xi d\eta,
	\label{eq:cavity-pressure-tensor}
\end{equation}
and
\begin{equation}
	q_i=\frac{1}{2}\iint c_i
	\left[(c_x^2+c_y^2)G+B\right]d\xi d\eta,
	\label{eq:cavity-heat-flux}
\end{equation}
where $P_{ij}$ and $q_i$ denote the in-plane stress tensor and heat flux, 
respectively. The DVM dataset provides the conservative-state vector 
$W=(\rho,\rho u,\rho v,\rho E)$, $P_{ij}$, $q_i$, and the two reduced distributions. 
The learned field representation predicts the four primitive variables $(\rho,u,v,T)$ 
together with $G$ and $B$; conservative and higher-order quantities can then be 
recomputed using the same velocity quadrature.

\subsubsection{DVM discretization and data partitions}

The reference fields were obtained with a finite-volume discrete velocity
method \citep{XiaoTB2021,WangX2026}. The physical domain uses a
$50\times50$ uniform cell-centered Cartesian grid, whose coordinates range
from $0.01$ to $0.99$ in each direction. The nominal velocity domain is
$[-5,5]\times[-5,5]$ and is represented by an $80\times80$ uniform
cell-centered grid with nodes from $-4.9375$ to $4.9375$. Its spacing is
$\Delta\xi=\Delta\eta=0.125$, and velocity moments are evaluated by the
composite midpoint Newton--Cotes rule,
\begin{equation}
 \iint \psi(\xi,\eta) d\xi d\eta
 \simeq
 \sum_{r=1}^{80}\sum_{s=1}^{80}
 \psi(\xi_r,\eta_s) \Delta\xi\Delta\eta,
 \quad \Delta\xi\Delta\eta=0.015625,
 \label{eq:cavity-dvm-quadrature}
\end{equation}
where $\psi$ is an arbitrary velocity-space integrand, and $r$ and $s$ index
the discrete nodes in the $\xi$ and $\eta$ directions, respectively.
Thus, one reduced distribution contains
$50\times50\times80\times80=1.6\times10^7$ values per condition, and the
pair $(G,B)$ contains $3.2\times10^7$ values. This dimensionality motivates a
continuous coordinate representation rather than direct generation of the
full physical--velocity tensor.

The representation and conditional generators use the following 16 training
conditions:
\begin{equation}
 \{0.05,0.10,0.15,0.25,0.30,0.35,0.40,0.45,
   0.55,0.60,0.65,0.70,0.75,0.85,0.90,0.95\}.
 \label{eq:cavity-train-kn}
\end{equation}
The three conditions $\{0.20,0.50,0.80\}$
are used for architecture, hyperparameter, and checkpoint selection and are
therefore validation rather than blind-test cases. For concise presentation,
the main-text field comparisons use the following four conditions from the
additional evaluation data:
\begin{equation}
 \{0.03,0.23,0.73,1.00\},
 \label{eq:cavity-shown-kn}
\end{equation}
where $\{0.23,0.73\}$ are interpolation cases inside the training support,
whereas $0.03$ and $1.00$ probe lower- and upper-side extrapolation,
respectively. These four conditions are selected as representative predictions
for the main-text comparison.

\subsection{Rarefied flow around a cylinder and DSMC data}
\label{sec:cylinder-data}

\subsubsection{Physical configuration and operating conditions}

The second benchmark considers a fixed circular cylinder in a high-speed
rarefied argon stream. This canonical configuration contains a detached bow
shock, a near-wall Knudsen layer, and a downstream wake whose structures vary
strongly with compressibility and rarefaction \citep{WangX2026}. The cylinder
is centered at the origin and has diameter $D=1$ and radius $R_c=0.5D$.
Coordinates are nondimensionalized by $D$, giving the computational box
\begin{equation}
 \Omega_c=[-5D,6D]\times[-5D,5D],
 \quad x^2+y^2>(D/2)^2,
 \label{eq:cylinder-domain}
\end{equation}
with a periodic spanwise direction in the underlying DSMC calculation. The
wall and freestream temperatures are both $273 \mathrm{K}$, and gas--surface
interaction at the cylinder is modeled by fully diffuse reflection with
complete thermal accommodation.

The operating condition is described by
\begin{equation}
 \mathrm{Kn}=\frac{\lambda_\infty}{D},
 \quad
 \mathrm{Ma}=\frac{U_\infty}{a_\infty},
 \quad
 a_\infty=\sqrt{\frac{\gamma k_B T_\infty}{m_{\mathrm{Ar}}}},
 \quad \gamma=\frac{5}{3},
 \label{eq:cylinder-conditions}
\end{equation}
where $\mathrm{Kn}$ and $\mathrm{Ma}$ are the Knudsen and Mach numbers;
$\lambda_\infty$, $U_\infty$, and $T_\infty$ are the freestream mean free
path, speed, and temperature, respectively; $a_\infty$ is the freestream
sound speed, $k_B$ is the Boltzmann constant, and $m_{\mathrm{Ar}}$ is the
argon atomic mass. The source simulation campaign spans
$0.06\leq\mathrm{Kn}\leq1.00$ in increments of $0.02$ and
$2.0\leq\mathrm{Ma}\leq6.9$ in increments of $0.1$, yielding $2400$
operating conditions \citep{WangX2026}. The present study uses the following
tensor-product subset of 64 cases:
\begin{equation}
 \begin{aligned}
 \mathrm{Kn}&\in\{0.06,0.10,0.14,0.20,0.30,0.44,0.66,1.00\},\\
 \mathrm{Ma}&\in\{2.0,2.7,3.4,4.1,4.8,5.5,6.2,6.9\}.
 \end{aligned}
 \label{eq:cylinder-condition-grid}
\end{equation}
This range spans substantial changes in the bow shock, near-wall Knudsen
layer, and downstream wake: decreasing $\mathrm{Kn}$ produces thinner and
sharper structures, while increasing rarefaction makes the macroscopic
features more diffuse.

\subsubsection{DSMC generation and macroscopic fields}

The reference fields were generated using the direct simulation Monte Carlo
method \citep{Bird1994,WangX2026}, with a variable-hard-sphere model for argon.
Hierarchically refined Cartesian cells resolve the cylinder surface,
near-wall rarefaction, bow shock, and wake. After a statistically steady state
is reached, particle statistics are time averaged to reduce DSMC sampling
noise. The DSMC data were generated using the open-source SPARTA solver with
the variable-hard-sphere (VHS) collision model for argon and the
no-time-counter collision scheme, following the simulation protocol of
TransportBench \citep{WangX2026}; the argon collision and reference
parameters are summarized in \cref{tab:argon-vhs}.

\begin{table}[htbp]
 \centering
 \caption{Argon collision and reference parameters of the DSMC benchmark
 \citep{Bird1994,WangX2026}.}
 \label{tab:argon-vhs}
 \begin{tabular}{lc}
  \toprule
  Quantity & Value\\
  \midrule
  Atomic mass $m_{\mathrm{Ar}}$ & $6.63\times10^{-26}$ kg\\
  VHS reference diameter $d_{\mathrm{ref}}$ & $4.17\times10^{-10}$ m\\
  VHS viscosity index $\omega$ & 0.81\\
  Reference temperature $T_{\mathrm{ref}}$ & 273 K\\
  Wall and freestream temperature $T_w=T_\infty$ & 273 K\\
  \bottomrule
 \end{tabular}
\end{table}

The nominally two-dimensional prediction task retains number density $n$, the
in-plane velocity components $(u,v)$, and translational temperature $T$.
Rotational and vibrational temperatures are absent for monatomic argon, and the
spanwise mean velocity is not included as a prediction target. Mass density
and pressure are not treated as independent learned variables but are
reconstructed through
\begin{equation}
 \rho=m_{\mathrm{Ar}}n,
 \quad
 p=nk_BT.
 \label{eq:cylinder-derived-fields}
\end{equation}

The native multilevel Cartesian mesh is retained rather than interpolating the
training fields to a common image grid. Its resolution varies with
$\mathrm{Kn}$ and is locally refined around the cylinder, bow shock, and wake.
Only fluid cells contribute to field losses and metrics. Regular grids used
later for common-resolution field comparisons are coordinate queries to the
learned neural field, not the representation on which the decoder is trained.
Concretely, all cylinder out-of-sample field comparisons are evaluated on a
regular common grid of $256\times192$ uniform points spanning
$[-3D,5D]\times[-3D,3D]$, the cavity error metrics are evaluated on the
native $50\times50$ DVM grid with its velocity quadrature
(\cref{eq:cavity-dvm-quadrature}), and the displayed cavity contour figures
use uniform $256\times256$ grids.

\subsubsection{Preprocessing and data partitions}

For each condition, the four learned targets are expressed as disturbances
relative to the prescribed freestream state,
\begin{equation}
 \bm\delta(x,y)=
 \left[
 \log\frac{n}{n_\infty},
 \frac{u-U_\infty}{U_\infty},
 \frac{v}{U_\infty},
 \log\frac{T}{T_\infty}
 \right]^T,
 \label{eq:cylinder-targets}
\end{equation}
where the subscript $\infty$ denotes the prescribed freestream value and the
superscript $T$ denotes vector transpose. Each
component is subsequently divided by a case-balanced root-mean-square (RMS)
scale fitted from training cases only. This construction makes zero
disturbance exactly equal to the prescribed freestream state.

The 64-case tensor grid is divided into 48 training, eight validation, and
eight internal test conditions. The validation set is
\begin{equation}
 \{(0.06,2.7),(0.10,4.1),(0.14,5.5),(0.20,6.9),(0.30,2.0),(0.44,3.4),
 (0.66,4.8),(1.00,6.2)\},
\end{equation}
and the internal test set is
\begin{equation}
 \{(0.06,4.8),(0.10,6.2),(0.14,2.0),(0.20,3.4),(0.30,5.5),(0.44,6.9),
 (0.66,2.7),(1.00,4.1)\},
\end{equation}
where each ordered pair denotes
$(\mathrm{Kn},\mathrm{Ma})$. Neither validation nor
test cases contribute to target normalization or neural-field parameter
optimization. For concise presentation of the out-of-sample predictions, the
main text displays the following representative conditions:
\begin{equation}
 \{(0.12,2.3),(0.38,3.7),(0.76,4.5)\},
 \label{eq:cylinder-shown-conditions}
\end{equation}
where each ordered pair denotes $(\mathrm{Kn},\mathrm{Ma})$.

\subsection{Unified prediction task and evaluation protocol}
\label{sec:unified-task}

\subsubsection{Conditional prediction and error decomposition}

Both benchmarks are formulated as conditional parameter-to-field prediction.
Let $\bm\mu$ denote the physical condition and $\bm s$ the coordinates at
which a field is queried:
\begin{equation}
 \bm\mu=
 \begin{cases}
  \mathrm{Kn}, & \text{cavity},\\
  (\mathrm{Kn},\mathrm{Ma}), & \text{cylinder},
 \end{cases}
 \quad
 \bm s=
 \begin{cases}
  (x,y)\ \text{or}\ (x,y,\xi,\eta), & \text{cavity},\\
  (x,y), & \text{cylinder}.
 \end{cases}
 \label{eq:unified-condition-coordinate}
\end{equation}
The cavity target comprises the primitive variables and reduced distributions
$\bm y_c=(\rho,u,v,T,G,B)$, whereas the cylinder target is
$\bm y_d=(n,u,v,T)$; the subscripts $c$ and $d$ identify the cavity and
cylinder benchmarks, respectively. Additional macroscopic quantities are obtained from
\cref{eq:cylinder-derived-fields}.
Throughout the remainder of the paper, boldface $\bm y$ denotes a
multichannel target or prediction, scalar $y$ remains the second Cartesian
coordinate, and $q_i$ is reserved exclusively for the heat-flux component
defined in \cref{eq:cavity-heat-flux}.

For either benchmark, let $\mathcal G_{\theta_G}$ denote the conditional
generator and $D_{\theta_D}$ the frozen neural-field decoder, with
$\theta_G$ and $\theta_D$ denoting their parameter sets. The generator maps
the $m$th reference or source sample $\bm r^{(m)}$ to a compact latent-chart
coordinate,
\begin{equation}
 \widehat{\bm a}^{(m)}(\bm\mu)
 =\mathcal G_{\theta_G}(\bm r^{(m)};\bm\mu).
 \label{eq:unified-generative-map}
\end{equation}
The data provide one converged steady reference field at each condition.
Accordingly, the primary point prediction is obtained by averaging the chart
endpoints before the fixed inverse transformation and nonlinear decoding,
\begin{equation}
 \overline{\bm a}(\bm\mu)
 =\frac{1}{M}\sum_{m=1}^{M}\widehat{\bm a}^{(m)}(\bm\mu),
 \quad
 \overline{\bm y}(\bm s;\bm\mu)
 =D_{\theta_D} \left[
 \bm s,\mathcal P^{-1} \left(\overline{\bm a}\right)
 \right],
 \label{eq:unified-ensemble-mean}
\end{equation}
where $\mathcal P^{-1}$ is the inverse latent-chart transformation, a hat
denotes a generated endpoint, and $M$ is the ensemble size. This convention is
fixed for all reported methods; because the decoder is nonlinear, it is not
equivalent to averaging independently decoded fields. The endpoint spread is
treated as a diagnostic of the learned transport, not as calibrated physical
uncertainty.

A frozen-decoder audit quantifies this noncommutativity (\cref{app:endpoint-averaging}): the relative difference between latent-space and
decoded-space averaging is at most $6.0\times10^{-6}$ for FM and, away from
the lower extrapolation condition, at most $5.9\times10^{-5}$ for diffusion;
at $\mathrm{Kn}=0.03$ latent averaging contributes a non-negligible part of
the extrapolation error.

Three sources of error are reported separately. The \textit{representation
error} is obtained by decoding the optimized lookup latent of an observed
case; it measures the neural field's attainable reconstruction floor. The
\textit{conditional-generation error} is obtained from
\cref{eq:unified-generative-map} with the decoder fixed and therefore includes
latent-transport error. The \textit{physics-correction effect} is the paired
change from a frozen generative prediction to its adapter-corrected counterpart.
Because the decoder and correction are nonlinear, these quantities are not
assumed to form an additive error decomposition.

\subsubsection{Field-error measures}

For the cavity, let $j$ index physical grid points, $k$ index discrete
velocity nodes, $\ell$ index the four conservative-state components, and
$\omega_k=\Delta\xi\Delta\eta$ denote the DVM quadrature weight. The
macroscopic and kinetic relative errors are
\begin{equation}
 E_{\mathrm{macro}}
 =\frac{\sum_{j,\ell}|\widehat W_{j,\ell}-W_{j,\ell}|}
 {\sum_{j,\ell}|W_{j,\ell}|+\epsilon},
 \label{eq:cavity-macro-error}
\end{equation}
\begin{equation}
 E_G=
 \frac{\sum_{j,k}\omega_k|\widehat G_{j,k}-G_{j,k}|}
 {\sum_{j,k}\omega_k|G_{j,k}|+\epsilon},
 \quad
 E_B=
 \frac{\sum_{j,k}\omega_k|\widehat B_{j,k}-B_{j,k}|}
 {\sum_{j,k}\omega_k|B_{j,k}|+\epsilon},
 \label{eq:cavity-kinetic-errors}
\end{equation}
with $E_{\mathrm{kin}}=(E_G+E_B)/2$ and a small $\epsilon>0$ preventing
division by zero.  For a derived tensor or vector
$\mathcal Q\in\{\bm P,\bm q\}$, the componentwise relative error is
\begin{equation}
 E_{\mathcal Q}=
 \frac{\displaystyle\sum_{j,c}
 |\widehat{\mathcal Q}_{j,c}-\mathcal Q_{j,c}|}
 {\displaystyle\sum_{j,c}|\mathcal Q_{j,c}|+\epsilon},
 \label{eq:cavity-derived-error}
\end{equation}
where $c$ indexes the pressure-tensor or heat-flux components defined in
\cref{eq:cavity-pressure-tensor,eq:cavity-heat-flux}.  The matched-grid reduced-BGK
diagnostic is evaluated with the same physical grid, velocity quadrature, and
discrete differentiation rule for every compared model. It is therefore used
for paired comparisons and is not interpreted as a rigorous bound on the
continuous BGK residual.

For the cylinder, each reported field is first nondimensionalized by its
physical reference scale. Let $\phi^*$ denote a generic dimensionless field
from the following set:
\begin{equation}
 \phi^*\in
 \left\{
 \frac{n}{n_\infty},\frac{p}{p_\infty},\frac{T}{T_\infty},
 \frac{u}{U_\infty},\frac{v}{U_\infty},
 \frac{\sqrt{u^2+v^2}}{U_\infty}
 \right\}.
 \label{eq:cylinder-dimensionless-fields}
\end{equation}
For a condition indexed by $i$, the area-weighted root-mean-square error
(RMSE) is
\begin{equation}
 E_{\phi,i}=
 \left[
 \frac{\sum_j a_{i,j}
 (\widehat\phi^*_{i,j}-\phi^*_{i,j})^2}
 {\sum_j a_{i,j}}
 \right]^{1/2},
 \label{eq:cylinder-field-rmse}
\end{equation}
where $j$ indexes spatial samples and $a_{i,j}$ is the corresponding spatial
quadrature weight, which is uniform on the common regular grid used for the
final field comparison.

\subsubsection{Data use and reporting conventions}

Training conditions determine the field decoder, latent chart, conditional
generator, and train-only normalization statistics. Validation conditions are
used for architecture, hyperparameter, and checkpoint selection. The final
out-of-sample conditions do not contribute to gradient updates or checkpoint
ranking. A condition is classified as interpolation when it lies within the
support of the training conditions and as extrapolation when it lies outside
that support; the two categories are reported separately.

Errors are first computed for each physical condition and are then aggregated
with equal case weight,
\begin{equation}
 \overline E=\frac{1}{N_{\mathrm{case}}}
 \sum_{i=1}^{N_{\mathrm{case}}}E_i,
 \label{eq:equal-case-mean}
\end{equation}
where $E_i$ is the error for the $i$th condition and $N_{\mathrm{case}}$ is the
number of conditions included in the reported aggregate.
This prevents conditions with finer DSMC meshes or more spatial samples from
dominating the summary. Frozen FM and Frozen Diffusion denote the unadapted
conditional generators. A physics-constrained label is reserved for an adapted
checkpoint that passes all benchmark-specific feasibility gates. The cavity FM
checkpoint satisfies its feasibility gates; the
corresponding diffusion checkpoint and both retained cylinder checkpoints pass
their respective gates. All models are compared against the same DVM or DSMC reference within
each benchmark. Physics residuals are reported relative to the corresponding
frozen generator and are not compared as absolute values between the cavity
and cylinder problems, because the available variables and evaluable physical
constraints differ.

\section{Conditional latent generative framework}\label{Sec3}

\subsection{Overall parameter-to-field architecture}
\label{sec:overall-architecture}

The proposed framework separates high-dimensional field representation from
conditional generation. For each benchmark, a neural-field auto-decoder first
assigns an independently optimized lookup code $\bm z_i$ to every training
condition and learns a coordinate decoder $D_{\theta_D}$. The physical
condition $\bm\mu_i$ is not supplied to this representation stage; hence, the
case-dependent information required for reconstruction must pass through
$\bm z_i$. A train-only latent transformation $\mathcal P$ then maps the
lookup codes to a lower-dimensional chart $\bm a_i=\mathcal P(\bm z_i)$ on
which the conditional generative models are trained. This separation avoids
requiring either flow matching or diffusion to generate the full spatial or
physical--velocity tensor directly.

Let $m\in\{\mathrm{FM},\mathrm{diff}\}$ index the generator family and
$\bm r^{(m)}$ denote its model-specific reference state. A
conditional generator $\mathcal G_{\theta_G}$ produces a latent-chart sample,
which is mapped back to the auto-decoder space and evaluated at arbitrary query
coordinates:
\begin{subequations}\label{eq:overall-frozen-pipeline}
\begin{align}
 \widehat{\bm a}^{(m)}(\bm\mu)
 &=\mathcal G_{\theta_G}(\bm r^{(m)};\bm\mu),
 \label{eq:overall-chart-generation}\\
 \widehat{\bm z}^{(m)}(\bm\mu)
 &=\mathcal P^{-1} \left(\widehat{\bm a}^{(m)}\right),
 \label{eq:overall-inverse-chart}\\
 \widehat{\bm y}^{(m)}_{\mathrm{frozen}}(\bm s;\bm\mu)
 &=D_{\theta_D} \left(\bm s,\widehat{\bm z}^{(m)}\right),
 \label{eq:overall-frozen-decode}
\end{align}
\end{subequations}
where $\bm a$ and $\bm z$ denote the chart and auto-decoder coordinates,
respectively. For diffusion and the cavity FM model, $\bm r^{(m)}$ is sampled
from a tractable Gaussian reference distribution. For the cylinder FM model,
it instead contains an observed source latent and its source condition; the
model transports this anchor to the requested target condition. These two
choices instantiate the same reference-to-target formulation but are not
assumed to define identical probability paths. Their training and inference
are specified separately below.

During conditional-generator training and evaluation, $\mathcal P^{-1}$ and
$D_{\theta_D}$ are frozen. Consequently, differences between FM and diffusion
cannot be absorbed by retraining the field representation. The latent-endpoint
ensemble mean is used to obtain the primary steady-field prediction, while the
sample spread is retained only as a transport-sensitivity diagnostic, as
specified in \cref{sec:unified-task}. The sampling dimension is therefore
preserved without interpreting it as a measured physical uncertainty
distribution.

The optional physics-correction stage also preserves the selected generator
and decoder. It augments designated decoder heads with a structured low-rank
adapter $\mathcal A_{\psi}$,
\begin{equation}
 \widehat{\bm y}^{(m)}_{\mathrm{phys}}(\bm s;\bm\mu)
 =\mathcal A_{\psi} \left[
 D_{\theta_D};\bm s,\widehat{\bm z}^{(m)},\bm\mu
 \right],
 \quad
 \theta_G,\theta_D,\mathcal P^{-1}\ \text{frozen},
 \label{eq:overall-physics-pipeline}
\end{equation}
where $\psi$ denotes the only learned correction parameters. The upward
adapter projections are initialized to zero, so the learned low-rank branches
initially make no change to the frozen prediction. The cylinder protocol may
additionally apply a prescribed hard wall-normal projection; the exact
identity at initialization refers to the learned adapter before this
deterministic map. Separate adapters are trained for FM and diffusion. They
share the same freezing principle and benchmark-specific output structure but
not necessarily the same optimization schedule or selected checkpoint.

For the cavity problem, the auto-decoder has a lookup code of dimension 24 and
a train-only PCA chart of dimension four. Its spatial heads predict
$(\log\rho,u,v,\log T)$, while coupled spatial--velocity heads predict the
coefficients and biases used to construct the positive reduced distributions
$G$ and $B$. The physics adapter acts in parallel on the four primitive
outputs, the $G/B$ velocity-basis coefficients, and the corresponding
nonequilibrium biases. Thus, both macroscopic and kinetic predictions remain
linked through one generated latent state.

For the cylinder problem, the auto-decoder uses a 32-dimensional lookup code
and a 12-dimensional latent chart. Its coordinate network
predicts the four standardized freestream disturbances in
\cref{eq:cylinder-targets}; density and pressure are recovered through the
hard algebraic relations in \cref{eq:cylinder-derived-fields}. The structured
adapter corrects these four macroscopic decoder channels and includes a
wall-localized normal-velocity branch. The retained cylinder FM generator is a
source-to-target condition transport over local observed anchors and uses a
decoder-metric PCA chart, whereas its diffusion counterpart generates an
ordinary PCA chart of the same retained dimension from a Gaussian reference.
Both are mapped back to the same 32-dimensional lookup-latent space and decoded
by the same frozen neural-field representation.

\begin{figure}[hbtp]
 \centering
 \includegraphics[width=\linewidth]{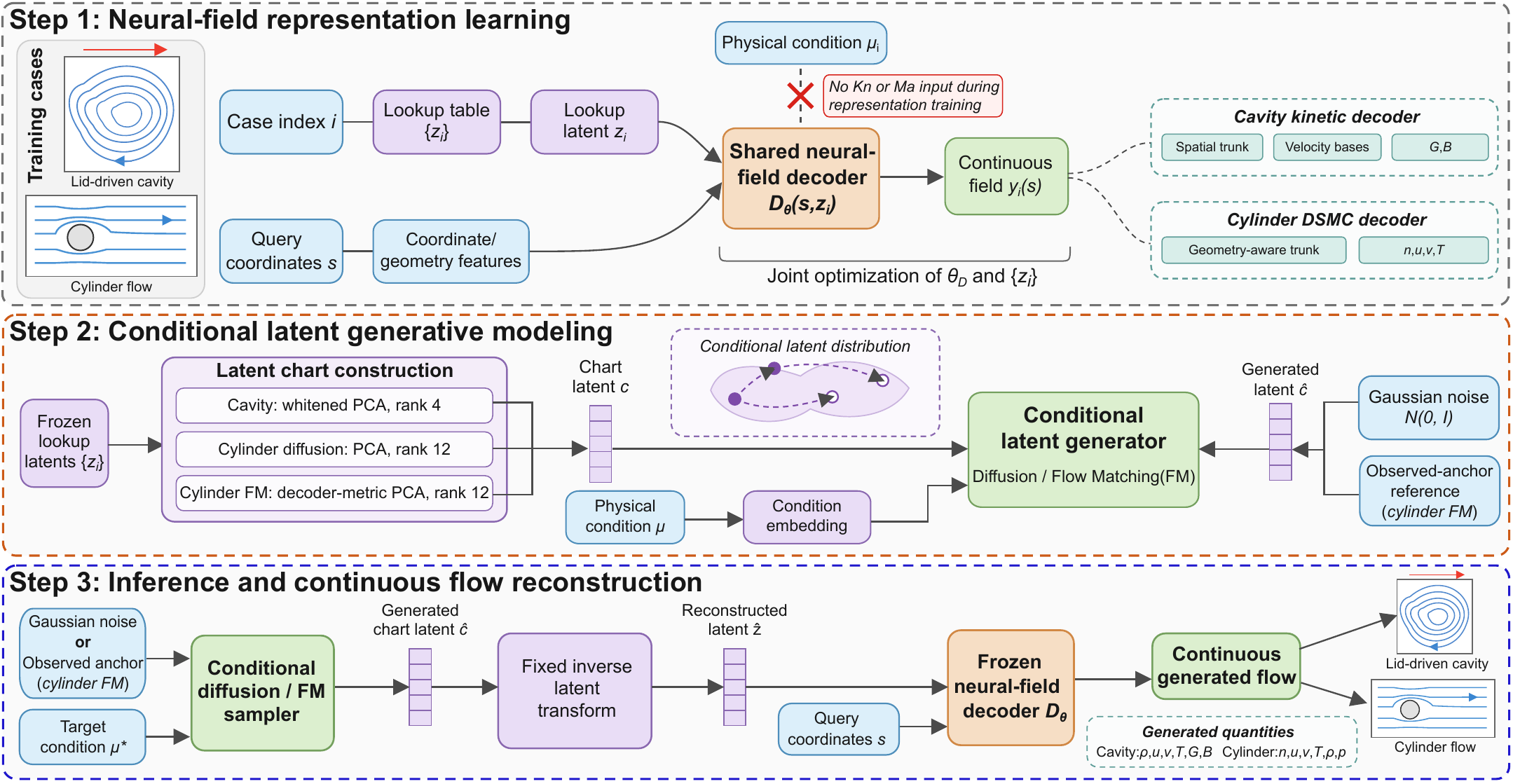}
 \caption{Conditional latent generative framework. Conditions are excluded
 from representation training. FM and diffusion operate in train-only latent
 charts using Gaussian references, except for the observed-anchor cylinder
 FM. Mean endpoints are inverse transformed and decoded before a structured
 adapter corrects selected channels with the upstream pipeline frozen.}
 \label{fig:conditional-latent-framework}
\end{figure}

No deterministic condition-to-latent regressor
$\bm z=f_{\vartheta}(\bm\mu)$ is used inside either generative pipeline. Such a
regressor is retained only as a controlled baseline and does not initialize,
supervise, or correct the generated latent samples. In particular, the term
``plain-concat MLP'' used for one FM architecture, where MLP denotes
multilayer perceptron, refers to a parameterization of the time-dependent
velocity field; it is not a direct
single-valued mapping from the physical condition to the terminal latent.
For the controlled deterministic baseline, an ensemble of regressors
$h_{\vartheta_j}$ maps the normalized physical condition directly to the
retained latent chart,
\begin{equation}
 \widehat{\bm a}_{\mathrm{MLP}}(\bm\mu)
 =\frac{1}{J}\sum_{j=1}^{J}h_{\vartheta_j}(\bm\mu),\qquad
 \widehat{\bm y}_{\mathrm{MLP}}(\bm s;\bm\mu)
 =D_{\theta_D}\!\left[\bm s,
 \mathcal P^{-1}(\widehat{\bm a}_{\mathrm{MLP}})\right],
 \label{eq:deterministic-latent-baseline}
\end{equation}
where $J$ is the number of independently initialized MLP members. The cavity
baseline has $J=3$, width 128, three hidden layers, and Knudsen Fourier
frequencies 0.5 and 1.0; it predicts the common whitened PCA-4 chart. The
cylinder baseline has $J=5$, width 64, and two residual blocks; it predicts the
ordinary unwhitened PCA-12 chart. Both models, their train-only chart
transformations, and the field decoders are frozen before evaluation on the
additional conditions. They are comparison models only and are not components
of either generative pipeline. The same protocol is additionally executed
with five independent seeds per configuration, at the parameter-matched
capacities, and, for the cylinder, on both retained charts; these controlled
extensions are reported in \cref{sec:deterministic-decoded-comparison}.
The complete separation between representation learning, latent generation,
and coordinate-based field reconstruction is summarized in
\cref{fig:conditional-latent-framework}.

\subsection{Neural-field auto-decoder and latent representation}
\label{sec:neural-field-representation}

\subsubsection{Coordinate-based field decoders}

For both benchmarks, the representation model is an auto-decoder rather than
an encoder--decoder \citep{Park2019}. A trainable lookup vector $\bm z_i$ identifies the
$i$th observed condition, while a shared coordinate network represents the
associated continuous field,
\begin{equation}
 \widehat{\bm y}_i(\bm s)
 =D_{\theta_D}(\bm s,\bm z_i).
 \label{eq:autodecoder-field}
\end{equation}
Neither $\mathrm{Kn}$ nor $\mathrm{Ma}$ is provided to
$D_{\theta_D}$. The condition labels are therefore unavailable as a shortcut
for field reconstruction, and all case dependence must pass through
$\bm z_i$. The two decoders share this principle but use different
coordinate features and output factorizations because the cavity includes
velocity-distribution functions whereas the cylinder data contain
macroscopic fields on nonuniform meshes.

Let $\gamma_{\mathcal F}(\bm\chi)$ denote a Fourier encoding of a normalized
coordinate vector $\bm\chi$ \citep{Tancik2020},
\begin{equation}
 \gamma_{\mathcal F}(\bm\chi)=
 \left[
 \bm\chi,\left\{
 \sin(\pi f \chi_j),\cos(\pi f \chi_j)
 \right\}_{f\in\mathcal F, j}
 \right].
 \label{eq:coordinate-fourier}
\end{equation}
The encoded coordinates are processed by residual multilayer perceptrons. In
a latent-conditioned block, feature-wise linear modulation (FiLM)
\citep{Perez2018} is applied after layer normalization,
\begin{equation}
 \widetilde{\bm h}_{l}
 =
 \left[\bm 1+\Delta\bm\gamma_l(\bm z_i)\right]
 \odot\operatorname{LN}(\bm h_l)
 +\bm\beta_l(\bm z_i),
 \quad
 \bm h_{l+1}
 =\bm h_l+c_l\mathcal R_l(\widetilde{\bm h}_l),
 \label{eq:decoder-film-block}
\end{equation}
where $l$ indexes the residual blocks, $\operatorname{LN}$ denotes layer
normalization, $\odot$ denotes componentwise multiplication,
$\Delta\bm\gamma_l(\bm z_i)$ and $\bm\beta_l(\bm z_i)$ are learned
feature-wise scale and shift maps, respectively,
$\mathcal R_l$ is a two-layer residual mapping with the sigmoid linear unit
(SiLU) activation, and $c_l$ is a residual scaling factor. The FiLM
projections are initialized so that
$\Delta\bm\gamma_l=\bm\beta_l=\bm 0$ at initialization.

For the cavity, the coordinates are normalized as
$\bar x=2x-1$, $\bar y=2y-1$,
$\bar\xi=\xi/5$, and $\bar\eta=\eta/5$. The principal spatial branch uses
frequencies $\{1,2,3,4,6,8,12\}$, supplemented by smooth indicators of
the moving lid, $\exp[-(1-y)/0.06]$, and of the stationary walls,
$1-(1-e^{-x/0.06})(1-e^{-(1-x)/0.06})(1-e^{-y/0.06})$.
A second spatial encoding with
frequencies $\{8,12,16,20\}$ supplies higher-frequency
features to the kinetic branch. The velocity encoder takes
$[\bar\xi,\bar\eta,\bar\xi^2+\bar\eta^2,\bar\xi\bar\eta]$ together with
Fourier frequencies
$\{1,2,3,4,6,8\}$. The resulting feature dimensions are 32,
18, and 28 for the principal spatial, kinetic spatial, and velocity encoders,
respectively.
A residual scale of $c_l=0.5$ is used throughout the cavity decoder. The
lookup code modulates the even-numbered blocks of the principal spatial trunk
and every block of the kinetic spatial branch.

A width-640 spatial trunk with 12 residual FiLM blocks predicts the
standardized primitive outputs
$(\log\rho,u,v,\log T)$. A width-768, eight-block kinetic spatial branch
produces coefficient vectors
$\bm\alpha_G,\bm\alpha_B\in\mathbb R^{256}$ and scalar biases
$\beta_G,\beta_B$. Independently, a condition-invariant width-512 velocity
trunk with 10 residual blocks produces the two velocity bases
$\bm b_G(\xi,\eta),\bm b_B(\xi,\eta)\in\mathbb R^{256}$. The raw
nonequilibrium corrections are
\begin{equation}
 \phi_F^{\mathrm{raw}}(x,y,\xi,\eta)
 =\beta_F(x,y)
 +\bm\alpha_F(x,y)^{T}\bm b_F(\xi,\eta),
 \quad F\in\{G,B\},
 \label{eq:cavity-low-rank-kinetic}
\end{equation}
and are smoothly bounded before being applied to the local equilibrium:
\begin{equation}
 \phi_F=4\tanh \left(\phi_F^{\mathrm{raw}}/4\right),
 \quad
 \widehat G=G_0\exp(\phi_G),
 \quad
 \widehat B=B_0\exp(\phi_B).
 \label{eq:cavity-positive-output}
\end{equation}
This construction preserves $\widehat G,\widehat B>0$, couples the kinetic
outputs to the predicted macroscopic state through $G_0$ and $B_0$, and
avoids an unconstrained exponential correction. The cavity decoder has a
24-dimensional lookup code and approximately $26.37$ million trainable
parameters.

For the cylinder, the coordinate encoder combines normalized Cartesian
coordinates with cylinder-centered polar geometry. Its 163 features include
geometric localization terms together with the Cartesian, polar-angle, and
wall-distance frequency sets and the wall-localization length scales (in
units of $D$),
\[
 \begin{aligned}
 &\{1,2,3,4,6,8,12,16,24\},\quad
 \{1,2,3,4,6,8,12,16,20,24\},\\
 &\{1,2,4,8,12,16,24,32\},\quad
 \{0.025,0.05,0.10,0.20,0.40,0.80\}.
 \end{aligned}
\]
The remaining geometric
inputs include the signed distance
$d_w=\sqrt{x^2+y^2}-R_c$, the polar angle, the wall-normal direction, and
exponential wall-distance localization features. These features distinguish the
near-wall Knudsen layer from the detached shock and wake without first
interpolating the native DSMC cells to an image grid.

The cylinder decoder uses a 32-dimensional lookup code, a width-512 trunk,
and 12 FiLM residual blocks, with the lookup code injected through a scaled
input projection and through FiLM in every block. Four independent width-192 heads predict the
standardized components of $\bm\delta$ in
\cref{eq:cylinder-targets}; separate heads reduce interference between density,
the two velocity components, and temperature. Their final layers are initialized
to zero, making the initial prediction the prescribed freestream state. The
physical variables are reconstructed by inverting
\cref{eq:cylinder-targets}, after which
$\rho=m_{\mathrm{Ar}}n$ and $p=nk_BT$ are imposed algebraically. This
decoder has approximately $7.22$ million trainable parameters. The principal
architectural settings of both neural fields are summarized in
\cref{tab:field-decoder-architecture}.

\begin{table}[htbp]
 \centering
 \caption{Neural-field auto-decoder architectures used for the two
 benchmarks. The listed parameters include the lookup table during
 representation training.}
 \label{tab:field-decoder-architecture}
 \footnotesize
 \begin{tabular}{lcc}
  \toprule
  Component & Cavity DVM & Cylinder DSMC \\
  \midrule
  Lookup-latent dimension & 24 & 32 \\
  Principal trunk & $640\times12$ blocks & $512\times12$ blocks \\
  Additional branch & $768\times8$ kinetic & Four $192$-wide heads \\
  Velocity trunk/basis & $512\times10$, rank 256 & Not applicable \\
  Coordinate feature dimensions & $32/18/28$ & 163 \\
  Trainable parameters & $26.37$ M & $7.22$ M \\
  \bottomrule
 \end{tabular}
\end{table}

\subsubsection{Lookup-latent optimization}

The decoder parameters and the lookup table are optimized jointly over the
training cases:
\begin{equation}
 \min_{\theta_D,\{\bm z_i\}_{i=1}^{N_{\mathrm{tr}}}}
 \frac{1}{N_{\mathrm{tr}}}
 \sum_{i=1}^{N_{\mathrm{tr}}}
 \mathcal L_{\mathrm{rep}}
 \left[D_{\theta_D}(\cdot,\bm z_i),\bm y_i\right].
 \label{eq:joint-autodecoder-objective}
\end{equation}
The physical condition is excluded from
\cref{eq:joint-autodecoder-objective}; it is used only after representation
training to learn conditional transport. Small latent-norm and centered-table
penalties remove unconstrained drift without assigning a physical meaning to
individual raw latent coordinates.

For the cavity, reconstruction combines normalized conservative and primitive
losses, the quadrature-weighted relative losses of $G$ and $B$, and
consistency between the conservative state predicted by the macroscopic head
and the moments of the kinetic heads:
\begin{equation}
 \begin{aligned}
 \mathcal L_{\mathrm{cav}}={}&
 0.75\mathcal L_W+0.25\mathcal L_{\mathrm{prim}}
 +\mathcal L_G+\mathcal L_B
 +10^{-2}\mathcal L_{\mathrm{mom}}\\
 &+10^{-5}\mathcal L_z+10^{-4}\mathcal L_{\mathrm{ctr}}
 +10^{-6}\mathcal L_{\mathrm{curv}}
 +\lambda_{\mathrm{basis}}\mathcal L_{\mathrm{basis}}
 +2\times10^{-5}\mathcal L_{\mathrm{bc}} ,
 \end{aligned}
 \label{eq:cavity-representation-loss}
\end{equation}
where $\mathcal L_W$ and $\mathcal L_{\mathrm{prim}}$ are normalized
smooth-$L_1$ losses; $\mathcal L_G$ and $\mathcal L_B$ have the
quadrature-weighted relative-$L_1$ form of
\cref{eq:cavity-kinetic-errors}; and $\mathcal L_{\mathrm{mom}}$ compares
the four discrete moments of $(\widehat G,\widehat B)$ with the macroscopic
head. The terms
$\mathcal L_z=\langle\|\bm z_i\|_2^2\rangle$ and
$\mathcal L_{\mathrm{ctr}}=\|\langle\bm z_i\rangle\|_2^2$ regularize the
lookup table. A normalized second-difference penalty
$\mathcal L_{\mathrm{curv}}$ is applied along $\log_{10}\mathrm{Kn}$, but
no first-difference penalty is used. The scale-invariant basis-correlation
penalty $\mathcal L_{\mathrm{basis}}$ is ramped from zero between 50\% and
75\% of training to a maximum coefficient of $10^{-5}$.
$\mathcal L_{\mathrm{bc}}$ provides a weak diffuse-wall preconditioner for
the incoming portions of $G$ and $B$.

Cavity representation training contains three phases within one 200,000-update
optimization. Updates 1--2,000 use
$3\mathcal L_W+\mathcal L_{\mathrm{prim}}$ and the lookup-code
regularizers. Updates 2,001--60,000 introduce the full-grid kinetic losses,
change the macroscopic weights to 0.75 and 0.25, and periodically evaluate
$\mathcal L_{\mathrm{mom}}$. The remaining updates additionally activate
the latent-curvature, basis-correlation, and wall-preconditioning terms in
\cref{eq:cavity-representation-loss}. Each kinetic update uses all 16 training
cases, all 2,500 spatial cells, and all 6,400 velocity nodes. Consequently, the
principal $G/B$ losses are full-grid quantities rather than Monte Carlo
estimates; only the moment and wall terms are evaluated at their prescribed
intervals to control memory and computation.

For the cylinder, the primary data loss is an area- and sampling-confidence-
weighted smooth-$L_1$ loss on the standardized freestream disturbances,
\begin{equation}
 \mathcal L_{\mathrm{rep,data}}^{\mathrm{cyl}}
 =
 \frac{1}{N_{\mathrm{el}}}\sum_{i,p,c}w_{ip} 
 \operatorname{SmoothL1}_{\beta=0.02}
  \left(
 \frac{\widehat\delta_{ipc}-\delta_{ipc}}{\sigma_{\delta,c}}
 \right),
 \label{eq:cylinder-representation-data-loss}
\end{equation}
where $i$, $p$, and $c$ index cases, sampled cells, and the four output
channels, respectively; the SmoothL1 term is the Huber (smooth $L_1$) loss
with threshold $0.02$; $N_{\mathrm{el}}$ is the corresponding number of averaged
elements; $\bm\sigma_\delta$ contains train-only, case-balanced
root-mean-square (RMS) scales;
and $w_{ip}$ is the normalized product of cell area and a clipped
square-root particle-confidence factor. Each update samples 16 cases and
18,432 points per case, with 20\% global, 25\% shock, 30\% near-wall, 20\%
wake, and 5\% far-field samples. This sampling distribution emphasizes
the shock and Knudsen layer during optimization, whereas validation remains
area weighted over the complete native mesh.

The cylinder representation objective before the late physical refinement is
\begin{equation}
 \mathcal L_{\mathrm{cyl,rep}}
 =\mathcal L_{\mathrm{rep,data}}^{\mathrm{cyl}}
 +5\times10^{-6}\mathcal L_z
 +10^{-4}\mathcal L_{\mathrm{ctr}}
 +\lambda_{\mathrm{surf}}(k)\mathcal L_{\mathrm{surf}},
 \label{eq:cylinder-representation-total}
\end{equation}
where $\lambda_{\mathrm{surf}}$ follows a half-cosine ramp from zero after
25\% of training to $5\times10^{-6}$ at the final update. The local
affine-surface term uses neighboring points in
$(\log\mathrm{Kn},\mathrm{Ma})$ only to regularize the geometry of the
lookup table; the labels still do not enter the decoder. A late, weak
representation-stage physical refinement applies normalized wall
no-penetration, inlet freestream, and control-volume mass terms in the ratio
0.30:0.15:0.55, with a combined peak coefficient of $8\times10^{-6}$.
Its multiplier is zero through 68\% of training, reaches one at 82\%, and,
after 92\%, decays smoothly to 0.10 at the final update. This
preconditioning is distinct from the generator-specific physics adapters
introduced later: it forms part of the common frozen field representation and
does not use generated latents. The cylinder model is trained for 180,000
updates using region-aware point sampling.

For both benchmarks, a held-out Oracle code is obtained by freezing
$D_{\theta_D}$ and optimizing only a new $\bm z$ against the observed
field. Oracle decoding measures the representational floor of the frozen
neural field; it is not a deployable inference procedure and is kept separate
from conditional-generation error.

\subsubsection{Train-only latent charts}

Raw auto-decoder codes possess a gauge freedom: an invertible change of latent
coordinates can be compensated by the decoder without changing the represented
fields. The generative models therefore operate on a train-only principal
component analysis (PCA) chart \citep{JolliffeCadima2016} rather than directly
on individual lookup coordinates. This choice removes weakly constrained
directions, restricts generated endpoints to the empirically supported latent
subspace, and balances coordinate scales for FM and diffusion optimization.
It thereby reduces the statistical and numerical burden of learning all 24 or
32 raw lookup coordinates from sparsely sampled physical conditions. Directly
transporting the raw lookup coordinates would force the generator to fit
directions that the decoder can absorb without changing the represented
field, which is ill-conditioned under 16 or 48 training cases; the chart
removes these directions before transport.
\textit{The chart is not a separate field representation: it is a
low-dimensional coordinate system for conditional generation, and every
generated endpoint is mapped back to the original auto-decoder coordinates
before evaluation by the frozen decoder.} Let $N_{\mathrm{tr}}$ be the number of
training conditions, $d_z$ the lookup dimension,
$\overline{\bm z}=N_{\mathrm{tr}}^{-1}\sum_i\bm z_i$, and
$Z_c\in\mathbb R^{N_{\mathrm{tr}}\times d_z}$ the matrix whose $i$th row
is $\bm z_i-\overline{\bm z}$. Its singular value decomposition is
\begin{equation}
 Z_c=U\Sigma V^T,
 \quad
 \bm \ell_i=(\bm z_i-\overline{\bm z})V_r,
 \label{eq:latent-pca}
\end{equation}
where $U$ and $V$ contain the left and right singular vectors,
$\Sigma$ contains the singular values, $V_r\in\mathbb R^{d_z\times r}$
contains the first $r$ right singular vectors, and
$\bm\ell_i\in\mathbb R^r$ is the retained, unscaled PCA score. Every centering,
scaling, metric, and basis quantity is fitted using training codes only and is
refitted when a training condition is omitted during model selection.

For the cavity, $r=4$. Its PCA scores are whitened using
\begin{equation}
 \bm a_i=\bm \ell_i\oslash\widetilde{\bm s},
 \quad
 \widetilde s_j
 =\max(s_j,10^{-3}s_1),
 \quad
 \mathcal P_{\mathrm{cav}}^{-1}(\bm a)
 =\overline{\bm z}
 +(\bm a\odot\widetilde{\bm s})V_r^T ,
 \label{eq:cavity-latent-chart}
\end{equation}
where $s_j=\Sigma_{jj}/\sqrt{N_{\mathrm{tr}}-1}$ is the empirical standard
deviation of the $j$th PCA score. The whitening floor prevents weak
lookup-code directions from being amplified.
Candidate ranks $r\in\{2,4,8\}$ are assessed by decoding the truncated
lookup codes, so rank selection depends on field fidelity in addition to
explained latent variance.

The cylinder diffusion branch uses the rank-12 unwhitened PCA scores in
\cref{eq:latent-pca}. Before generative optimization, each retained score is
standardized by its train-only standard deviation with a lower scale of
$2\times10^{-2}$. This scaling is an optimizer preconditioner and is undone
before the ordinary PCA inverse maps the generated endpoint to the
32-dimensional lookup-latent space.

The retained cylinder FM branch uses a separate rank-12 decoder-metric PCA chart.
Frozen-decoder sensitivities are evaluated at global, shock, near-wall, wake,
and far-field points to estimate one Gauss--Newton matrix per training case.
Their mean is symmetrized and divided by its mean eigenvalue (equivalently,
its trace divided by $d_z$) to obtain $G_D$ in the 32-dimensional lookup
space. It is blended with the Euclidean metric and factorized by its symmetric
positive-definite square root,
\begin{equation}
 \widetilde G_D=(1-\lambda_D)I+\lambda_DG_D,
 \quad \lambda_D=0.5,
 \quad \widetilde G_D=R_DR_D^T.
 \label{eq:decoder-latent-metric}
\end{equation}
PCA is then applied to
$(\bm z_i-\overline{\bm z})R_D$, followed by the same score-scale floor of
$2\times10^{-2}$. Euclidean displacement in this chart consequently
approximates, locally, a standardized decoded-field change and gives greater
weight to lookup directions to which the frozen decoder is sensitive. The
metric and PCA basis are recomputed from retained cases within every strict
leave-one-condition-out fold.

Thus, the cylinder FM and diffusion models share the same frozen decoder,
32-dimensional lookup space, retained chart dimension, data partitions, and
decoded-field evaluation, but they do not use an identical chart basis. The
distinction is retained because decoder-metric coordinates belong to the
retained curved-transport FM architecture, whereas the retained diffusion
baseline uses the ordinary PCA chart. In all cases,
$\mathcal P^{-1}$ is a fixed linear transformation during generator and
physics-adapter training.

\subsection{Conditional flow matching}
\label{sec:conditional-fm}

Flow matching (FM) learns a conditional velocity field that transports a
reference state to a target latent without requiring likelihood evaluation
\citep{Lipman2023,Tong2024}. Let
$\bm a_1$ denote a target chart coordinate, $\bm a_0$ a reference or
source coordinate, and $t\in[0,1]$ the transport time. For a prescribed
conditional path
\begin{equation}
 \bm a_t=A(t)\bm a_0+B(t)\bm a_1,
 \quad
 \bm u_t=\dot A(t)\bm a_0+\dot B(t)\bm a_1,
 \label{eq:fm-general-path}
\end{equation}
the conditional vector field $v_{\theta_F}$ is trained by
\begin{equation}
 \mathcal L_{\mathrm{FM}}
 =
 \mathbb E_{\bm a_0,\bm a_1,t,\bm\mu}
 \left[
 \left\|
 v_{\theta_F}(\bm a_t,t,\bm\mu)-\bm u_t
 \right\|_2^2 / d_a
 \right],
 \label{eq:fm-objective}
\end{equation}
where $\theta_F$ denotes the FM-network parameters, overdots denote
derivatives with respect to $t$, and $d_a$ is the chart dimension.
Generation integrates the neural
ordinary differential equation
\begin{equation}
 \frac{d\bm a_t}{dt}
 =v_{\theta_F}(\bm a_t,t,\bm\mu),
 \quad t:0\rightarrow1.
 \label{eq:fm-neural-ode}
\end{equation}
The cavity and cylinder models use different couplings because the former has
a one-dimensional condition space with a Gaussian reference, whereas the
latter transports observed solution anchors over a two-dimensional condition
manifold.

\subsubsection{Cavity Gaussian-reference flow matching}

For the cavity, $\bm a_0=\bm\epsilon\sim\mathcal N(\bm0,I_4)$ and
$\bm a_1$ is the four-dimensional whitened PCA coordinate of an observed
lookup latent. The selected probability path is
\begin{equation}
 \begin{aligned}
 A(t)&=\sigma_{\min}+(1-\sigma_{\min})
       \cos\left(\frac{\pi t}{2}\right),\\
 B(t)&=\sin\left(\frac{\pi t}{2}\right),
 \quad \sigma_{\min}=10^{-3}.
 \end{aligned}
 \label{eq:cavity-fm-path}
\end{equation}
The residual Gaussian scale at $t=1$ avoids numerically forcing a continuous
reference law to a strictly singular endpoint.

The condition is normalized using training-set endpoints,
\begin{equation}
 s(\mathrm{Kn})=
 2\frac{\log_{10}\mathrm{Kn}-\log_{10}\mathrm{Kn}_{\min}}
 {\log_{10}\mathrm{Kn}_{\max}-\log_{10}\mathrm{Kn}_{\min}}-1,
 \label{eq:cavity-condition-normalization}
\end{equation}
so $s\in[-1,1]$ on the training interval.
Fixed Fourier encodings use
$\mathcal F_t=\{1,2,4,8\}$ for time and
$\mathcal F_s=\{0.5,1,2\}$ for the condition. The vector-field input
\[
 [\bm a_t;\gamma_{\mathcal F_t}(t);
   \gamma_{\mathcal F_s}(s)]
 \in\mathbb R^{20}
\]
is processed by a plain-concatenation MLP comprising an input layer of width
704, six width-704 SiLU layers, layer normalization, and a four-dimensional
zero-initialized output. It contains $2{,}996{,}932$ parameters. A weak
second-difference penalty in the normalized Knudsen coordinate is
\begin{equation}
 \mathcal L_{\mathrm{Kn,curv}}
 =\mathbb E \left[
 \left\|
 \frac{v_{\theta_F}(\bm a_t,t,s_+)
 -2v_{\theta_F}(\bm a_t,t,s)
 +v_{\theta_F}(\bm a_t,t,s_-)}{(\Delta s)^2}
 \right\|_2^2
 \right],
 \label{eq:cavity-fm-curvature}
\end{equation}
where $\Delta s=0.02$ and
$s_\pm=\operatorname{clip}(s\pm\Delta s,-1,1)$.
This term is evaluated in 32-bit floating-point arithmetic (FP32), giving
\begin{equation}
 \mathcal L_{\mathrm{FM}}^{\mathrm{cav}}
 =
 \mathcal L_{\mathrm{FM}}
 +10^{-6}\mathcal L_{\mathrm{Kn,curv}}.
 \label{eq:cavity-fm-total}
\end{equation}
The principal FM regression is evaluated with 16-bit brain floating-point
(BF16) mixed precision. Inference
uses a 48-step explicit Heun solver and antithetic Gaussian initial states.
For even ensemble size $M$, antithetic sampling draws $M/2$ independent
states and appends their negatives, so the empirical reference mean is exactly
zero before transport.
The ensemble-mean endpoint is inverse transformed and decoded; 64 endpoints
are used for validation and 256 for the final field comparison.

\subsubsection{Cylinder curved condition-manifold transport}

For the cylinder, the source is an observed chart coordinate
$\bm a_a$ at
$\bm c_a=(\log\mathrm{Kn}_a,\mathrm{Ma}_a)$, and the target is
$\bm a_b$ at $\bm c_b$. Each condition is standardized componentwise as
$\bar{\bm c}=(\bm c-\overline{\bm c}_{\mathrm{tr}})
\oslash\bm s_{\bm c,\mathrm{tr}}$, using training-set means and standard
deviations. Training uses bidirectional edges between four nearest
conditions and the rectified path
\begin{equation}
 \bm a_t=(1-t)\bm a_a+t\bm a_b,
 \quad
 \bm u_t=\bm a_b-\bm a_a.
 \label{eq:cylinder-fm-path}
\end{equation}
To represent curvature in the two-dimensional condition manifold without an
unrestricted source--target lookup, the velocity is decomposed as
\begin{equation}
 \begin{aligned}
 v_{\theta_F}={}&
 J_{\theta_F}(\bm a_t,t,\bar{\bm c}_t)\Delta\bar{\bm c}\\
 &+\sum_{r=1}^{8}\sum_{k=1}^{3}
 a_{rk,\theta_F}(\bm a_t,t,\bar{\bm c}_t)
 m_k(\Delta\bar{\bm c}) \bm b_r ,
 \end{aligned}
 \label{eq:cylinder-quadratic-transport}
\end{equation}
where
\[
 \Delta\bar{\bm c}=\bar{\bm c}_b-\bar{\bm c}_a,\quad
 \bar{\bm c}_t=\bar{\bm c}_a+t\Delta\bar{\bm c},
 \quad
 \bm m(\Delta\bar{\bm c})
 =(\Delta\bar c_1^2,\Delta\bar c_1\Delta\bar c_2,
   \Delta\bar c_2^2).
\]
Here $J_{\theta_F}\in\mathbb R^{12\times2}$ supplies the complete
first-order tangent response, $a_{rk,\theta_F}$ is a state-, time-, and
condition-dependent scalar coefficient, and the eight learned vectors
$\bm b_r\in\mathbb R^{12}$ span a low-rank quadratic correction. Both output
heads are zero initialized, and
$v_{\theta_F}=0$ identically for $\bm c_a=\bm c_b$. The shared residual
MLP has width 128, three residual blocks, expansion factor two, SiLU
activation, layer normalization, and time frequencies
$\{1,2,4,8\}$; no condition Fourier features are used. It contains 208,144
trainable parameters.

Let $\mathcal T_{\theta_F}(\bm a_a,\bm c_a,\bm c_b)$ denote the endpoint
obtained by numerically integrating \cref{eq:fm-neural-ode} from source to
target condition. The endpoint and consensus terms used for cylinder training
are
\begin{subequations}\label{eq:cylinder-fm-auxiliary}
\begin{align}
 \mathcal L_{\mathrm{end}}
 &=\mathbb E_{a\to b} \left[
 \|\mathcal T_{\theta_F}(\bm a_a,\bm c_a,\bm c_b)-\bm a_b\|_2^2/d_a
 \right],\\
 \mathcal L_{\mathrm{con}}
 &=\mathbb E_b \left[
 \frac{1}{M_b d_a}\sum_{j=1}^{M_b}
 \|\widehat{\bm a}_{j\to b}-\overline{\bm a}_b\|_2^2
 \right],\\
 \mathcal L_{\mathrm{con,end}}
 &=\mathbb E_b \left[
 \frac{1}{M_b d_a}\sum_{j=1}^{M_b}
 \|\widehat{\bm a}_{j\to b}-\bm a_b\|_2^2
 \right],
\end{align}
\end{subequations}
where $\widehat{\bm a}_{j\to b}=\mathcal T_{\theta_F}
(\bm a_j,\bm c_j,\bm c_b)$, $M_b$ is the number of sampled source
anchors for target $b$, and $\overline{\bm a}_b$ is their endpoint mean.
For a mini-batch of target cases, the decoded-field term is
\begin{equation}
 \mathcal L_{\mathrm{field}}
 =\frac{1}{N_{\mathrm{case}}}\sum_{i=1}^{N_{\mathrm{case}}}e_i
 +0.25\max_i e_i,
 \quad
 e_i=\frac{1}{4N_p}\sum_{p=1}^{N_p}
 \|D^{\mathrm{std}}_{\theta_D}(\bm s_{ip},\widehat{\bm z}_i)
 -\bm y^{\mathrm{std}}_{ip}\|_2^2,
 \label{eq:cylinder-fm-field-loss}
\end{equation}
where the inverse chart used to obtain $\widehat{\bm z}_i$ and the decoder
are fixed; $N_{\mathrm{case}}$ and $N_p$ are the numbers of cases and
sampled spatial points in this auxiliary batch, and superscript
$\mathrm{std}$ denotes the standardized cylinder outputs. The complete
objective is
\begin{equation}
 \mathcal L_{\mathrm{FM}}^{\mathrm{cyl}}
 =
 \mathcal L_{\mathrm{FM}}
 +0.20\mathcal L_{\mathrm{end}}
 +0.50\mathcal L_{\mathrm{field}}
 +0.05\mathcal L_{\mathrm{con}}
 +0.10\mathcal L_{\mathrm{con,end}}.
 \label{eq:cylinder-fm-total}
\end{equation}
The auxiliary terms are estimated every 10 updates after the first 25\% of
optimization, using a differentiable four-step Heun integration. The field
term samples global, shock, near-wall, wake, and far-field regions
with respective fractions 30\%, 25\%, 25\%, 15\%, and 5\%, and includes a worst-case
contribution in addition to the mean. Gradients pass through the frozen
decoder to the generator, but no decoder parameter is updated.

At inference, a local nondegenerate triangle containing the query is selected
in condition space. Its three vertices independently generate target latents
with a 24-step Heun solver, after which their nonnegative barycentric weighted
mean is used. Outside the convex hull, three nearest anchors and positive
distance-softmax weights are used. Small centered source perturbations of
scale $0.01$ provide a transport-sensitivity ensemble; they do not represent
measured DSMC uncertainty.

\subsection{Conditional diffusion}
\label{sec:conditional-diffusion}

The conditional diffusion models act on standardized chart coordinates
\citep{SohlDickstein2015,Ho2020,Song2021}. A continuous linear
variance-preserving (VP) schedule is defined by
\begin{equation}
 \begin{aligned}
 \beta(t)&=\beta_{\min}
  +t(\beta_{\max}-\beta_{\min}),\\
 \alpha(t)&=\exp\left[
 -\frac12\beta_{\min}t
 -\frac14(\beta_{\max}-\beta_{\min})t^2\right],\\
 \sigma(t)&=\sqrt{1-\alpha(t)^2},\quad
 \bm a_t=\alpha(t)\bm a_{\mathrm{data}}+\sigma(t)\bm\epsilon,
 \end{aligned}
 \label{eq:vp-forward-process}
\end{equation}
where $\bm a_{\mathrm{data}}$ is a clean target chart coordinate and
$\bm\epsilon\sim\mathcal N(\bm0,I)$. Using $\bm\tau_p$ to distinguish a
diffusion regression target from the physical field $\bm y$, the three
parameterizations are
\begin{equation}
\bm \tau_x=\bm a_{\mathrm{data}},\quad
\bm \tau_\epsilon=\bm\epsilon,\quad
\bm \tau_v=\alpha(t)\bm\epsilon-\sigma(t)\bm a_{\mathrm{data}},
\label{eq:diffusion-targets}
\end{equation}
with
\begin{equation}
 \mathcal L_{\mathrm{diff}}^{(p)}
 =
 \mathbb E\left[
 \left\|
 \widehat{\bm\tau}_{\theta_\Delta}^{(p)}(\bm a_t,t,\bm\mu)-\bm \tau_p
 \right\|_2^2/d_a
 \right],
\quad p\in\{x,\epsilon,v\},
\label{eq:diffusion-objective}
\end{equation}
where $\theta_\Delta$ denotes the denoiser parameters. The subscript $x$
follows the conventional name ``$x$-prediction'' and means prediction of the
clean latent $\bm a_{\mathrm{data}}$; it is unrelated to the spatial
coordinate $x$. Although these parameterizations describe the same forward perturbation
analytically, their loss weighting across noise levels and their numerical
conditioning differ.

Both denoisers concatenate the noisy chart state, a Fourier time encoding, and
the normalized physical condition. The cavity model additionally applies
Fourier features at condition frequencies $\{0.5,1,2\}$; the cylinder model
uses standardized $(\log\mathrm{Kn},\mathrm{Ma})$ directly. The cavity
denoiser has the same width-704, six-layer plain-concat architecture and
parameter count as its FM counterpart. Its retained formulation uses
$v$-prediction, FP32 training, and
$(\beta_{\min},\beta_{\max})=(0.05,40)$. The cylinder denoiser is a
203,404-parameter residual MLP with width 128, three blocks, expansion factor
two, and time frequencies $\{1,2,4,8\}$. It uses $x$-prediction and
$(\beta_{\min},\beta_{\max})=(0.1,20)$.

Because each cylinder condition has one reference solution, the cylinder fit
also constrains the endpoint mean and suppresses artificial conditional
spread. For $M=4$ antithetic samples
$\widehat{\bm a}_{ij}$ at each of four sampled conditions,
\begin{equation}
 \mathcal L_{\mathrm{diff}}^{\mathrm{cyl}}
 =\mathcal L_{\mathrm{diff}}^{(x)}
 +0.10\left\langle
 \left\|\frac{1}{M}\sum_{j=1}^{M}\widehat{\bm a}_{ij}-\bm a_i\right\|_2^2
 \right\rangle_i
 +0.01\left\langle
 \operatorname{Var}_{j}(\widehat{\bm a}_{ij})\right\rangle_i.
 \label{eq:cylinder-diffusion-total}
\end{equation}
These endpoint terms are estimated with four-step denoising diffusion implicit
model (DDIM) sampling \citep{DDIM} every 20 updates after 60\% of optimization.
The cavity diffusion fit does not use
endpoint regularization. During cylinder sampling, the predicted clean chart
coordinate is smoothly clipped at magnitude 6 before each reverse update.

The ablation protocol includes the denoising diffusion probabilistic model
(DDPM) reverse chain \citep{Ho2020}, deterministic DDIM sampling, and the
second-order multistep (2M) DPM-Solver++ method \citep{LuC2025}. The frozen
cavity model is sampled with
35 DPM-Solver++(2M) steps, using 64 antithetic samples for validation and 256
for final evaluation. The frozen cylinder model uses 40 deterministic DDIM
steps and 40 antithetic samples. The clean chart endpoint is averaged before
the fixed inverse transform and nonlinear field decoding. Generated spread is
reported as a sampler and transport diagnostic, not as calibrated physical
uncertainty.

\subsection{Controlled FM--diffusion comparison}
\label{sec:controlled-generator-comparison}

Within each benchmark, FM and diffusion use the same representation-stage lookup table,
field decoder, training/validation partition, target normalization, and
decoded-field metrics. For the cavity, the comparison is additionally
parameter matched: both conditional networks contain $2{,}996{,}932$
parameters, operate on the same rank-four whitened PCA chart, use the same condition
features, and are evaluated with antithetic ensembles. The remaining
differences are the training objective, probability path or noise schedule,
numerical precision selected by ablation, and deployment integrator.

For the cylinder, the two conditional networks have similar capacity and the
same retained chart dimension, but the comparison is not representation
identical. Diffusion uses the ordinary rank-12 PCA chart, whereas the retained
FM uses the decoder-metric rank-12 chart in
\cref{eq:decoder-latent-metric}. FM also transports local observed anchors and
uses barycentric aggregation, while diffusion starts from a Gaussian reference.
These differences are retained because they belong to the frozen retained
models; they are disclosed rather than absorbed into a claim of perfectly
matched generative computation. All endpoints nevertheless return to the same
32-dimensional lookup-latent space and the same frozen cylinder decoder.

\subsection{Structured physics adaptation}
\label{sec:structured-physics-adaptation}

\subsubsection{Frozen-backbone low-rank correction}

After generator selection, both $\theta_G$ and $\theta_D$, as well as the
latent inverse transform, are frozen. For a decoder-derived adapter input
feature $\bm h$, the parallel low-rank adapter
\citep{Houlsby2019,Hu2022} has the generic form
\begin{equation}
 \bm h_a=\operatorname{SiLU} \left[
 W_{\mathrm{down}}\operatorname{LN}(\bm h)\right],
 \quad
 \Delta\bm y
 =
 s_{\mathrm{out}} g(\bm\mu)
 \frac{\alpha_a}{r_a}
 W_{\mathrm{up}}\bm h_a,
 \label{eq:generic-low-rank-adapter}
\end{equation}
where $r_a$ is the adapter rank, $\alpha_a$ its scale, and
$g(\bm\mu)$ an optional condition gate. For an input feature of dimension
$d_h$ and a corrected head of dimension $d_o$,
$W_{\mathrm{down}}\in\mathbb R^{r_a\times d_h}$ and
$W_{\mathrm{up}}\in\mathbb R^{d_o\times r_a}$; $s_{\mathrm{out}}$ is a
fixed branch-specific output scale. $W_{\mathrm{up}}$ is initialized to
zero so the learned branch begins with no correction. Data anchors,
frozen-field teachers, trust regions, and parameter regularization restrict
the correction to the neighborhood of the pretrained field manifold.

\subsubsection{Cavity reduced-BGK adaptation}

The cavity adapters act on the four standardized primitive outputs, the two
256-dimensional kinetic coefficient vectors, and the two scalar
nonequilibrium biases. Their output scales are $0.10$, $0.10$, and
$0.02$, respectively. A width-32 condition network with Knudsen Fourier
frequencies $\{0.5,1,2\}$ produces five sigmoid gates: one shared by the
four primitive corrections, one for each of the $G$- and $B$-coefficient
branches, and one for each of their scalar-bias branches.
Because the corrections are inserted before
\cref{eq:cavity-positive-output}, positivity and the bounded
nonequilibrium parameterization are preserved.

The steady reduced-BGK residual is evaluated in logarithmic distribution form,
\begin{equation}
 \begin{aligned}
 r_G={}&\tau(\xi\partial_x\log G+\eta\partial_y\log G)
       -[\exp(-\phi_G)-1],\\
 r_B={}&\tau(\xi\partial_x\log B+\eta\partial_y\log B)
       -[\exp(-\phi_B)-1],
 \end{aligned}
 \label{eq:adapter-reduced-bgk}
\end{equation}
with $\tau$ from \cref{eq:cavity-relaxation}. Central differences use
$h=2\times10^{-3}$, and the residual is integrated over all 6,400 velocity
nodes. Unless stated otherwise, training and checkpoint probes use a
near-wall-biased collocation measure; a separate uniform-area
sensitivity audit is reported in \cref{sec:bgk-numerical-sensitivity}. With normalized velocity weights
$\bar\omega_k=|\omega_k|/\sum_{k'}|\omega_{k'}|$, the reported matched-grid
BGK diagnostic is
\begin{equation}
 \mathcal L_{\mathrm{BGK}}
 =\left\langle
 \sum_k\bar\omega_k\left(r_{G,k}^2+r_{B,k}^2\right)
 \right\rangle_{\bm x,\mathrm{Kn}},
 \label{eq:cavity-bgk-diagnostic}
\end{equation}
where the brackets average the sampled spatial locations and conditions.
Additional terms measure diffuse-wall consistency, conservation of
collision invariants, and agreement between kinetic moments and the
macroscopic head.

For each component $j\in\{\mathrm{BGK},\mathrm{cons},\mathrm{mom},
\mathrm{bc}\}$, eight independently sampled frozen-decoder probe batches
$\mathcal B_p$ define a detached scale
\begin{equation}
 s_j=\max \left\{
 \operatorname{median}_{p=1,\ldots,8}
 \mathcal L_j^{\mathrm{frozen}}(\mathcal B_p),s_{j,\min}
 \right\}.
 \label{eq:cavity-adapter-normalization}
\end{equation}
The normalized physical objective is
\begin{equation}
 \widetilde{\mathcal L}_{\mathrm{phys}}^{(m)}
 =\frac{\mathcal L_{\mathrm{BGK}}}{s_{\mathrm{BGK}}}
 +0.25\frac{\mathcal L_{\mathrm{bc}}}{s_{\mathrm{bc}}}
 +w_{\mathrm{inv}}^{(m)} \sum_{j\in\{\mathrm{cons},\mathrm{mom}\}}
 \left[\max \left(
 \frac{\mathcal L_j}{s_j}-\kappa_m,0\right)\right]^2,
 \label{eq:cavity-normalized-physics}
\end{equation}
where $m\in\{\mathrm{FM},\mathrm{diff}\}$ identifies the frozen generator
family, and $(\kappa_m,w_{\mathrm{inv}}^{(m)})=(1.05,0.01)$ for FM and
$(1.02,0.05)$ for diffusion. Thus, conservation and moment consistency are
protected by one-sided guards rather than minimized without limit.
The floors $(s_{\mathrm{BGK},\min},s_{\mathrm{cons},\min},
s_{\mathrm{mom},\min},s_{\mathrm{bc},\min})$ are
$(10^{-3},10^{-8},10^{-8},10^{-6})$.

The DVM anchor loss is the equal mixture of losses evaluated with optimized
lookup latents and generator-produced latents. Each constituent is the sum of
a standardized primitive-field mean-square error and the two
quadrature-weighted relative $L_1$ errors of $G$ and $B$. At continuously
sampled Knudsen numbers, a frozen-decoder teacher penalizes changes in the
primitive outputs, $\phi_G,\phi_B$, and the actual $G/B$ distributions
under the same generated latent. The total objective is
\begin{equation}
 \mathcal L_{\mathrm{cav,adapt}}^{(m)}
 =
 2\times10^{-3}r(k)\widetilde{\mathcal L}_{\mathrm{phys}}^{(m)}
 +2\mathcal L_{\mathrm{DVM}}^{\mathrm{mix}}
 +4\mathcal L_{\mathrm{teacher}}
 +\mathcal L_{\mathrm{reg}},
 \label{eq:cavity-adapter-total}
\end{equation}
where $k$ is the optimizer-update index and $r(k)=\min(k/400,1)$. The
regularizer $\mathcal L_{\mathrm{reg}}$
contains squared hinges beyond trust radii 0.05 for the macroscopic RMS,
relative coefficient RMS, and bias RMS, plus $10^{-6}$ times the mean
squared composed weight $W_{\mathrm{up}}W_{\mathrm{down}}$. The FM and diffusion adapters
use ranks 8 and 16, respectively, reflecting their separately frozen
model-selection protocols.

For reporting and checkpoint confirmation, the frozen and adapted models are
evaluated on the same $N_{\mathrm{pr}}$ independently seeded probe batches.  Let
$Q_{p,\nu}^{(m)}$ be \cref{eq:cavity-normalized-physics} evaluated for
$\nu\in\{\mathrm{fr},\mathrm{ad}\}$ on probe $p$, using the same detached
scales.  The quantities reported in \cref{Sec4} are
\begin{equation}
 \begin{aligned}
 R_{\mathrm{phys}}^{(m)}
 &=1-\frac{N_{\mathrm{pr}}^{-1}\sum_pQ_{p,\mathrm{ad}}^{(m)}}
 {\max(N_{\mathrm{pr}}^{-1}\sum_pQ_{p,\mathrm{fr}}^{(m)},10^{-12})},\\
 R_{\mathrm{BGK}}
 &=1-\frac{\overline{\mathcal L}_{\mathrm{BGK}}^{\mathrm{ad}}}
 {\max(\overline{\mathcal L}_{\mathrm{BGK}}^{\mathrm{fr}},10^{-12})},
 \qquad
 \rho_j=\frac{\overline{\mathcal L}_{j}^{\mathrm{ad}}}
 {\max(\overline{\mathcal L}_{j}^{\mathrm{fr}},10^{-12})},
 \end{aligned}
 \label{eq:cavity-physics-audit}
\end{equation}
where an overbar denotes the mean over paired probes and
$j\in\{\mathrm{cons},\mathrm{mom}\}$.  A ``worst-probe'' reduction is the
minimum paired-probe reduction, whereas a worst-probe ratio is the maximum
paired-probe ratio.  These frozen-relative quantities are audit statistics,
not additional optimization losses.

A reduction in \cref{eq:cavity-bgk-diagnostic} is interpreted only as improved
consistency under the matched variables, quadrature, and differentiation rule.
It neither proves satisfaction of the exact continuous BGK equation nor
implies simultaneous improvement of every moment and boundary diagnostic.

\subsubsection{Cylinder macroscopic adaptation}

For the cylinder, the $\bm h$ supplied to
\cref{eq:generic-low-rank-adapter} concatenates the final frozen trunk feature
in $\mathbb R^{512}$ and generated lookup latent
$\bm z\in\mathbb R^{32}$. A rank-16 branch corrects the four standardized
macroscopic outputs, while a scalar branch supplies a localized normal-velocity
correction:
\begin{equation}
 \Delta\bm u_n
 =
 \exp[-(d_w/D)/0.20] 
 \frac{\alpha_a}{r_a}(W_n\bm h_a+b_n)\bm n,
 \quad r_a=\alpha_a=16 ,
 \label{eq:cylinder-wall-adapter}
\end{equation}
where $W_n\in\mathbb R^{1\times r_a}$, $b_n\in\mathbb R$, and, because
the cylinder is centered at the origin,
$\bm n=(x,y)/\sqrt{x^2+y^2}$ is the outward unit normal. The retained
protocol also uses
\begin{equation}
 \bm u^{\mathrm{hard}}
 =
 \bm u_{\mathrm{tan}}+
 \tanh\left[d_w/(0.03D)\right]\bm u_n,
 \label{eq:cylinder-hard-wall}
\end{equation}
which enforces zero normal velocity at the cylinder surface without imposing
no-slip. Here $\bm u_n=(\bm u \cdot \bm n)\bm n$ and
$\bm u_{\mathrm{tan}}=\bm u-\bm u_n$ are the normal and tangential velocity
components. This exact wall behavior is structural and must be distinguished
from a residual reduction learned by the adapter. Each generator-specific
cylinder adapter contains 9,893 trainable parameters.

Only constraints closed by the four available DSMC fields are used:
wall no-penetration, upstream inlet consistency, local weak control-volume (CV)
mass balance, and global boundary mass balance. For a fluid-only control volume
$V$,
\begin{equation}
 R_V=
 \left(
\displaystyle\oint_{\partial V}
  n\bm u\cdot\bm n_V dS
 \right) / \left( 
\displaystyle\oint_{\partial V}
 n|\bm u\cdot\bm n_V| dS+\epsilon 
 \right) ,
 \quad
 \mathcal L_{\mathrm{CV}}=\langle R_V^2\rangle,
 \label{eq:cylinder-control-volume}
\end{equation}
where $\bm n_V$ is the outward unit normal to $\partial V$, and
$\epsilon>0$ prevents division by zero. An analogous normalized flux is
evaluated on the outer domain boundary. For a
sampled mini-batch, define
\begin{equation}
 \mathcal L_{\mathrm{adapt,data}}
 =\frac{1}{N_c}\sum_{i=1}^{N_c}e_i+0.25\max_i e_i,
 \quad
 \mathcal L_{\mathrm{anchor}}
 =\left\langle\|\Delta\bm y^{\mathrm{std}}\|_2^2\right\rangle,
 \label{eq:cylinder-adapter-data}
\end{equation}
where $e_i$ is the mean-square standardized four-field error for case
$i$, $N_c$ is the number of sampled cases, and
$\Delta\bm y^{\mathrm{std}}$ is the complete adapter correction relative to
the frozen decoder. The trust loss is the squared positive part of
the ratio between correction RMS and frozen-field RMS plus 0.10, beyond a
radius of 0.05; $\mathcal L_{\mathrm{param}}$ is the mean squared magnitude
of the trainable adapter parameters. The complete objective is
\begin{equation}
 \begin{aligned}
 \mathcal L_{\mathrm{cyl,adapt}}={}&
 \mathcal L_{\mathrm{adapt,data}}
 +0.20\mathcal L_{\mathrm{anchor}}
 +0.20\mathcal L_{\mathrm{trust}}
 +10^{-8}\mathcal L_{\mathrm{param}}\\
 &+5\times10^{-4}r(k)
 \left(0.25\widetilde{\mathcal L}_{\mathrm{wall}}
 +0.10\widetilde{\mathcal L}_{\mathrm{inlet}}
 +0.45\widetilde{\mathcal L}_{\mathrm{CV}}
 +0.20\widetilde{\mathcal L}_{\mathrm{global}}\right).
 \end{aligned}
 \label{eq:cylinder-adapter-total}
\end{equation}
For each raw physical loss $\mathcal L_j$, the detached normalization scale
is initialized by its first observation and subsequently updated as
\begin{equation}
 s_j^{(k)}=\max \left\{
 0.99s_j^{(k-1)}+0.01\operatorname{stopgrad}
 (\mathcal L_j^{(k)}),s_{j,\min}\right\},
 \quad
 \widetilde{\mathcal L}_j=\mathcal L_j/s_j^{(k)}.
 \label{eq:cylinder-adapter-ema}
\end{equation}
The component floors for wall, inlet, local CV, and global mass losses are
$(10^{-6},10^{-6},10^{-7},10^{-7})$, respectively.
The half-cosine ramp $r(k)$ is zero for the first 10\% of training and
reaches one at 30\%. The FM adapter is exposed to optimized lookup latents;
the diffusion adapter draws lookup and diffusion-generated latents with equal
probability.

The scalar used for checkpoint selection and post-training reporting is a
separate frozen-relative audit statistic, not the training term normalized by
the exponential-moving-average (EMA) loss scales in
\cref{eq:cylinder-adapter-total}.  Let
$\mathcal L_j^{\mathrm{fr}}$ and $\mathcal L_j^{\mathrm{ad}}$ denote the raw
loss of component $j$ evaluated for the frozen and adapted predictions under
the same audit samples and quadrature.  The component ratio and aggregate
physics score are
\begin{equation}
 \rho_j=\frac{\mathcal L_j^{\mathrm{ad}}}
 {\max(\mathcal L_j^{\mathrm{fr}},10^{-12})},
 \quad
 S_{\mathrm{phys}}= \displaystyle\sum_j w_j\rho_j / \displaystyle\sum_j w_j.
 \label{eq:cylinder-physics-score}
\end{equation}
The weights $(w_{\mathrm{wall}},w_{\mathrm{inlet}},w_{\mathrm{CV}},
w_{\mathrm{global}})=(0.25,0.10,0.45,0.20)$ match those in
\cref{eq:cylinder-adapter-total}.
Thus, $S_{\mathrm{phys}}=1$ indicates no aggregate change from the paired
frozen generator, and $S_{\mathrm{phys}}<1$ indicates a weighted reduction.
Because each generator has its own frozen denominator, this score measures
relative improvement within a generator and is not an absolute residual for
ranking FM against diffusion.

Momentum and total-energy residuals are not evaluated because the DSMC
targets do not contain the pressure tensor, viscous stress, or heat flux
required to close them. Complete BGK or Boltzmann residuals, Navier--Stokes
constitutive terms, no-slip, gas-temperature-equals-wall-temperature, strong
shock continuity, and symmetry penalties are likewise disabled.

\subsubsection{Feasibility and checkpoint gates}

A physics-adapted checkpoint is eligible only if it preserves the frozen
decoded-field accuracy within the stated tolerance. Among eligible
checkpoints, a normalized physics score ranks the correction. Cavity selection
also guards kinetic accuracy, teacher consistency, collision invariants,
moments, and the wall diagnostic. Cylinder selection constrains validation
mean error to remain within a fixed feasibility margin of its frozen value;
the retained diffusion protocol additionally applies the same margin to the
worst case. A
positive-step checkpoint is required, so a zero-initialized model cannot be
reported as a trained physical improvement. These are multi-objective
feasibility rules, not a claim that data and physics losses form a single
universal scalar metric.

\subsection{Training and evaluation protocols}
\label{sec:training-evaluation-protocols}

\subsubsection{Generator optimization settings}

The retained conditional generators are summarized in
\cref{tab:generator-training-settings}. AdamW \citep{LoshchilovHutter2019},
cosine learning-rate decay, gradient clipping at one, and an EMA of model
weights are used in all generator fits, but the numerical precision and
deployment solvers are model specific.

\begin{table}[htbp]
 \centering
 \caption{Frozen conditional-generator configurations. Ensemble counts refer
 to the validation protocol; larger counts are used for final cavity field
 evaluation.}
 \label{tab:generator-training-settings}
 \footnotesize
 \begin{tabular}{lcccc}
  \toprule
  Setting & Cavity FM & Cavity diffusion & Cylinder FM & Cylinder diffusion\\
  \midrule
  Chart dimension & 4 & 4 & 12 & 12\\
  Parameters & 2.997 M & 2.997 M & 0.208 M & 0.203 M\\
  Prediction/path & Trigonometric FM & VP $v$-prediction & Quadratic FM & VP $x$-prediction\\
  Peak learning rate & $3 \times 10^{-4}$ & $3 \times 10^{-4}$
   & $2 \times 10^{-4}$ & $2 \times 10^{-4}$\\
  Batch size & 4096 & 4096 & 512 & 512\\
  Training updates & 30000 & 8000 & 8000 & 12000\\
  Precision & BF16/FP32 & FP32 & FP32 & FP32\\
  Solver & Heun-48 & DPM-Solver++(2M)-35 & Heun-24 & DDIM-40\\
  Ensemble/anchors & 64 & 64 & 3 anchors & 40\\
  \bottomrule
 \end{tabular}
\end{table}

In the cavity FM fit, BF16 is used for the principal regression and FP32 for
the curvature term; all other entries use FP32 throughout. A notation such as
Heun-48 or DDIM-40 specifies the solver followed by its number of integration
steps.

\subsubsection{Ablation and checkpoint selection}

Generator ablations change one principal modeling choice at a time and rank
checkpoints by held-out decoded-field error rather than by training loss. The
cavity FM study compares the trigonometric FiLM reference, a rectified path,
the capacity-matched plain-concatenation network, a smaller network, removal of
Knudsen Fourier features, and removal of the Knudsen-curvature term. Separate
follow-up controls test velocity-field ensembling and differentiable endpoint
mean/variance regularization. The cavity diffusion study first compares
$x$-, $\epsilon$-, and $v$-prediction, BF16 and FP32 arithmetic,
plain-concatenation and Knudsen-FiLM conditioning, and endpoint
regularization. After fixing the selected denoiser, it scans
$\beta_{\max}$ and the DDPM, DDIM, and DPM-Solver++ sampler families,
including the deployment step count.

For the cylinder, the initial six-way comparison varies the FM path,
plain versus FiLM conditioning, and the three diffusion prediction targets.
The subsequent FM study holds the local four-neighbor training graph fixed and
then tests, in sequence, the rank-eight quadratic condition correction,
simplex rather than distance aggregation, decoder-metric rather than ordinary
PCA, decoded-field endpoint supervision, multi-source consensus, and optional
composition/loop terms. This staged design separates improvements in latent
geometry from changes in the transport network and numerical aggregation.
Strict leave-one-condition-out refits are reserved for the selected candidates;
in each fold, all latent-chart and graph quantities are reconstructed without
the omitted condition.

Physics-adapter ablations are performed only after the corresponding frozen
generator is selected. The cavity study varies adapter rank and the normalized
BGK, boundary, conservation, moment, anchor, and teacher components. The
cylinder study compares a data-only adapter, wall/inlet terms, mass-balance
terms, ranks eight and 16, their combined objective, and the optional hard
wall-normal projection. Eligibility is always determined by the feasibility
gates in \cref{sec:structured-physics-adaptation}; a smaller physics residual
cannot compensate for violation of the decoded-field accuracy guard.
These experiments provide component-removal and coarse-strength evidence, but
they are not an exhaustive continuous search over every scalar loss weight and
ramp location. The reported weights should therefore be interpreted as one
validated feasible setting rather than a uniquely identified optimum.

\subsubsection{Physics-adapter training settings}

The structured-adapter protocols are summarized in
\cref{tab:adapter-training-settings}. All listed runs begin from a new
zero-output adapter and do not resume an earlier adapter checkpoint.

\begin{table}[htbp]
 \centering
 \caption{Physics-adapter configurations. The two cylinder columns share the
 same corrected outputs and physics definitions but differ in latent exposure
 and per-update sampling.}
 \label{tab:adapter-training-settings}
 \footnotesize
 \begin{tabular}{lcccc}
  \toprule
  Setting & Cavity FM & Cavity diffusion & Cylinder FM & Cylinder diffusion\\
  \midrule
  Adapter rank & 8 & 16 & 16 & 16\\
  Trainable parameters & 41,685 & 75,525 & 9,893 & 9,893\\
  Training updates & 1500 & 1500 & 12000 & 12000\\
  Maximum branch learning rate & $5 \times 10^{-5}$ & $5 \times 10^{-5}$
   & $2 \times 10^{-4}$ & $2 \times 10^{-4}$\\
  Latent exposure & Lookup/FM, 1:1 & Lookup/diffusion, 1:1 & Lookup only & Lookup/diffusion, 1:1\\
  Conditions or cases/update & 32 & 32 & 8 & 16\\
  Field-anchor points/case & 384 & 384 & 4096 & 8192\\
  Interior collocation points & 512 & 512 & -- & --\\
  Boundary quadrature count & 96/wall & 96/wall & 512 & 1024\\
  Velocity nodes (physics/anchor) & 6400/4096 & 6400/4096 & -- & --\\
  CVs $\times$ face points & -- & -- & $12\times48$ & $24\times64$\\
  \bottomrule
 \end{tabular}
\end{table}

Here ``lookup'' denotes the optimized representation-stage code for an
observed training case; it is not an Oracle code fitted to a held-out case.

\subsubsection{Data separation and reproducibility}

Architecture and checkpoint selection use only the development partitions
defined in \cref{Sec2}. Additional out-of-sample cases do not affect normalization,
latent-chart fitting, optimizer updates, or checkpoint ranking. During
leave-one-condition-out studies, the omitted lookup code is also excluded from
PCA or decoder-metric fitting, score normalization, graph construction, and
training pairs. All aggregate errors first assign equal weight to each
condition as in \cref{eq:equal-case-mean}.

Reproducibility artifacts store the resolved configuration, random seed,
normalization and chart state, selected model weights, upstream checkpoint
paths, and cryptographic hashes. Field evaluation reloads the frozen artifacts
rather than reconstructing architectures from manually transcribed
hyperparameters. The reported ensemble size, numerical solver, and checkpoint
must therefore be treated as part of each frozen model definition rather than
as interchangeable post-processing choices. Training wall times and a
controlled single-GPU inference benchmark are reported in
\cref{app:computational-cost}; because the distributed fields include no solver job
records, no solver-to-surrogate acceleration ratio is reported. A
20-sampling-seed audit of the frozen cavity inference (\cref{app:cavity-sampling-seeds}) shows
sampling variability far below the method differences reported below; it does
not quantify retraining uncertainty, which the five-seed refits of
\cref{sec:deterministic-decoded-comparison} address separately.

\section{Model selection and controlled ablation studies}\label{Sec4}

This section fixes the representation, generator, and adapter configurations
used in the out-of-sample study.  All choices are made on the development
partitions defined in \cref{Sec2} or by strict leave-one-condition-out (LOCO)
refitting of the training set.  The additional cases used in \cref{Sec5} are not
used for the comparisons below. The data partitions, reporting metrics, and
equal-case aggregation were fixed before final checkpoint scoring. The
architecture candidates were developed iteratively on the
development data and should not be interpreted as an exhaustively
preregistered search space. Retention therefore follows the stated
multi-criterion protocol---decoded-field accuracy, endpoint stability, and
gate feasibility---rather than selecting the minimum entry of each table.
Cavity results are reported using the
relative errors in \cref{eq:cavity-macro-error,eq:cavity-kinetic-errors},
whereas cylinder model selection uses the area-weighted standardized RMSE of
the four decoder outputs.  For condition $i$, define
\begin{equation}
 \begin{aligned}
 E_{i,\mathrm{std}}
 &=\left[
\displaystyle\sum_p a_{i,p}
 \left\|\widehat{\bm y}_{i,p}^{\mathrm{std}}
 -\bm y_{i,p}^{\mathrm{std}}\right\|_2^2
 /( \displaystyle4\sum_p a_{i,p} )
 \right]^{1/2},\\
 \overline E_{\mathrm{std}}&=\frac{1}{N}\sum_iE_{i,\mathrm{std}},
 \qquad
 E_{\mathrm{std}}^{\max}=\max_iE_{i,\mathrm{std}}.
 \end{aligned}
 \label{eq:cylinder-selection-rmse}
\end{equation}
where $p$ indexes spatial samples, $a_{i,p}$ is the spatial quadrature weight,
and $\bm y^{\mathrm{std}}$ contains the four standardized disturbances in
\cref{eq:cylinder-targets}.  This selection metric differs from the
dimensionless per-field RMSE in \cref{eq:cylinder-field-rmse}, which is used
for physical-field reporting.  Neither metric is numerically comparable with
the cavity relative $L_1$ errors.  Throughout this section, boldface in
ablation and confirmation tables denotes a configuration retained for later
evaluation, not necessarily the minimum in every reported column.

An \textit{Oracle} result freezes all decoder parameters and optimizes only one
latent code directly against the complete held-out reference field.  It is a
representation-floor diagnostic, not a deployable conditional prediction;
the converged cavity and cylinder values below use 200 and 6,000 latent-only
updates, respectively.  In a LOCO fold, condition $i$ and its lookup code are
removed from every fitted chart and conditional model before predicting that
condition.  For any per-condition error $E_i^{(-i)}$, the reported aggregates
are
\begin{equation}
 E_{\mathrm{LOCO}}^{\mathrm{mean}}
 =\frac{1}{N_{\mathrm{fold}}}\sum_i E_i^{(-i)},
 \qquad
 E_{\mathrm{LOCO}}^{\mathrm{worst}}=\max_i E_i^{(-i)}.
 \label{eq:loco-aggregation}
\end{equation}
The validation mean and worst columns use the same mean and maximum operations
on the fixed validation partition, without refitting.  Unless stated
otherwise, a reported reduction of a nonnegative diagnostic $A$ is
$1-A_{\mathrm{new}}/A_{\mathrm{ref}}$, and a signed percentage change is
$A_{\mathrm{new}}/A_{\mathrm{ref}}-1$.

For $M$ generated chart endpoints $\bm a_i^{(m)}\in\mathbb R^r$ at condition
$i$, define
\begin{equation}
 \bar a_{ik}=\frac{1}{M}\sum_{m=1}^{M}a_{ik}^{(m)},
 \qquad
 s_i=\left[
 \frac{1}{rM}\sum_{k=1}^{r}\sum_{m=1}^{M}
 (a_{ik}^{(m)}-\bar a_{ik})^2
 \right]^{1/2}.
 \label{eq:latent-spread}
\end{equation}
The ``spread range'' is $\min_i s_i$--$\max_i s_i$ over the stated cases.
``Training spread'' is the root mean square of the componentwise population
standard deviations over all observed training conditions and chart
coordinates, equivalently the aggregate form of \cref{eq:latent-spread}.
Both are evaluated in the transformed chart and diagnose numerical endpoint
dispersion rather than physical uncertainty.

\subsection{Representation fidelity and latent-chart dimension}
\label{sec:ablation-representation}

The cavity auto-decoder attains an Oracle validation mean of
$E_{\mathrm{macro}}=4.308\times10^{-5}$ and
$5.138\times10^{-5}$ for $E_{\mathrm{kin}}$.  In contrast, direct linear
interpolation of the 24-dimensional lookup codes in $\log\mathrm{Kn}$ gives
$E_{\mathrm{kin}}=8.890\times10^{-5}$.  The frozen decoder can therefore
represent the held-out solutions more accurately than a simple condition-space
interpolation can locate them.  The gap between these two errors is attributable
to latent inference under a fixed decoder, although the nonzero Oracle error
remains part of the total prediction error.

The retained cavity chart in \cref{eq:cavity-latent-chart} is selected by
decoding each truncated latent rather than by explained variance alone.  If
$\bm z_i^{(r)}$ is the inverse-transformed rank-$r$ reconstruction of lookup
code $\bm z_i$, define
\begin{equation}
 \Delta E_{i,\mathrm{kin}}^{(r)}=
 E_{\mathrm{kin}}[D(\bm z_i^{(r)})]
 -E_{\mathrm{kin}}[D(\bm z_i)].
 \label{eq:pca-decoded-increment}
\end{equation}
The mean and worst increments in \cref{tab:cavity-pca-ablation} are the mean
and maximum of $\Delta E_{i,\mathrm{kin}}^{(r)}$ over the training cases.  For
each rank, the LOCO column comes from the same fixed-capacity conditional MLP
refitted after excluding the held-out lookup code; this MLP is used only as a
train-only chart-selection probe and is not one of the deployed generators.
As shown in
\cref{tab:cavity-pca-ablation}, rank four reduces the worst truncation-induced
kinetic increment to $3.02\times10^{-7}$ and gives the smallest decoded LOCO
error.  Rank eight has a slightly smaller reconstruction increment but a
larger LOCO error, demonstrating that additional weak latent directions can
make conditional learning harder without materially improving field
representation.  The small negative mean increment at rank four is a
finite-precision cancellation effect.

Because the chart is estimated from only 16 lookup codes, a
leave-one-condition-out principal-angle audit (\cref{app:loco-chart-stability})
additionally shows that the dominant four-dimensional span is stable
for most omissions (mean largest angle $2.10^{\circ}$) but not at the
training-support boundary (maximum $11.12^{\circ}$), limiting claims about an
intrinsic four-dimensional physical manifold.

\begin{table}[htbp]
 \centering
 \caption{Cavity latent-chart ablation under decoded LOCO evaluation.}
 \label{tab:cavity-pca-ablation}
 \footnotesize
 \begin{tabular}{ccccc}
  \toprule
  PCA rank & Explained variance & Mean increment & Worst increment & LOCO $E_{\mathrm{kin}}$\\
  \midrule
  2 & 99.5975\% & $1.294 \times 10^{-5}$ & $3.250 \times 10^{-5}$ & $1.158 \times 10^{-4}$\\
  \textbf{4} & \textbf{99.9958\%} & \textbf{$-8.292 \times 10^{-9}$} & \textbf{$3.022 \times 10^{-7}$} & \textbf{$9.361 \times 10^{-5}$}\\
  8 & 99.9999\% & $4.426 \times 10^{-9}$ & $2.036 \times 10^{-7}$ & $1.152 \times 10^{-4}$\\
  \bottomrule
 \end{tabular}
\end{table}

For the cylinder, converged frozen-decoder Oracle optimization gives final
mean and worst values of $E_{i,\mathrm{std}}$ equal to 0.01521 and 0.02083,
respectively; restricting the same metric to cells with
$0\le d_w/D<0.20$ gives a mean near-wall value of 0.05370.  A common chart dimension
of 12 is fixed before generator training by complete train-only LOCO
decoded-field screening.  The ordinary unwhitened PCA-12 chart used by the
diffusion baseline retains 95.51\% of the training-latent variance.  The
retained FM instead uses the separate decoder-metric PCA-12 chart in
\cref{eq:decoder-latent-metric}; its dimension is the same, but its metric and
coordinate scaling are not.  Whitening is rejected because it gives weak,
poorly predictable latent directions the same regression weight as dominant
directions.  Condition Fourier features are rejected separately because they
reduce training error while degrading omitted-condition field prediction.
The nonzero Oracle floor, especially near the cylinder, is retained in all
subsequent error accounting rather than attributed to the conditional
generator.

\subsection{Flow-matching ablations}
\label{sec:ablation-fm}

\subsubsection{Cavity: conditioning and probability path}

The six cavity FM candidates use the path and objective in
\cref{eq:cavity-fm-path,eq:cavity-fm-total}, the same rank-four chart, training
cases, 30,000-update budget, and decoded-field evaluation. The FiLM reference
and the plain-concatenation candidate contain 2.970 and 2.997 million
parameters, respectively, so their comparison is effectively capacity
matched.  \Cref{tab:cavity-fm-ablation} shows that the simpler plain
concatenation of the state, time, and Knudsen features is substantially more
accurate than repeated FiLM conditioning.  Removing Knudsen Fourier features
also increases endpoint spread, while the original curvature penalty is not
beneficial for the FiLM reference.  The trigonometric FiLM reference has
width 256 and six residual blocks.  The candidate labeled linear optimal
transport (linear-OT) replaces only its
probability path by $\bm a_t=(1-t)\bm a_0+t\bm a_1$; the reduced FiLM uses
width 128 and three blocks.  The remaining two controls remove, respectively,
the Knudsen Fourier lifting and \cref{eq:cavity-fm-curvature}.

\begin{table}[htbp]
 \centering
 \caption{Cavity FM architecture ablation on three held-out development
 Knudsen numbers; spread is measured in the transformed rank-four chart.}
 \label{tab:cavity-fm-ablation}
 \footnotesize
 \begin{tabular}{lccc}
  \toprule
  Candidate & Mean $E_{\mathrm{kin}}$ & Maximum $E_{\mathrm{kin}}$ & Spread range\\
  \midrule
  Trigonometric FiLM & $3.583 \times 10^{-4}$ & $7.006 \times 10^{-4}$ & 0.0175--0.0761\\
  Linear-OT FiLM & $4.642 \times 10^{-4}$ & $1.155 \times 10^{-3}$ & 0.0109--0.0667\\
  \textbf{Plain concatenation} & \textbf{$1.046 \times 10^{-4}$} & \textbf{$1.153 \times 10^{-4}$} & \textbf{0.0145--0.0210}\\
  Reduced FiLM (width 128, 3 blocks) & $4.531 \times 10^{-4}$ & $9.337 \times 10^{-4}$ & 0.0158--0.0831\\
  No Knudsen Fourier features & $3.292 \times 10^{-4}$ & $6.808 \times 10^{-4}$ & 0.0515--0.1507\\
  No curvature penalty & $2.092 \times 10^{-4}$ & $3.362 \times 10^{-4}$ & 0.0107--0.0335\\
  \bottomrule
 \end{tabular}
\end{table}

Plain concatenation lowers the mean error by 70.8\% relative to the
trigonometric FiLM reference. A plausible mechanism is that repeated
feature-wise modulation gives the sparsely sampled condition signal multiple
opportunities to rescale hidden activations, so small condition-fitting errors
can be amplified across residual blocks. Plain concatenation instead exposes
the state, time, and condition features directly to one smooth vector field.
This is an interpretation of the controlled ablation, not a general claim
that FiLM is inferior in larger or more densely sampled data sets. Increasing
the antithetic ensemble from 64 to
4096 changes its mean only from $1.046\times10^{-4}$ to
$1.027\times10^{-4}$, so the residual is not primarily Monte Carlo error.
A four-member velocity-field ensemble instead gives
$1.234\times10^{-4}$; this control trains four independently initialized
vector fields and averages their velocities at every ODE evaluation.
Differentiable endpoint mean matching and variance contraction, respectively
penalizing $\|\bar{\bm a}_i-\bm a_i\|_2^2$ and excess component variance above
$10^{-6}$, also increase the mean to $1.113\times10^{-4}$; they correct
statistics at observed Knudsen numbers but do not supply the missing mean at
an unobserved condition.  These controls support freezing the single plain
concatenation FM without endpoint regularization.

\subsubsection{Cylinder: decoder-aware curved transport}

The cylinder study uses strict LOCO refits because ordinary validation alone
can favor a model that interpolates the particular split.  Starting from the
local source-to-target FM in
\cref{eq:cylinder-quadratic-transport,eq:cylinder-fm-total}, the ablation
successively adds a rank-eight
quadratic condition correction, simplex aggregation, decoder-metric latent
coordinates, decoded-field supervision, source consensus, and loop
regularization.  The selected model is the simplest candidate containing the
decoder metric, field term, and consensus term; the loop penalty is rejected.
Here distance aggregation uses the nearest anchors with positive
distance-softmax weights, whereas simplex aggregation uses the barycentric
weights of a local enclosing triangle when one exists.  The loop control adds
$\mathbb E\|\mathcal T(\mathcal T(\bm a_a,\bm c_a,\bm c_b),
\bm c_b,\bm c_a)-\bm a_a\|_2^2/d_a$ to penalize round-trip inconsistency.

\begin{table}[htbp]
 \centering
 \caption{Cylinder curved-transport ablation under the LOCO criterion in
 \cref{eq:cylinder-selection-rmse}.}
 \label{tab:cylinder-fm-ablation}
 \footnotesize
 \begin{tabular}{lcccc}
  \toprule
  Candidate & LOCO mean & LOCO worst & Validation mean & Validation worst\\
  \midrule
  Quadratic PCA, distance anchors & 0.05835 & 0.27137 & 0.03584 & 0.13565\\
  Quadratic PCA, simplex & 0.05608 & 0.26740 & 0.03825 & 0.15473\\
  Quadratic metric, simplex & 0.05301 & 0.22635 & 0.04003 & 0.18258\\
  Metric, field, consensus, loop & 0.05222 & 0.20741 & 0.04054 & 0.18709\\
  \textbf{Metric, field, consensus} & \textbf{0.05114} & \textbf{0.20322} & \textbf{0.04095} & \textbf{0.18766}\\
  \bottomrule
 \end{tabular}
\end{table}

Relative to the best preceding first-order transport, the selected curved
model reduces the LOCO mean and worst errors by 16.66\% and 22.22\%.  Its
largest remaining error occurs at the parameter-domain boundary case
$(\mathrm{Kn},\mathrm{Ma})=(0.06,2.0)$ and is concentrated in the shock and
near-wall regions.  The fact that simpler candidates have smaller ordinary
validation errors in \cref{tab:cylinder-fm-ablation} is not used to reverse
the selection: the architecture was fixed by the more stringent omitted-case
protocol.

\subsection{Diffusion ablations}
\label{sec:ablation-diffusion}

\subsubsection{Cavity: target, precision, schedule, and sampler}

The most consequential cavity diffusion choices are numerical precision and
the regression targets defined in \cref{eq:diffusion-targets}.  As summarized in
\cref{tab:cavity-diffusion-target-ablation}, changing only BF16 to FP32 lowers
the $v$-prediction mean by 37.5\%.  In FP32, $v$-prediction is markedly more
accurate than clean-latent ($x$) prediction or stabilized noise
($\epsilon$) prediction.  Native $\epsilon$-prediction is numerically
ill-conditioned near the high-noise endpoint and is not retained.
The stabilized $\epsilon$ control truncates diffusion time at
$t_{\max}=0.98$ and applies minimum signal-to-noise-ratio (SNR) weighting with
$\gamma=5$, i.e., weight
$\min(\mathrm{SNR},5)/\mathrm{SNR}$ for the $\epsilon$-prediction
mean-squared error (MSE).
The native control uses the unweighted objective and $t_{\max}=0.999$.

\begin{table}[htbp]
 \centering
 \caption{Cavity diffusion precision and prediction-target ablation with a
 common decoder and rank-four chart.  Training spread is the RMS endpoint
 spread in the transformed chart.}
 \label{tab:cavity-diffusion-target-ablation}
 \footnotesize
 \begin{tabular}{lccc}
  \toprule
  Target and arithmetic & Mean $E_{\mathrm{kin}}$ & Maximum $E_{\mathrm{kin}}$ & Training spread\\
  \midrule
  $v$, BF16 & $1.0783 \times 10^{-4}$ & $1.1620 \times 10^{-4}$ & --\\
  \textbf{$v$, FP32} & \textbf{$6.7375 \times 10^{-5}$} & \textbf{$7.9126 \times 10^{-5}$} & \textbf{$3.939 \times 10^{-3}$}\\
  $x$, FP32 & $1.1921 \times 10^{-4}$ & $1.9277 \times 10^{-4}$ & $5.483 \times 10^{-5}$\\
  $\epsilon$, FP32, $t_{\max}=0.98$, Min-SNR & $1.2492 \times 10^{-4}$ & $2.5053 \times 10^{-4}$ & $1.238 \times 10^{-2}$\\
  $\epsilon$, FP32, unweighted & $1.6939 \times 10^{-2}$ & $4.9101 \times 10^{-2}$ & $1.055 \times 10^{1}$\\
  \bottomrule
 \end{tabular}
\end{table}

Endpoint mean and variance losses change the FP32 $v$-prediction mean from
$6.7375\times10^{-5}$ to $6.7642\times10^{-5}$ and
$6.8012\times10^{-5}$, respectively, and Knudsen-FiLM conditioning gives
$6.7804\times10^{-5}$.  None improves on plain input concatenation.  An
8,000-update model is also retained over 20,000 updates because the latter
reduces training spread but increases validation $E_{\mathrm{kin}}$ by
1.17\%.

After fixing FP32 $v$-prediction, the development scan of
\cref{eq:vp-forward-process} identifies
$20\leq\beta_{\max}\leq40$ as an empirical low-error range and selects
$(\beta_{\min},\beta_{\max})=(0.05,40)$.  Under the 8,000-update fit,
changing the cosine VP schedule
$\alpha(t)=\cos(\pi t/2)$, $\sigma(t)=\sin(\pi t/2)$ to the continuous
linear-$\beta$ VP schedule lowers the
DDIM-48 kinetic mean from $6.7375\times10^{-5}$ to
$6.5243\times10^{-5}$.  The subsequent sampler study is summarized in
\cref{tab:cavity-diffusion-sampler}.  DPM-Solver++(2M)-35 is frozen because
the primary selection metric is the mean error; the 44-step alternative is
retained only as a worst-case diagnostic.

\begin{table}[htbp]
 \centering
 \caption{Cavity diffusion sampler comparison over 20 paired random seeds.
 Uncertainty is the population standard deviation, and the last column is the
 seed--condition maximum.}
 \label{tab:cavity-diffusion-sampler}
 \footnotesize
 \begin{tabular}{lccc}
  \toprule
  Sampler & Steps & Mean $E_{\mathrm{kin}}$ & Maximum\\
  \midrule
  DDIM & 48 & $(6.5053 \pm 0.0377) \times 10^{-5}$ & $7.4290 \times 10^{-5}$\\
  \textbf{DPM-Solver++(2M)} & \textbf{35} & \textbf{$(5.9073 \pm 0.0268) \times 10^{-5}$} & \textbf{$6.7510 \times 10^{-5}$}\\
  DPM-Solver++(2M) & 44 & $(6.1714 \pm 0.0432) \times 10^{-5}$ & $6.3870 \times 10^{-5}$\\
  \bottomrule
 \end{tabular}
\end{table}

Both DPM-Solver++ configurations outperform DDIM in all 20 paired seeds.  The seed
variation measures numerical Monte Carlo stability of the ensemble mean; it
does not imply multiple physical steady states at a fixed Knudsen number.
The 37.5\% BF16-to-FP32 gap in
\cref{tab:cavity-diffusion-target-ablation} is also a practical result:
low-dimensional latent diffusion is inexpensive enough that full-precision
training is preferable here, and mixed precision should not be assumed benign
merely because the decoded fields are smooth.

\subsubsection{Cylinder: Gaussian-reference diffusion baseline}

The cylinder target ablation produces a different ordering.  With identical
rank-12 charts and 12 held-out training conditions, $x$-prediction gives the
smallest LOCO mean, $v$-prediction the smallest LOCO worst error, and
$\epsilon$-prediction is inferior on both metrics:

\begin{table}[htbp]
 \centering
 \caption{Cylinder Gaussian-reference diffusion target ablation in the
 train-only screen using the preregistered LOCO-mean criterion.}
 \label{tab:cylinder-diffusion-ablation}
 \footnotesize
 \begin{tabular}{lcccc}
  \toprule
  Target & LOCO mean & LOCO worst & Validation mean & Validation worst\\
  \midrule
  \textbf{$x$-prediction} & \textbf{0.15537} & \textbf{0.59472} & \textbf{0.05326} & \textbf{0.17129}\\
  $v$-prediction & 0.17196 & 0.48652 & 0.03809 & 0.09293\\
  $\epsilon$-prediction & 0.24199 & 0.63904 & 0.04441 & 0.11545\\
  \bottomrule
 \end{tabular}
\end{table}

The predeclared train-only mean criterion selects $x$-prediction without
consulting the validation ordering.  In the complete 48-fold confirmation it
obtains LOCO mean/worst errors of 0.07284/0.46326, compared with
0.03896/0.16429 for the contemporaneous trigonometric FM\@. This comparison
belongs to the initial Gaussian-reference screen; the subsequently retained
curved FM is selected separately in \cref{sec:ablation-fm}.  No further
diffusion schedule or sampler search is conducted because the candidate fails
the preregistered promotion criterion.  The retained cylinder diffusion
comparison model, linear VP with
$(\beta_{\min},\beta_{\max})=(0.1,20)$ and DDIM-40, serves as a controlled
generative comparison rather than the principal accuracy model.

\subsection{Physics-adapter ablations}
\label{sec:ablation-physics}

\subsubsection{Cavity reduced-BGK correction}

For cavity FM, the unnormalized physics objective reduces the matched-grid BGK
diagnostic by 47.4\% but increases kinetic error by 16.0\% and approximately
doubles the conservation and moment diagnostics.  The normalized,
manifold-anchored objective in
\cref{eq:cavity-normalized-physics,eq:cavity-adapter-total} instead retains
the frozen prediction accuracy while
reducing BGK by 28.65\%; its principal results are compared with the formal
diffusion adapter in \cref{tab:cavity-physics-ablation}.
Here the BGK and composite-physics reductions are
$R_{\mathrm{BGK}}$ and $R_{\mathrm{phys}}^{(m)}$ from
\cref{eq:cavity-physics-audit}; conservation and moment values are the
corresponding frozen-relative ratios $\rho_j$.

\begin{table}[htbp]
 \centering
 \caption{Cavity physics-adapter selection on held-out development cases.
 Reductions are relative to each frozen generator and are not compared across
 generator families.}
 \label{tab:cavity-physics-ablation}
 \footnotesize
 \begin{tabular}{lcccc}
  \toprule
  Model & Frozen $E_{\mathrm{kin}}$ & Adapted $E_{\mathrm{kin}}$ & BGK reduction & Physics reduction\\
  \midrule
  FM, unnormalized objective & $5.659 \times 10^{-5}$ & $6.566 \times 10^{-5}$ & 47.4\% & --\\
  \textbf{FM, normalized manifold anchor} & \textbf{$5.659 \times 10^{-5}$} & \textbf{$5.652 \times 10^{-5}$} & \textbf{28.65\%} & \textbf{22.06\%}\\
  \textbf{Diffusion, rank 16} & \textbf{$5.846 \times 10^{-5}$} & \textbf{$5.606 \times 10^{-5}$} & \textbf{23.59\%} & \textbf{19.35\%}\\
  \bottomrule
 \end{tabular}
\end{table}

The normalized FM adapter changes the kinetic mean by $-0.12$\%, but its
independent-probe
conservation and moment ratios are both 1.0574. The checkpoint is therefore retained as Physics-constrained
FM. This terminology denotes a reduction of
the registered composite and BGK diagnostics subject to bounded invariant
degradation; it does not mean that the invariant diagnostics themselves
improve. A subsequent
positivity-preserving
hard-moment projection reduces its internal moment defect below
$7.25\times10^{-8}$, where the defect is the RMS relative mismatch between
the projected kinetic moments and their macroscopic-head targets.  It also
increases kinetic error and the pressure-tensor error in
\cref{eq:cavity-derived-error} by
4.08\% and 30.10\%, respectively; this hard-projection variant fails the joint
gates and is rejected.

For diffusion, the controlled rank/weight/anchor ablation selects rank 16.
During selection it changes $E_{\mathrm{kin}}$ from
$5.896\times10^{-5}$ to $5.500\times10^{-5}$, while the mean/worst
composite-physics reductions are 19.21\%/15.42\%.  Rank four produces a
larger 23.29\% mean physics reduction but increases kinetic error by 1.49\%;
rank eight gives a 21.08\% reduction and a 3.17\% kinetic improvement.
The independent-seed rank-16 confirmation in
\cref{tab:cavity-physics-ablation} passes every registered gate, including
worst-probe conservation and moment ratios of 0.9331.  Removing the invariant
guards or using physics weights of $10^{-3}$ and $4\times10^{-3}$ fails the
joint physical-reduction gates, confirming that physical loss magnitude alone
is not a sufficient checkpoint criterion. A robustness audit (\cref{sec:adapter-robustness})
repeats this adapter with two additional training seeds
and with the physics total weight scaled by $0.5\times$ and $2\times$ under a
common sampling protocol: the seed variants span BGK reductions of
24.5--32.6\% with invariant ratios 1.027--1.069, the weight variants yield
smaller reductions of 10.2--14.2\% with invariant ratios of 1.014--1.045, and
field accuracy is preserved in every run, so the adapter conclusion is stable
to seed and weight perturbations.

An independent frozen-checkpoint audit of the differentiation step and
query-grid refinement (\cref{sec:bgk-numerical-sensitivity}) shows that the BGK
reduction is stable to the difference step (0.223 and 0.104 percentage-point
spans for FM and diffusion) while its magnitude depends on the spatial
sampling measure; the near-wall-biased values in
\cref{tab:cavity-physics-ablation} should not be read as uniform-area
residual reductions.

\subsubsection{Cylinder boundary and weak-mass constraints}

The cylinder screen isolates adapter capacity from the individual evaluable
physics terms in \cref{eq:cylinder-hard-wall,eq:cylinder-control-volume}.
Its complete result is given in
\cref{tab:cylinder-adapter-screen}.  The mass-only candidate returns to the
zero-output checkpoint because its trained checkpoints do not improve the
independent composite audit.  Rank eight and rank 16 are nearly tied without
the hard map; the rank-16 hard-wall candidate is retained because it gives the
lowest feasible physics score without consuming the field-error budget.

\begin{table}[htbp]
 \centering
 \caption{Cylinder FM-adapter screen using the frozen-relative physics score
 in \cref{eq:cylinder-physics-score}.}
 \label{tab:cylinder-adapter-screen}
 \footnotesize
 \begin{tabular}{lccc}
  \toprule
  Candidate & $\overline E_{\mathrm{std}}$ & $E_{\mathrm{std}}^{\max}$ & Physics score\\
  \midrule
  Frozen/data-only control & 0.04095 & 0.18766 & 1.00000\\
  Wall and inlet, rank 16 & 0.04105 & 0.18755 & 0.88803\\
  Mass balance only, rank 16 & 0.04095 & 0.18766 & 1.00000\\
  Joint boundary--mass objective, rank 8 & 0.04093 & 0.18747 & 0.90959\\
  Joint boundary--mass objective, rank 16 & 0.04099 & 0.18748 & 0.90672\\
  \textbf{Joint objective + hard wall, rank 16} & \textbf{0.04094} & \textbf{0.18737} & \textbf{0.69893}\\
  \bottomrule
 \end{tabular}
\end{table}

The 12,000-update confirmations are reported in
\cref{tab:cylinder-adapter-formal}.  The FM adapter changes the exhaustive
mean/worst field errors by $+0.094$\%/$-0.029$\%; its wall, inlet, local-CV, and global-mass
ratios are $6.94\times10^{-11}$, 0.2769, 1.0218, and 0.9630.  A paired
counterfactual audit in \cref{tab:cylinder-hard-wall-attribution}, evaluated
on an independent fixed sample set, separates
inference-time use of the learned correction and analytic hard map while
holding the decoder, generated latents, adapter weights, quadrature, and audit
seeds fixed. The hard map alone already drives the wall ratio to
$6.75\times10^{-11}$ with the remaining ratios unchanged; the jointly
deployed model additionally reduces the inlet ratio to 0.2769 and the
global-mass ratio to 0.9629, while
the local-CV diagnostic does not improve.  The diffusion adapter improves its
mean error by 1.63\%, increases the worst error by 2.62\%, and reduces all
four registered physical diagnostics, with ratios
$5.14\times10^{-11}$, 0.4885, 0.9174, and 0.9602.
Here ``exhaustive'' means that \cref{eq:cylinder-selection-rmse} is evaluated
over all available native-mesh cells in every validation case rather than an
audit subsample.

\begin{table}[htbp]
 \centering
 \caption{Formal cylinder adapter confirmation.  Field-error columns use
 \cref{eq:cylinder-selection-rmse}; the physics score in
 \cref{eq:cylinder-physics-score} is evaluated independently for each
 generator.}
 \label{tab:cylinder-adapter-formal}
 \footnotesize
 \begin{tabular}{lccccc}
  \toprule
  Generator & Frozen mean & Adapted mean & Frozen worst & Adapted worst & Physics score\\
  \midrule
  \textbf{FM} & \textbf{0.04260} & \textbf{0.04264} & \textbf{0.19955} & \textbf{0.19949} & \textbf{0.68011}\\
  Diffusion & 0.05866 & 0.05770 & 0.23020 & 0.23623 & 0.65369\\
  \bottomrule
 \end{tabular}
\end{table}

The lower relative diffusion physics score means that its adapter produces a
slightly larger fractional change from its own baseline.  It does not mean
that diffusion has smaller absolute residuals.  Under the common exhaustive
field audit, the Physics-constrained Diffusion model has 35.31\% higher mean
error than Physics-constrained FM, which therefore remains the principal
cylinder model.

\begin{table*}[htbp]
 \centering
 \caption{Paired attribution of the formal cylinder FM correction. All rows
 use the same checkpoint and audit samples. Component entries are
 frozen-relative ratios from \cref{eq:cylinder-physics-score};
 $\overline E_{\mathrm{std}}$ is the exhaustive mean field error.}
 \label{tab:cylinder-hard-wall-attribution}
 \footnotesize
 \begin{tabular}{lrrrrrr}
  \toprule
  Inference variant & $\overline E_{\mathrm{std}}$ & Wall & Inlet & Local CV & Global & $S_{\mathrm{phys}}$\\
  \midrule
  Frozen decoder & 0.04260 & 1.0000 & 1.0000 & 1.0000 & 1.0000 & 1.0000\\
  Hard map only & 0.04263 & $6.75\times10^{-11}$ & 1.0000 & 1.0035 & 1.0000 & 0.7516\\
  Learned adapter only & 0.04273 & 58.9336 & 0.2769 & 1.1430 & 0.9629 & 15.4680\\
  \textbf{Learned adapter + hard map} & \textbf{0.04264} & \textbf{$6.98\times10^{-11}$} & \textbf{0.2769} & \textbf{1.0219} & \textbf{0.9629} & \textbf{0.6801}\\
  \bottomrule
 \end{tabular}
\end{table*}

The learned-only row is a post-training counterfactual, not a separately
trained no-map model. Its wall ratio of 58.93 shows that the learned correction
and analytic projection are not additive modules: during joint training, the
hard map removes wall-normal velocity after the adapter, so the adapter alone
is not constrained to remain feasible at the wall. Consequently, neither the
joint reduction nor its difference from the hard-only case should be
attributed entirely to learning. The defensible conclusion is narrower: the
analytic map supplies exact no-penetration, while the learned correction adds
inlet and global-balance improvements only when deployed with that map.

Because the two pipelines differ in latent chart, reference state, latent
exposure, and per-update sampling, this result ranks the retained pipelines
rather than isolating the generator algorithm; the controlled differences are
summarized in
\cref{sec:controlled-generator-comparison,tab:adapter-training-settings}.
Neither result establishes the complete Boltzmann equation: only wall
no-penetration, inlet consistency, and weak local/global mass balance are
closed by the available cylinder variables.

\subsection{Frozen configurations for out-of-sample evaluation}
\label{sec:frozen-configurations}

The ablations above define the four configurations in
\cref{tab:final-frozen-models}.  No architecture, sampler, ensemble size, or
adapter checkpoint is changed after examining the additional out-of-sample
cases in \cref{Sec5}.

\begin{table*}[htbp]
 \centering
 \caption{Final frozen conditional generative configurations.  The cylinder
 diffusion branch is retained as a controlled comparison, whereas the curved
 FM branch is the principal cylinder accuracy model.}
 \label{tab:final-frozen-models}
 \footnotesize
 \begin{tabular}{@{}ll
  p{0.13\textwidth}
  p{0.22\textwidth}
  p{0.19\textwidth}
  p{0.17\textwidth}@{}}
  \toprule
  Problem & Generator & Latent chart & Conditional dynamics & Deployment & Physics adapter\\
  \midrule
  Cavity & FM & whitened PCA-4 & plain-concat trigonometric FM & Heun-48, 256 samples & rank 8, normalized manifold BGK\\
  Cavity & Diffusion & whitened PCA-4 & FP32 VP $v$-prediction, $(0.05,40)$ & DPM-Solver++(2M)-35, 256 samples & rank 16, normalized BGK\\
  Cylinder & FM & decoder-metric PCA-12 & quadratic local anchor transport & Heun-24, 3 anchors, 21 samples & rank 16, hard wall\\
  Cylinder & Diffusion & unwhitened PCA-12 & VP $x$-prediction, $(0.1,20)$ & DDIM-40, 40 samples & rank 16, hard wall\\
  \bottomrule
 \end{tabular}
\end{table*}

In summary, the retained configurations are not selected by training loss
alone.  Cavity FM favors a simpler conditional vector field, cavity diffusion
requires FP32 $v$-prediction and a higher-order deterministic sampler, and
cylinder FM benefits from local decoder-aware curved transport. The reported
adapters pass their benchmark-specific final acceptance rules; for cavity FM,
the retained checkpoint satisfies the invariant gate. This
separation permits \cref{Sec5} to evaluate generalization without further model
selection.
The deterministic baselines in
\cref{eq:deterministic-latent-baseline} are frozen under the same data-separation
rule and are evaluated only after generator selection. Because that comparison
uses additional conditions, its decoded-field results are reported in
\cref{sec:deterministic-decoded-comparison}, rather than used as another
development-set ablation.

\section{Out-of-sample predictions and discussion}\label{Sec5}

The configurations frozen in \cref{sec:frozen-configurations} are evaluated
without further fitting or checkpoint selection. They remain accurate over a
range of out-of-sample conditions; the following discussion focuses on the
representative cases declared in \cref{Sec2}. Throughout this section,
Frozen FM and Frozen Diffusion denote the unadapted conditional generators.
Physics-constrained FM and Physics-constrained Diffusion denote the
corresponding frozen generators followed by their accepted structured
adapters.
Neither the generator nor the latent transform is refitted on these additional
cases.
Conditional generative surrogates for partial differential equations learn a
condition-dependent distribution of solution fields
\citep{Shysheya2024,GaoH2025}; freezing every learned
component here ensures that the following results measure that conditional
generalization rather than test-time adaptation. Sampling can reduce numerical
Monte Carlo variability, but it cannot add missing condition-space evidence.

\subsection{Deterministic baseline under frozen decoded-field evaluation}
\label{sec:deterministic-decoded-comparison}

To determine whether conditional generation improves steady point prediction,
\cref{tab:deterministic-decoded-comparison} compares the direct MLP in
\cref{eq:deterministic-latent-baseline} with the two frozen generative
pipelines. No predictor, chart, decoder, or normalization statistic is refitted
on these cases. For the cavity, all three predictors share the whitened PCA-4
chart and the same decoder. For the cylinder, the MLP and diffusion share the
ordinary PCA-12 chart, whereas FM retains the decoder-metric PCA-12 chart
selected in \cref{Sec4}; the latter is therefore a pipeline-level comparison.
Each MLP entry reports the mean over five independent training seeds with the
population standard deviation in parentheses. Two capacity tiers are trained
on every chart: the compact ensembles (203,532 cavity and 89,340 cylinder
parameters) and single-member parameter-matched controls with 2,989,188
(cavity) and 209,772 (cylinder) parameters, matching the retained generators
to within about 3\%. The cylinder MLPs are evaluated on both the ordinary and
the decoder-metric PCA-12 charts, so the deterministic baselines bracket the
chart choices of the two generators.
For the cylinder rows, define the descriptive six-field average
\begin{equation}
 E_6=\frac{1}{6}\left(
 E_{\rho,\mathrm{RMSE}}+E_{p,\mathrm{RMSE}}+E_{T,\mathrm{RMSE}}
 +E_{u,\mathrm{RMSE}}+E_{v,\mathrm{RMSE}}+E_{|\bm u|,\mathrm{RMSE}}
 \right),
 \label{eq:cylinder-six-field-error}
\end{equation}
where every component is the dimensionless field RMSE from
\cref{eq:cylinder-field-rmse}. This equal weighting is a compact reporting
statistic, not a thermodynamic norm.

\begin{table*}[htbp]
 \centering
 \caption{Frozen deterministic and generative predictors under common
 decoded-field evaluation. Cavity errors are $E_{\mathrm{kin}}$ from
 \cref{eq:cavity-kinetic-errors}; cylinder errors are $E_6$ from
 \cref{eq:cylinder-six-field-error}. MLP entries are five-seed means with
 population standard deviations in parentheses; compact and
 parameter-matched capacity tiers are listed separately. Cylinder entries are
 evaluated over the three displayed conditions, not the complete
 additional evaluation set. Bold entries are
 the minimum mean within each problem and subset.}
 \label{tab:deterministic-decoded-comparison}
 \footnotesize
 \resizebox{\textwidth}{!}{%
 \begin{tabular}{lllrrr}
  \toprule
  Problem/subset & Predictor & Chart & Parameters & Mean & Worst case\\
  \midrule
  Cavity, interpolation & Deterministic MLP (compact) & whitened PCA-4 & 203\,532 & $(5.5896\pm0.0092)\times10^{-5}$ & $(6.0017\pm0.0544)\times10^{-5}$\\
  $\mathrm{Kn}=0.23,0.73$ & \textbf{Deterministic MLP (matched)} & whitened PCA-4 & 2\,989\,188 & \textbf{$(5.4725\pm0.0052)\times10^{-5}$} & \textbf{$(5.9806\pm0.0105)\times10^{-5}$}\\
   & Frozen FM & whitened PCA-4 & 2\,996\,932 & $5.5023\times10^{-5}$ & $5.7763\times10^{-5}$\\
   & Frozen Diffusion & whitened PCA-4 & 2\,996\,932 & $5.9051\times10^{-5}$ & $6.4356\times10^{-5}$\\
  \addlinespace
  Cavity, extrapolation & Deterministic MLP (compact) & whitened PCA-4 & 203\,532 & $(1.4998\pm0.2041)\times10^{-3}$ & $(2.9240\pm0.4086)\times10^{-3}$\\
  $\mathrm{Kn}=0.03,1.00$ & \textbf{Deterministic MLP (matched)} & whitened PCA-4 & 2\,989\,188 & \textbf{$(1.4952\pm0.4751)\times10^{-3}$} & \textbf{$(2.8593\pm0.9881)\times10^{-3}$}\\
   & Frozen FM & whitened PCA-4 & 2\,996\,932 & $3.0933\times10^{-3}$ & $5.2636\times10^{-3}$\\
   & Frozen Diffusion & whitened PCA-4 & 2\,996\,932 & $1.6087\times10^{-3}$ & $3.1348\times10^{-3}$\\
  \addlinespace
  Cylinder, three shown & \textbf{Deterministic MLP (compact)} & ordinary PCA-12 & 89\,340 & \textbf{$0.02862\pm0.00252$} & \textbf{$0.03385\pm0.00272$}\\
   & Deterministic MLP (matched) & ordinary PCA-12 & 209\,772 & $0.03107\pm0.00253$ & $0.04109\pm0.00908$\\
   & Deterministic MLP (compact) & decoder-metric PCA-12 & 89\,340 & $0.03523\pm0.00191$ & $0.05320\pm0.00484$\\
   & Deterministic MLP (matched) & decoder-metric PCA-12 & 209\,772 & $0.02998\pm0.00081$ & $0.04002\pm0.00180$\\
   & Frozen FM & decoder-metric PCA-12 & 208\,144 & 0.04103 & 0.06211\\
   & Frozen Diffusion & ordinary PCA-12 & 203\,404 & 0.04814 & 0.06242\\
  \bottomrule
 \end{tabular}}
\end{table*}

Repeating the generator fits with five independent seeds places the
single-fit rows in context: the cavity FM and diffusion interpolation means,
$(6.72\pm0.79)\times10^{-5}$ and $(6.03\pm0.22)\times10^{-5}$, are
statistically indistinguishable and remain above the MLP means
(5.47--5.59$\times10^{-5}$); on extrapolation the diffusion mean
$(1.75\pm0.38)\times10^{-3}$ is level with the MLPs and clearly better than
FM, $(3.56\pm0.30)\times10^{-3}$. On the cylinder the FM five-seed mean
$E_6$, $0.0365\pm0.0017$ against $0.0531\pm0.0037$ for diffusion, confirms a
seed-robust pipeline ranking, while every MLP variant (seven-condition means
0.0288--0.0311 over the complete additional evaluation set; the cylinder rows
of \cref{tab:deterministic-decoded-comparison} report only the three displayed
conditions) outperforms both generators. Thus, conditional generation does not provide a universal
point-accuracy advantage for these single-reference steady problems. Its role
here is to model cross-condition evolution as transport on a compact solution
manifold and to retain a sampling-based inference mechanism on which the frozen
physics-adaptation protocol can operate. The deterministic baseline also shows
that this machinery is not required merely to obtain a strong conditional mean:
for a single-valued steady map, a much smaller regressor can provide equal or
lower decoded-field error. The generative formulation becomes more distinctive
when a condition admits multiple states or irreducible variability, because it
can in principle represent a conditional distribution instead of only its
point estimate. That extension is architectural rather than empirical evidence
from the present datasets, which contain one steady reference per condition.
Accordingly, the current endpoint spread remains a transport-sensitivity
diagnostic, not calibrated uncertainty, and the results do not show that the
additional sampling machinery dominates a compact deterministic surrogate.

\subsection{Lid-driven cavity flow}
\label{sec:cavity-generalization}

\subsubsection{Interpolation within the training support}
\label{sec:cavity-interpolation}

\begin{figure}[htbp]
 \centering
 \includegraphics[width=0.96\textwidth,height=0.78\textheight,keepaspectratio]{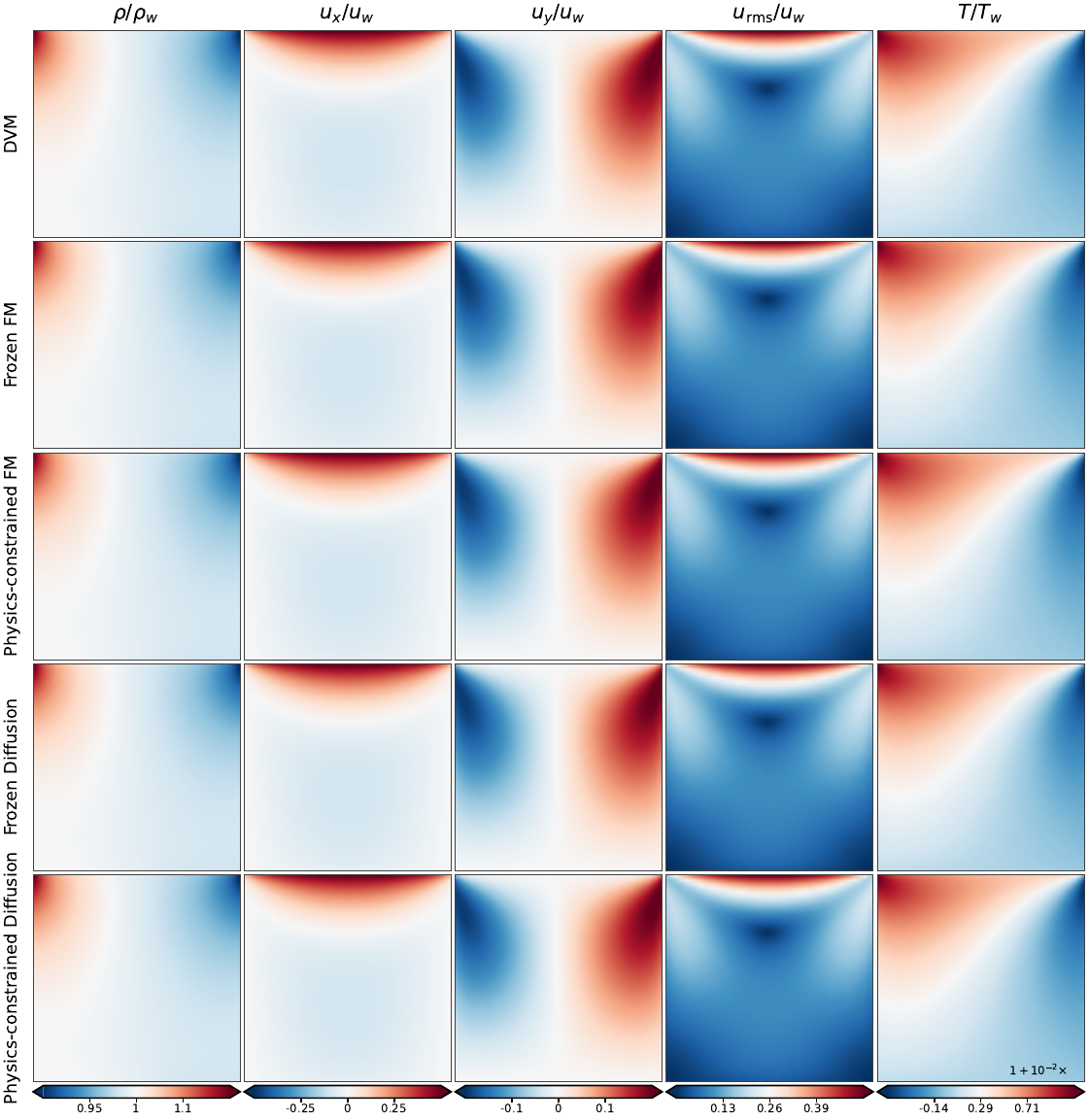}
 \caption{Dimensionless cavity fields at $\mathrm{Kn}=0.23$ from DVM and the
 four generative models. Columnwise color limits are shared across rows.}
 \label{fig:cavity-kn023-fields}
\end{figure}

\begin{figure}[htbp]
 \centering
 \includegraphics[width=0.96\textwidth,height=0.78\textheight,keepaspectratio]{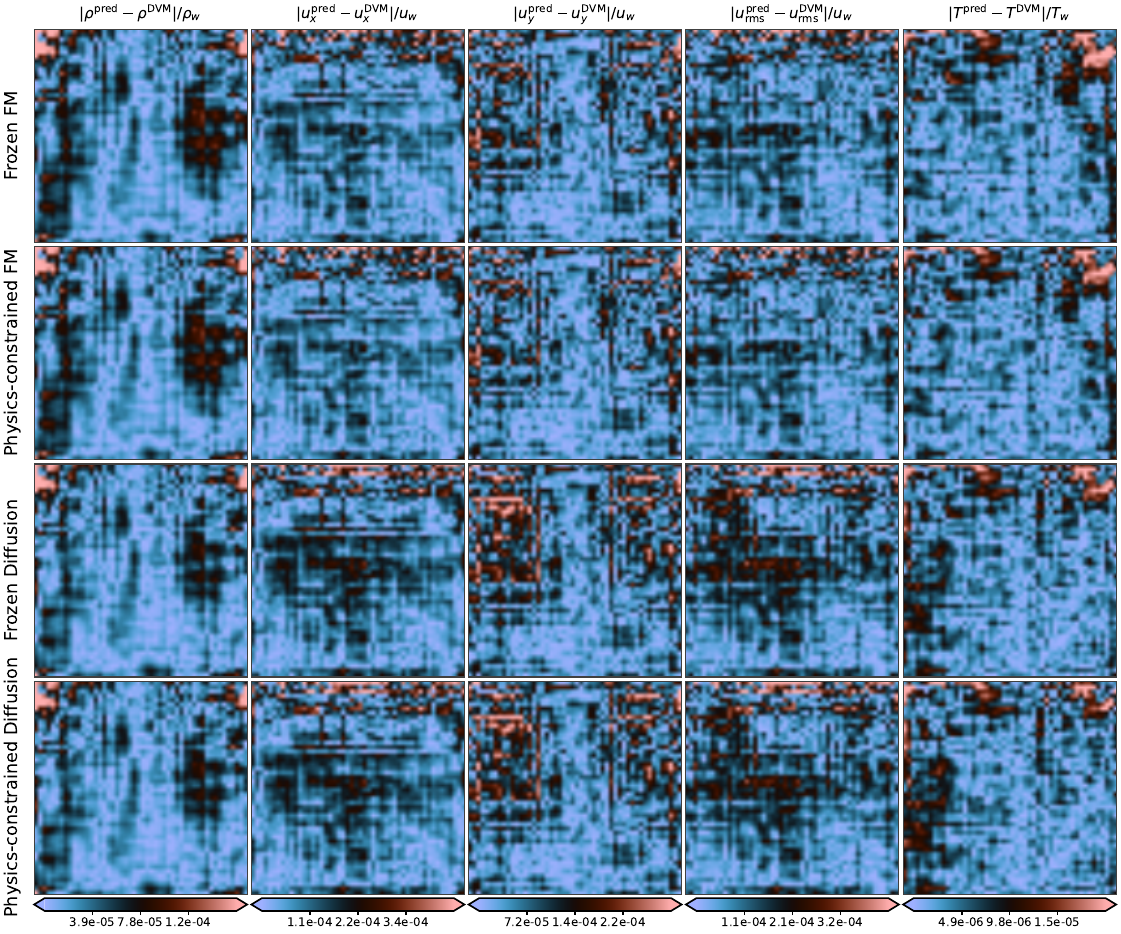}
 \caption{Direct dimensionless errors at $\mathrm{Kn}=0.23$. Each column uses
 99th-percentile upper clipping.}
 \label{fig:cavity-kn023-errors}
\end{figure}

\begin{figure}[htbp]
 \centering
 \includegraphics[width=0.96\textwidth,height=0.78\textheight,keepaspectratio]{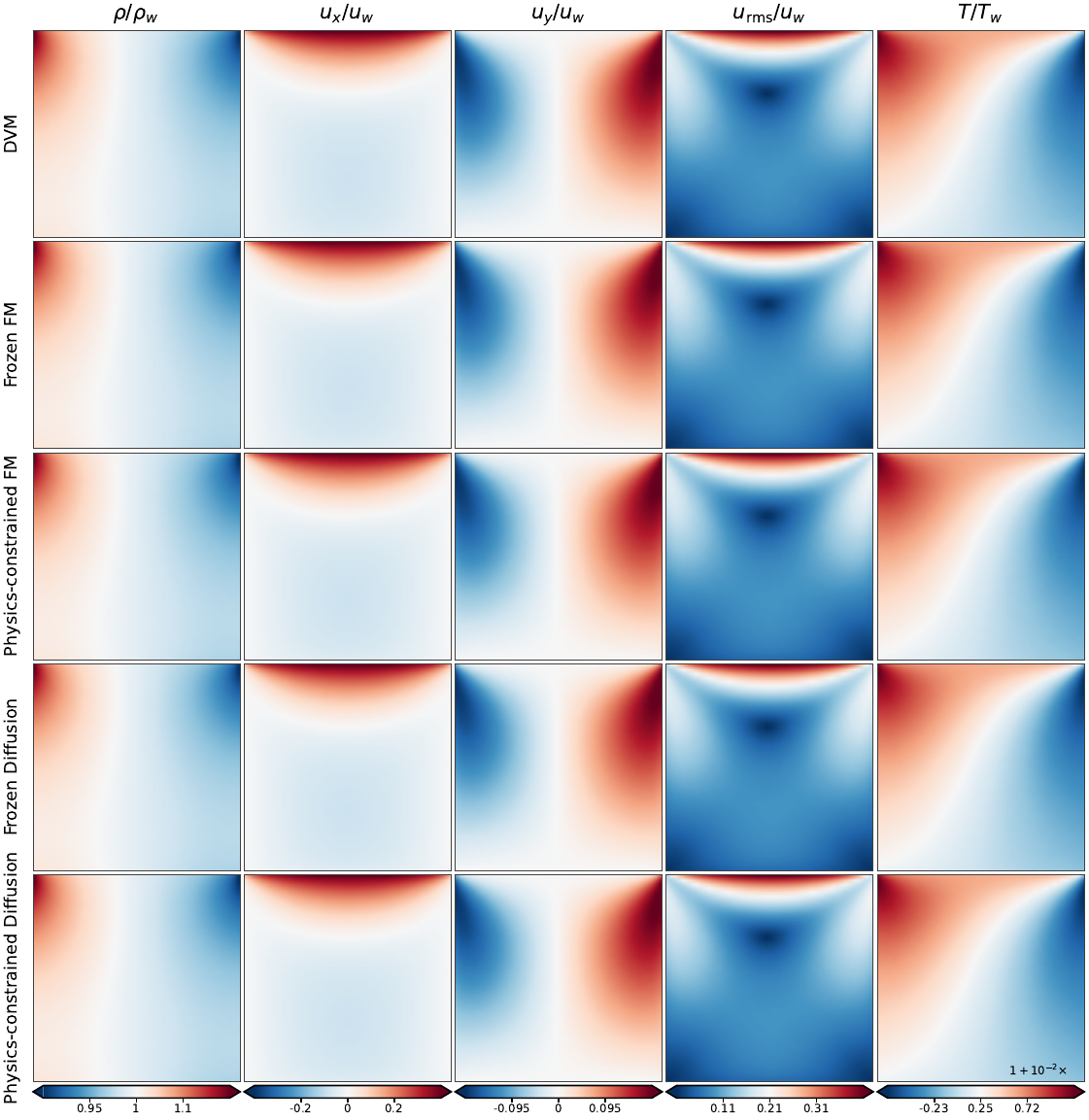}
 \caption{Dimensionless cavity fields at $\mathrm{Kn}=0.73$ from DVM and the
 four generative models. Columnwise color limits are shared across rows.}
 \label{fig:cavity-kn073-fields}
\end{figure}

\begin{figure}[htbp]
 \centering
 \includegraphics[width=0.96\textwidth,height=0.78\textheight,keepaspectratio]{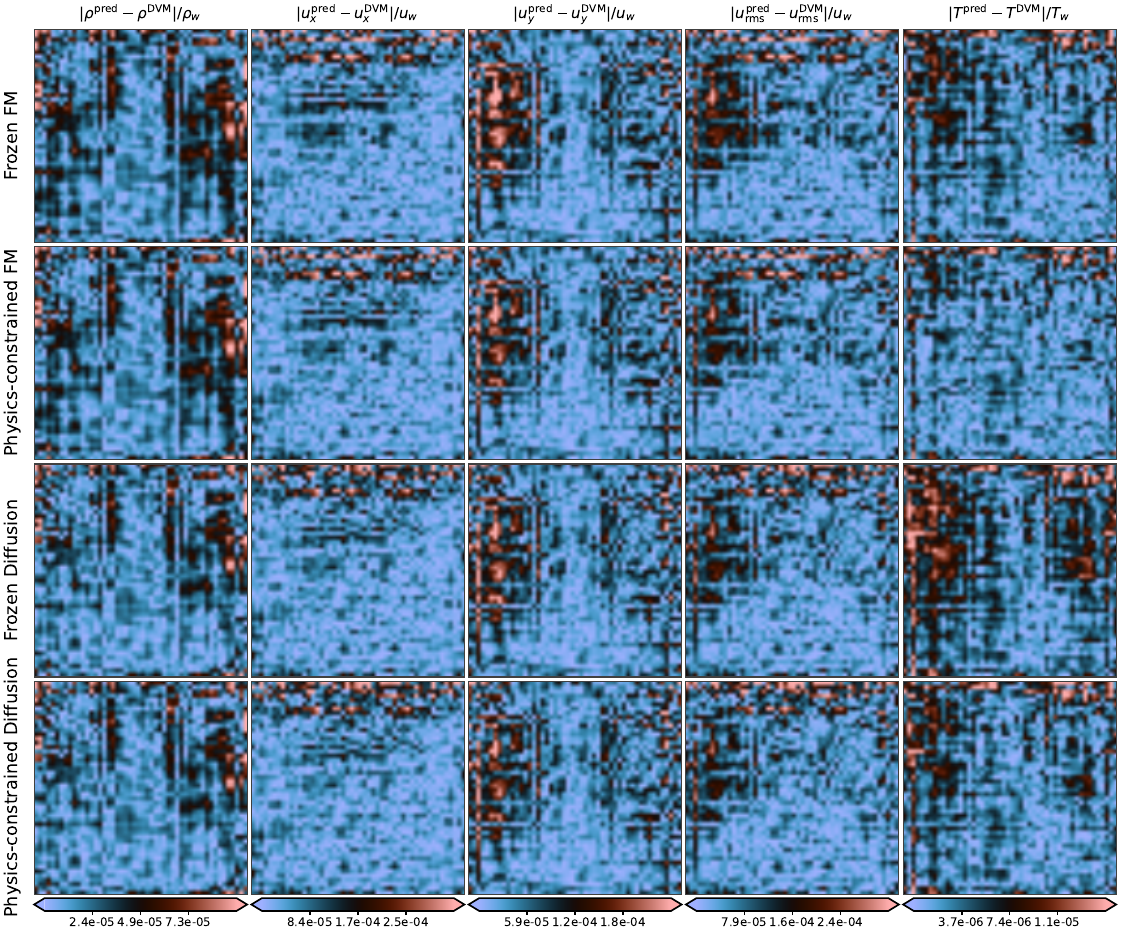}
 \caption{Direct dimensionless errors at $\mathrm{Kn}=0.73$. Each column uses
 99th-percentile upper clipping.}
 \label{fig:cavity-kn073-errors}
\end{figure}

The conditions $\mathrm{Kn}=0.23$ and 0.73 probe distinct locations inside the
training support. 
\Cref{fig:cavity-kn023-fields,fig:cavity-kn023-errors,fig:cavity-kn073-fields,fig:cavity-kn073-errors}
compare the DVM fields with Frozen FM, Physics-constrained FM, Frozen
Diffusion, and Physics-constrained Diffusion.
At both conditions, the density, velocity, speed, and temperature contours are
visually coincident at the plotted scale. The direct-error panels localize the
remaining discrepancy near the moving lid, the upper corners, and the recirculating 
core rather than revealing a generator-specific large-scale distortion.
The change from $\mathrm{Kn}=0.23$ to 0.73 is nevertheless resolved: increasing
rarefaction broadens the lid-driven response and reduces the penetration of
the high-speed layer into the cavity. The concentration of error near the
lid--sidewall junctions is consistent with the simultaneous presence of a
discontinuous wall velocity and a kinetic boundary layer, both of which make
the local solution more sensitive to spatial and velocity-space resolution
\citep{WangX2026,XinZY2025}.

At $\mathrm{Kn}=0.23$, the four macroscopic errors lie between
$4.88\times10^{-5}$ and $5.29\times10^{-5}$, while the kinetic errors lie
between $5.21\times10^{-5}$ and $5.37\times10^{-5}$. The FM and diffusion
physics adapters reduce their matched-grid BGK
diagnostics from 0.644 to 0.628 and from 0.644 to 0.629, respectively, without
a material change in field error.
At $\mathrm{Kn}=0.73$, the four macroscopic errors lie between
$3.33\times10^{-5}$ and $3.85\times10^{-5}$ and the kinetic errors between
$5.69\times10^{-5}$ and $6.44\times10^{-5}$; the discrepancy is again
concentrated near the moving lid and upper corners but spreads more deeply
into the core, consistent with the thicker lid-driven kinetic layer of the
more rarefied flow. Diffusion gives the smaller macroscopic error, whereas
FM gives the smaller kinetic error. Physics correction reduces the FM kinetic
error from $5.78\times10^{-5}$ to $5.69\times10^{-5}$ and the diffusion error
from $6.44\times10^{-5}$ to $5.96\times10^{-5}$; the corresponding BGK
diagnostics decrease from 5.812 to 5.186 and from 5.810 to 5.391. Thus, for
both displayed interpolation conditions, both adapted models improve the
registered discrete BGK diagnostic while retaining the accuracy of the frozen
fields. The FM invariant diagnostics increase by about 6\%, consistent with
the selection-time audit in \cref{sec:ablation-physics}; the
diffusion invariant diagnostics decrease.
This does not establish exact satisfaction of the continuous BGK equation:
the diagnostic inherits the finite spatial differentiation and velocity
quadrature of the matched DVM grid \citep{BGK,Mieussens2000}.

\begin{figure}[htbp]
 \centering
 \includegraphics[width=0.78\textwidth]{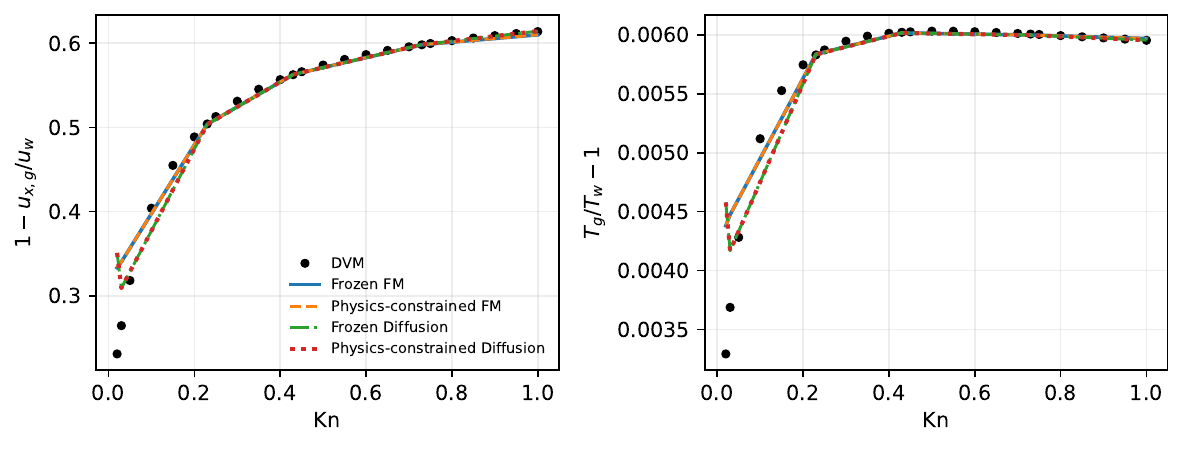}
 \caption{Wall-adjacent moving-lid slip and temperature jump. DVM markers span
 the available conditions; model curves use the additional evaluation cases.}
 \label{fig:cavity-wall-rarefaction}
\end{figure}

As a pointwise rarefaction check, \cref{fig:cavity-wall-rarefaction} reports
the mean gas slip and temperature jump in the central half of the first
cell-center row below the moving lid. The DVM slip
$1-u_{x,g}/u_w$ increases monotonically from 0.231 at
$\mathrm{Kn}=0.02$ to 0.614 at $\mathrm{Kn}=1$, while
$T_g/T_w-1$ rises from $3.29\times10^{-3}$, reaches approximately
$6.03\times10^{-3}$ near $\mathrm{Kn}=0.5$, and then decreases slightly.
The frozen and physics-constrained models reproduce both trends at
$\mathrm{Kn}=0.23$, 0.73, and 1.00; their visible discrepancy at 0.03 is
consistent with the low-side latent extrapolation error discussed below.
Because these quantities are sampled at $y/L_0=0.99$ rather than extrapolated
to the material wall, they are wall-adjacent diagnostics and are not presented
as asymptotic slip coefficients. Their trend provides a physical check of the
resolved rarefaction response without imposing no-slip or zero-temperature-
jump behavior \citep{XinZY2025,WangX2026}.

\subsubsection{Extrapolation beyond the training support}
\label{sec:cavity-extrapolation}

\begin{figure}[htbp]
 \centering
 \includegraphics[width=0.96\textwidth,height=0.78\textheight,keepaspectratio]{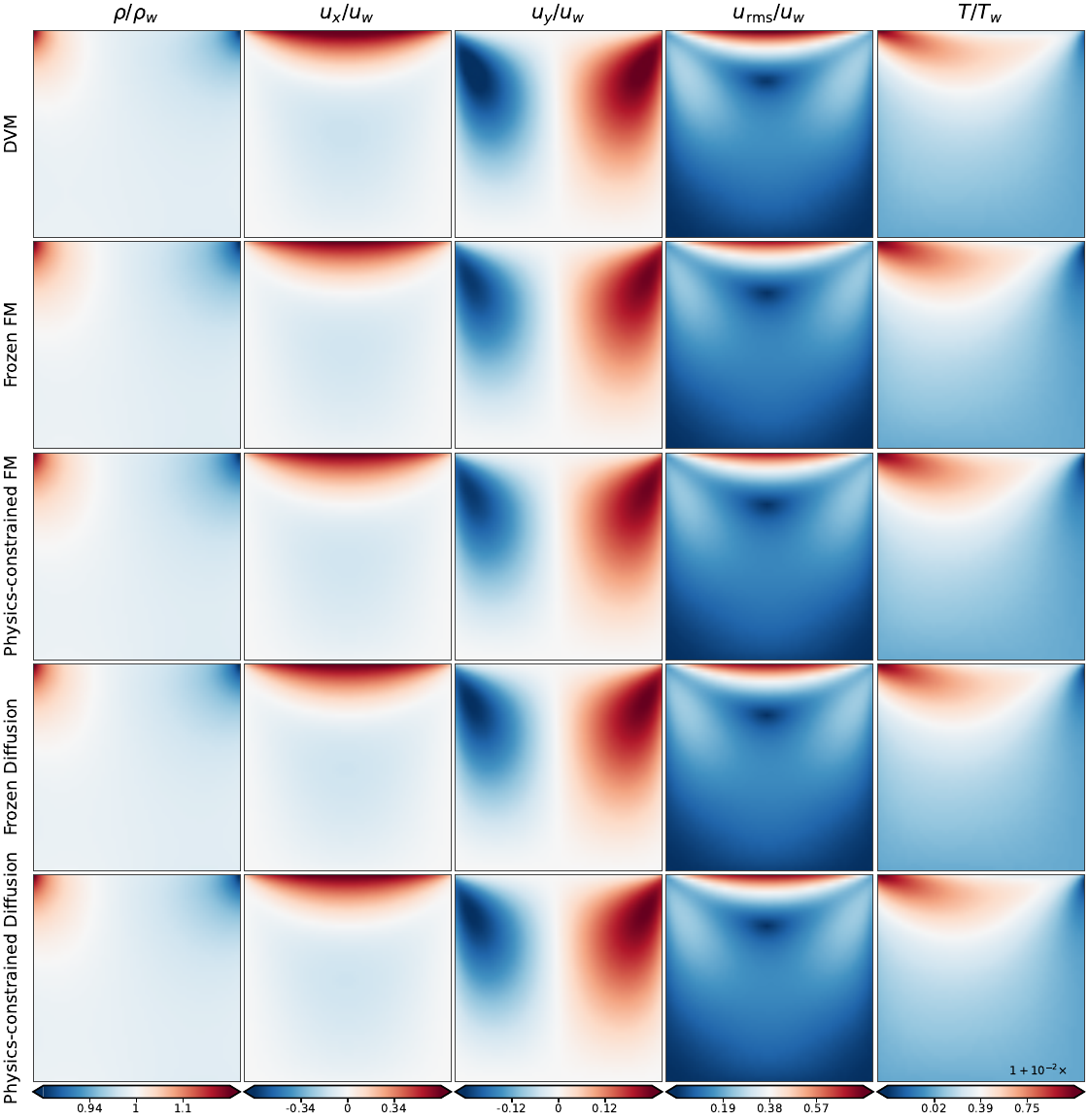}
 \caption{Dimensionless cavity fields for lower-side extrapolation at
 $\mathrm{Kn}=0.03$. Columnwise color limits are shared across rows.}
 \label{fig:cavity-kn003-fields}
\end{figure}

\begin{figure}[htbp]
 \centering
 \includegraphics[width=0.96\textwidth,height=0.78\textheight,keepaspectratio]{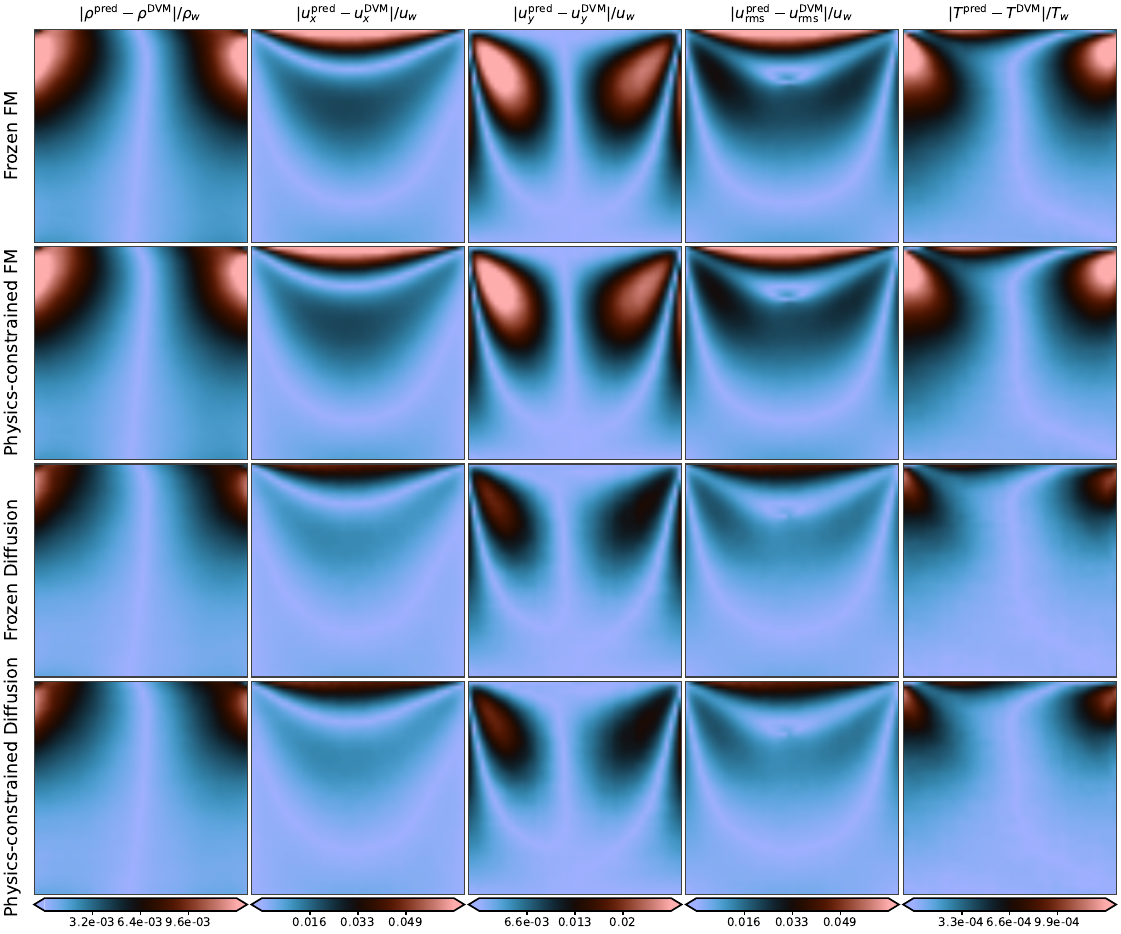}
 \caption{Direct dimensionless errors for lower-side cavity extrapolation at
 $\mathrm{Kn}=0.03$, using 99th-percentile upper clipping.}
 \label{fig:cavity-kn003-errors}
\end{figure}

\begin{figure}[htbp]
 \centering
 \includegraphics[width=0.96\textwidth,height=0.78\textheight,keepaspectratio]{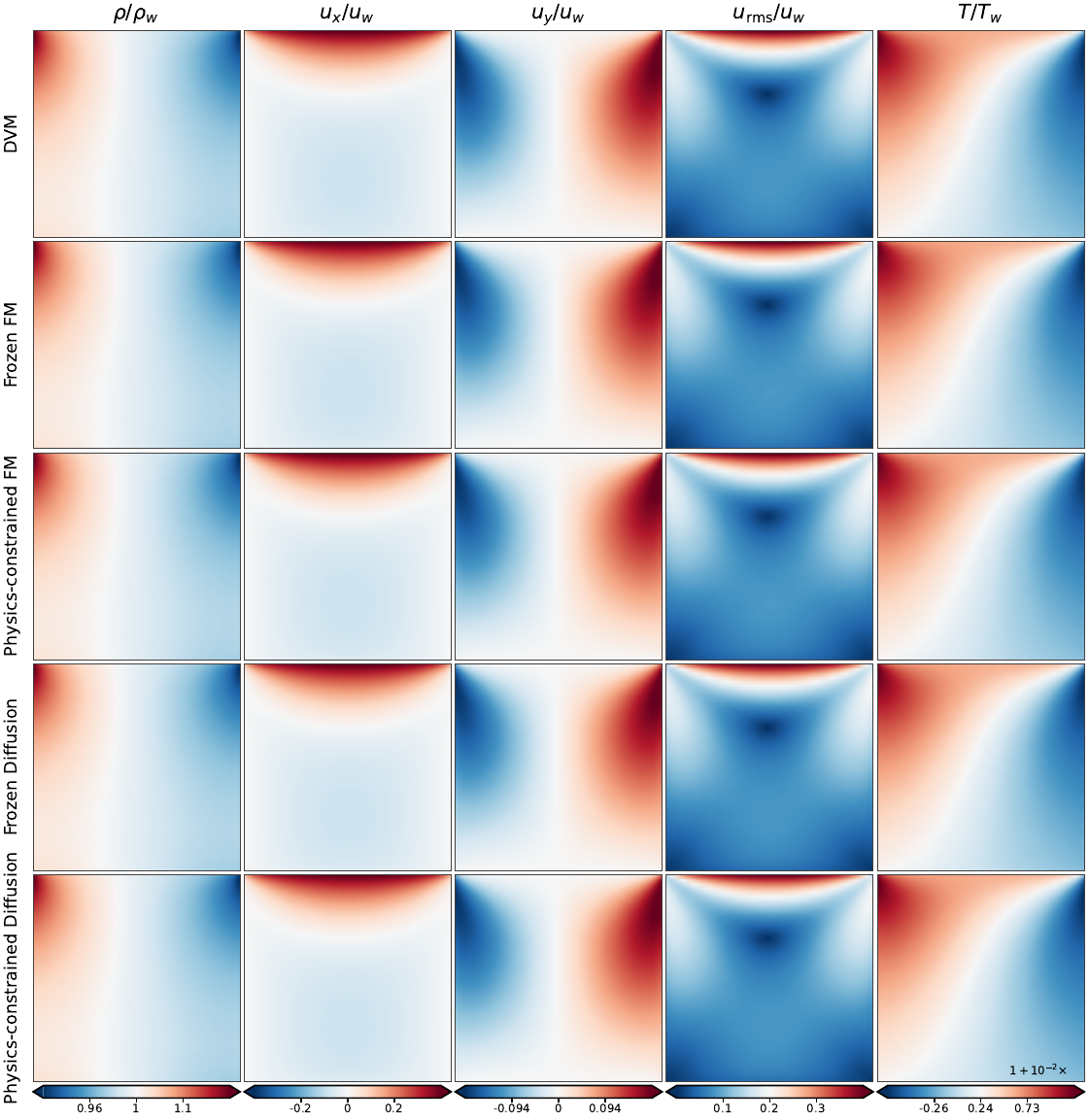}
 \caption{Dimensionless cavity fields for upper-side extrapolation at
 $\mathrm{Kn}=1.00$. Columnwise color limits are shared across rows.}
 \label{fig:cavity-kn100-fields}
\end{figure}

\begin{figure}[htbp]
 \centering
 \includegraphics[width=0.96\textwidth,height=0.78\textheight,keepaspectratio]{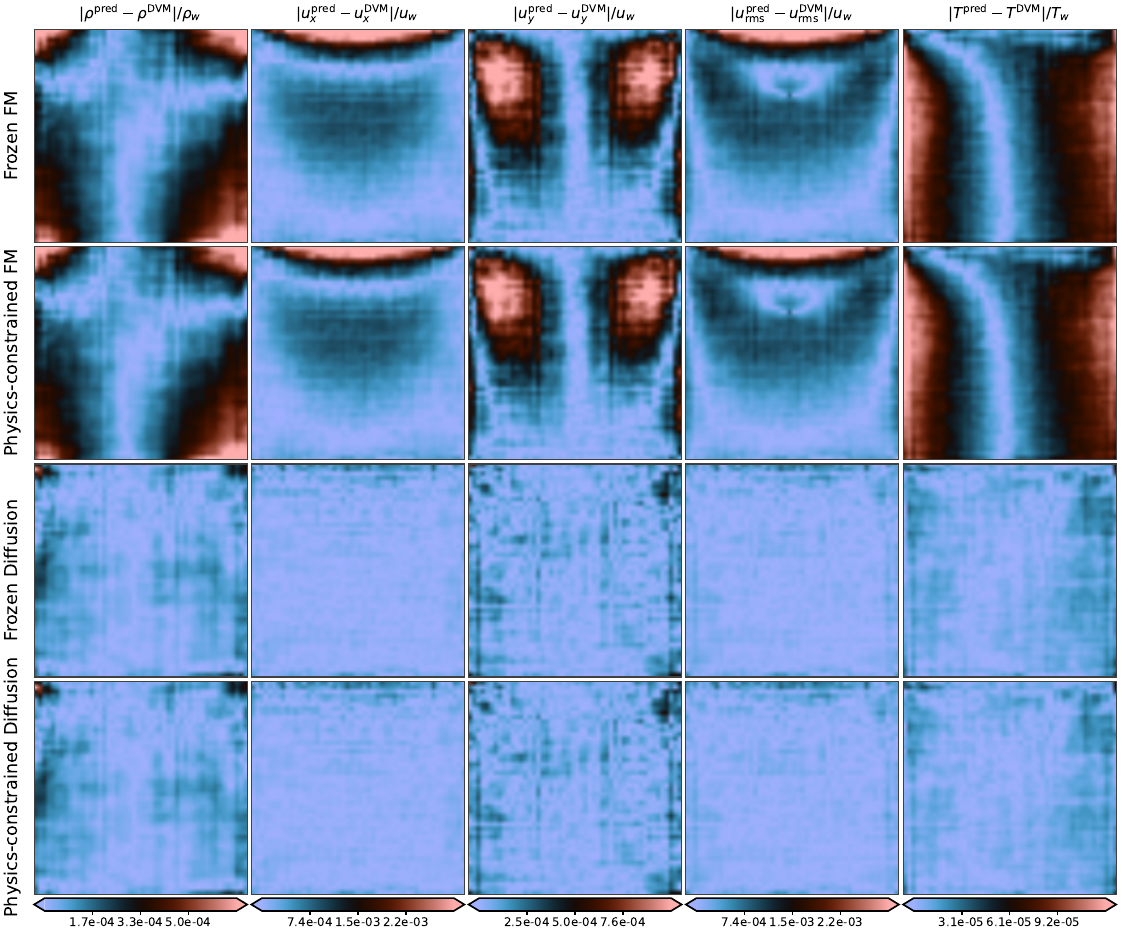}
 \caption{Direct dimensionless errors for upper-side cavity extrapolation at
 $\mathrm{Kn}=1.00$, using 99th-percentile upper clipping.}
 \label{fig:cavity-kn100-errors}
\end{figure}

The lower-side case $\mathrm{Kn}=0.03$ is shown in
\cref{fig:cavity-kn003-fields,fig:cavity-kn003-errors}. Its errors remain
spatially structured and are much
larger than those of the interpolation cases. Frozen FM and Frozen Diffusion
macroscopic errors are $5.01\times10^{-3}$ and $3.10\times10^{-3}$,
respectively, with kinetic errors of $5.26\times10^{-3}$ and
$3.13\times10^{-3}$. The structured adapters change these values by less than
$10^{-6}$ in absolute terms. Thus, the dominant limitation lies upstream in
the extrapolated chart endpoint; a capacity-limited local decoder correction
cannot replace missing low-Knudsen condition coverage.
The physical fields remain smooth, but the error maps show coherent rather
than sample-like deviations across the domain. This supports the
latent-endpoint diagnosis: the model produces a plausible cavity state, but
one associated with the wrong position on the rapidly varying low-Knudsen
solution chart.

Upper-side extrapolation behaves differently. At $\mathrm{Kn}=1.00$,
\cref{fig:cavity-kn100-fields,fig:cavity-kn100-errors} show that both model
families retain the global cavity
structure. Diffusion gives macroscopic and kinetic errors of
$7.19\times10^{-5}$ and $8.26\times10^{-5}$, whereas FM gives
$2.91\times10^{-4}$ and $9.23\times10^{-4}$. The diffusion adapter lowers the
kinetic error to $7.81\times10^{-5}$ and the matched-grid BGK diagnostic by
8.25\%; the FM physics adapter lowers the BGK diagnostic by 12.20\% but leaves
its kinetic error essentially unchanged. The marked asymmetry between lower- and
upper-side extrapolation shows that distance from the training interval alone
does not determine difficulty; the regularity and continuation of the learned
solution chart also matter.
In particular, upper-side continuation preserves the broad spatial structures
already present at the largest training Knudsen numbers, whereas lower-side
continuation moves toward thinner wall layers and sharper gradients. The
comparison therefore concerns the direction of continuation in solution
space, not merely the scalar distance in $\mathrm{Kn}$.

\subsection{Rarefied cylinder flow}
\label{sec:cylinder-generalization}

\subsubsection{Aggregate results over additional conditions}

The additional cylinder cases lie within the two-dimensional condition domain
described in \cref{sec:cylinder-data}. The descriptive statistic $E_6$ is
defined in \cref{eq:cylinder-six-field-error}; model selection remains that of
\cref{Sec4}.

\Cref{tab:cylinder-generalization-aggregate} shows that FM is more accurate than
diffusion on this benchmark. Frozen FM gives $E_6=0.03655$, compared with
0.04748 for Frozen Diffusion, a 29.9\% increase relative to FM\@. Five
independent refits of each generator yield $0.0365\pm0.0017$ (FM) and
$0.0531\pm0.0037$ (diffusion), so this ranking is robust to training seed. Pressure is
the largest-error field for every model, followed by temperature and density,
whereas both velocity components remain below $1.4\times10^{-2}$ in RMSE\@.
Unlike the four decoder outputs, pressure is reconstructed as $p=nk_BT$;
its error therefore combines correlated density and temperature errors and
should not be interpreted as the error of a separately trained output head.
Moreover, $E_6$ gives every reported field equal numerical weight, so it is a
compact comparison statistic rather than a thermodynamic norm.
Physics-constrained FM changes the equal-case mean from 0.03655 to 0.03670,
while Physics-constrained Diffusion changes it from 0.04748 to 0.05008. Thus,
neither physics-constrained model produces an aggregate field-accuracy gain on
these additional conditions.

\begin{table}[htbp]
 \centering
 \caption{Equal-case mean RMSE over the additional cylinder conditions.
 $E_6$ is the descriptive six-field average in
 \cref{eq:cylinder-six-field-error}.}
 \label{tab:cylinder-generalization-aggregate}
 \footnotesize
 \begin{tabular}{lccccccc}
  \toprule
  Model & $\rho$ & $p$ & $T$ & $u$ & $v$ & $|\bm u|$ & $E_6$\\
  \midrule
  \textbf{Frozen FM} & \textbf{0.03769} & 0.11649 & \textbf{0.04089} &
  \textbf{0.00951} & \textbf{0.00571} & \textbf{0.00901} & \textbf{0.03655}\\
  Physics-constrained FM & 0.03852 & \textbf{0.11636} & 0.04106 &
  0.00952 & 0.00571 & 0.00901 & 0.03670\\
  Frozen Diffusion & 0.04235 & 0.15207 & 0.06003 &
  0.01207 & 0.00684 & 0.01149 & 0.04748\\
  Physics-constrained Diffusion & 0.04518 & 0.15968 & 0.06205 &
  0.01351 & 0.00715 & 0.01293 & 0.05008\\
  \bottomrule
 \end{tabular}
\end{table}

This out-of-sample evaluation measures field error only. The DSMC output does not
contain the higher-order fluxes required for a closed momentum, energy, BGK,
or Boltzmann residual, and the common-seed physical audit used for checkpoint
selection in \cref{sec:ablation-physics} was not repeated on these additional
conditions. Consequently, \cref{tab:cylinder-generalization-aggregate} neither
confirms nor refutes whether the reductions in wall, inlet, and weak-mass
diagnostics observed during physics-adapter selection transfer to these
additional conditions. Moreover, the tabulated RMSE contains an unknown DSMC
sampling-noise floor because independent realizations or averaging blocks are
not available. The field-accuracy guard consequently limits degradation
relative to this sampled reference; it is not a bound relative to a
noise-free continuum mean.

\subsubsection{Representative condition-dependent behavior}

\begin{figure}[htbp]
 \centering
 \includegraphics[width=0.96\textwidth,height=0.78\textheight,keepaspectratio]{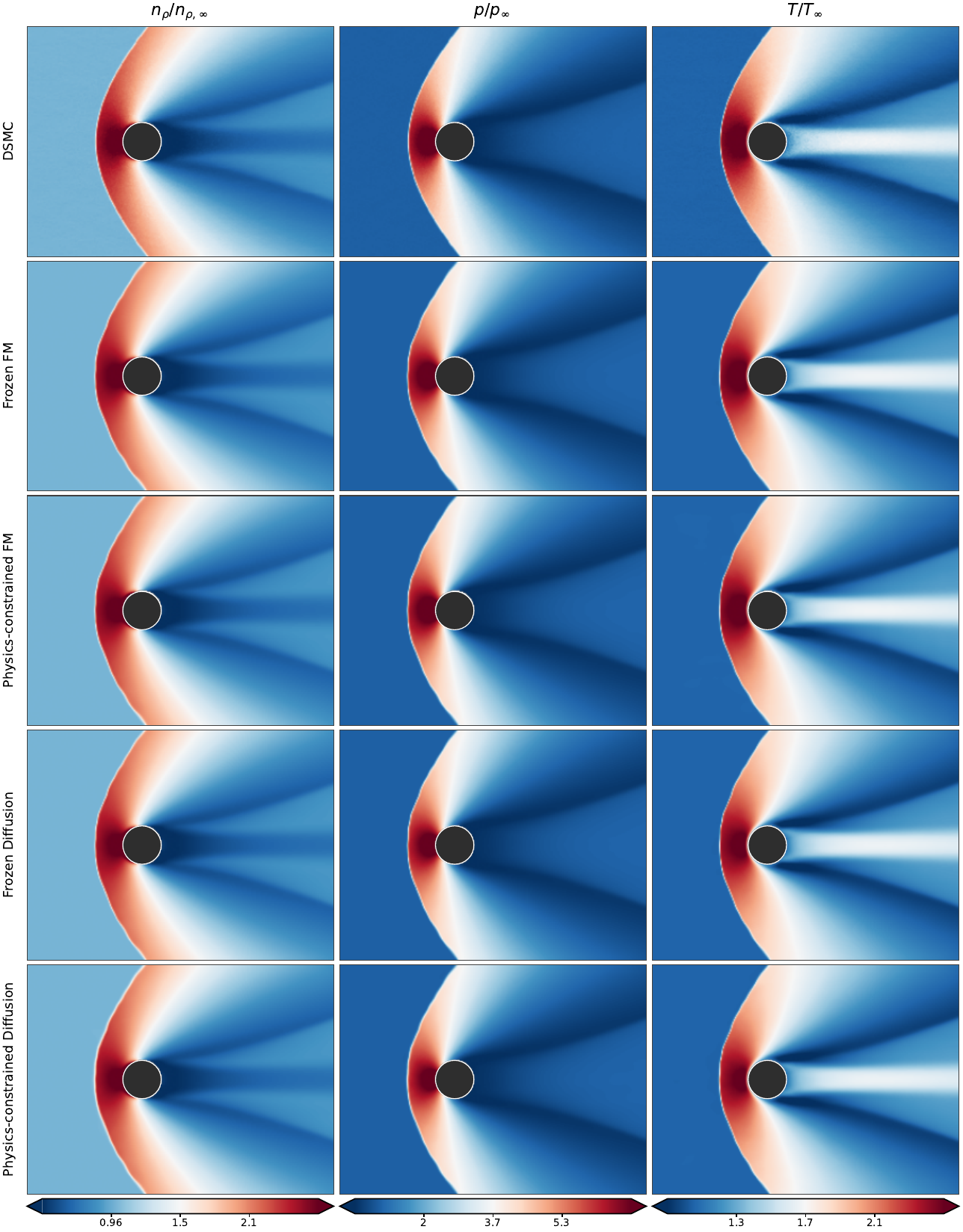}
 \caption{Dimensionless cylinder fields at
 $(\mathrm{Kn},\mathrm{Ma})=(0.12,2.3)$. Columnwise color limits are shared
 across rows and set independently for this condition. All cylinder panels in
 Figs.~10--15 show the window $x/D\in[-3,5]$, $y/D\in[-3,3]$ of the domain
 defined in \cref{eq:cylinder-domain}, with the cylinder centered at the
 origin.}
 \label{fig:cylinder-kn012-ma23-fields}
\end{figure}

\begin{figure}[htbp]
 \centering
 \includegraphics[width=0.96\textwidth,height=0.78\textheight,keepaspectratio]{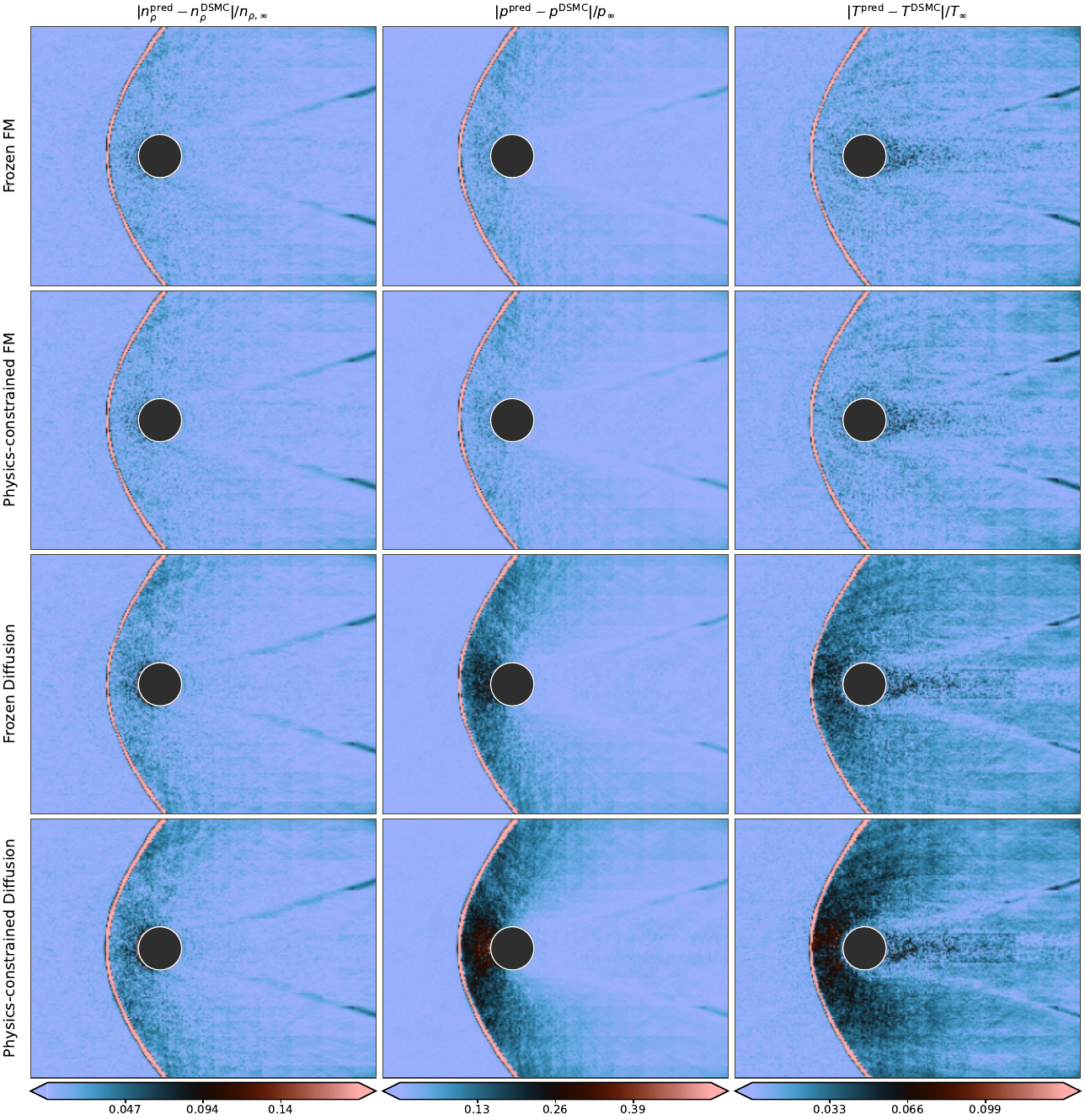}
 \caption{Direct dimensionless cylinder errors at $(\mathrm{Kn},\mathrm{Ma})
 =(0.12,2.3)$, using 99th-percentile upper clipping.}
 \label{fig:cylinder-kn012-ma23-errors}
\end{figure}

The representative cases are discussed in ascending Knudsen-number order.
Because their Mach numbers also differ, the sequence illustrates the joint
condition dependence and is not a one-parameter rarefaction sweep.  At the
lowest displayed condition, $(\mathrm{Kn},\mathrm{Ma})=(0.12,2.3)$,
\cref{fig:cylinder-kn012-ma23-fields,fig:cylinder-kn012-ma23-errors}, the flow
is closest to the slip regime: the bow shock is thin and stands close to the
cylinder, the stagnation compression is sharp, and the wake is comparatively
narrow.  The change
from Frozen FM to Physics-constrained FM is negligible ($E_6=0.06211$ to
0.06206), whereas the change from Frozen Diffusion to Physics-constrained
Diffusion increases the error from 0.06242 to 0.07243.  The degradation occurs
across density, pressure, temperature, and velocity rather than in one isolated
scalar.  This case demonstrates that passing the frozen-field accuracy guard
during adapter selection does not guarantee a field-error improvement at every
additional condition.

\begin{figure}[htbp]
 \centering
 \includegraphics[width=0.96\textwidth,height=0.78\textheight,keepaspectratio]{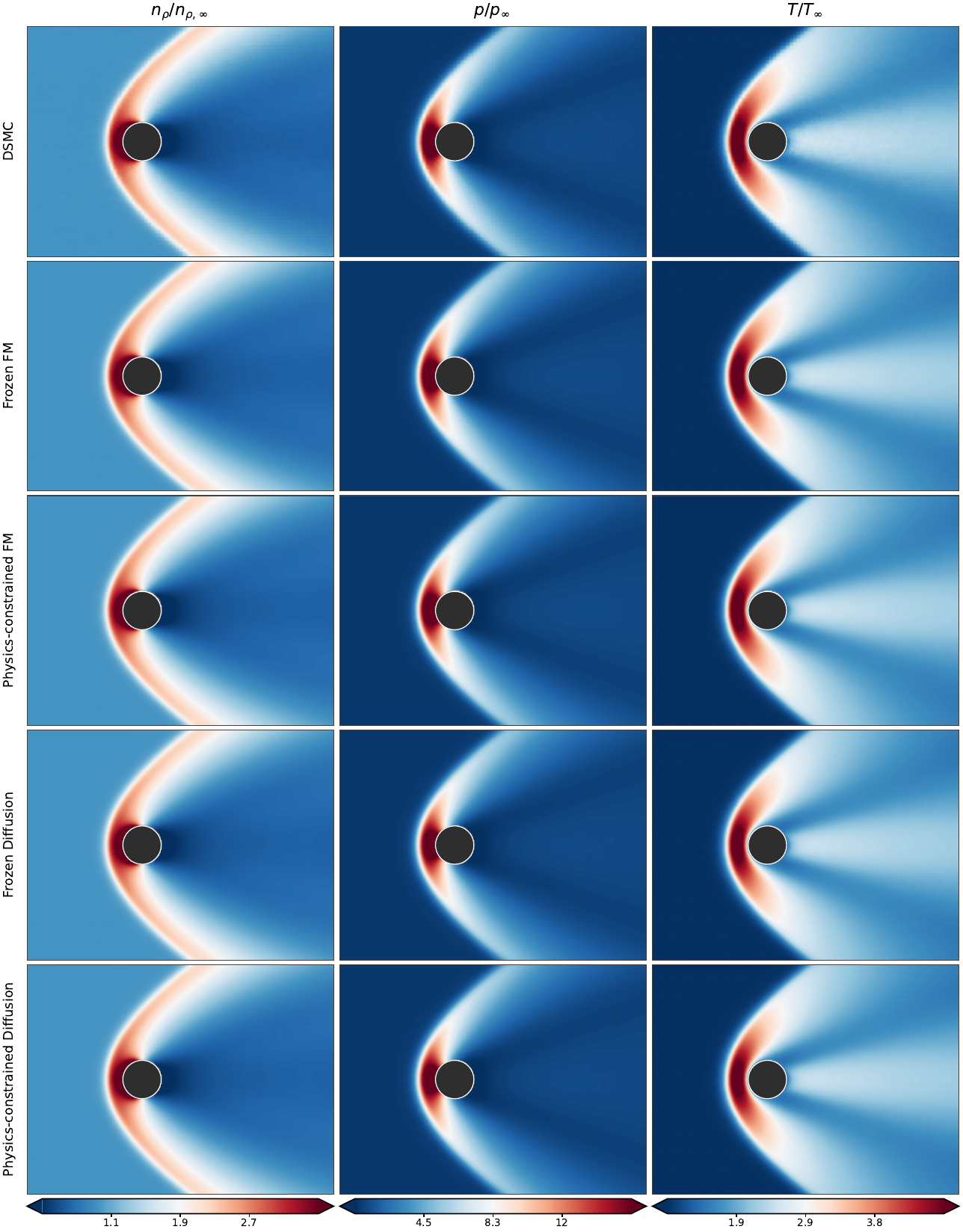}
 \caption{Dimensionless cylinder fields at
 $(\mathrm{Kn},\mathrm{Ma})=(0.38,3.7)$. Columnwise color limits are shared
 across rows and set independently for this condition.}
 \label{fig:cylinder-kn038-ma37-fields}
\end{figure}

\begin{figure}[htbp]
 \centering
 \includegraphics[width=0.96\textwidth,height=0.78\textheight,keepaspectratio]{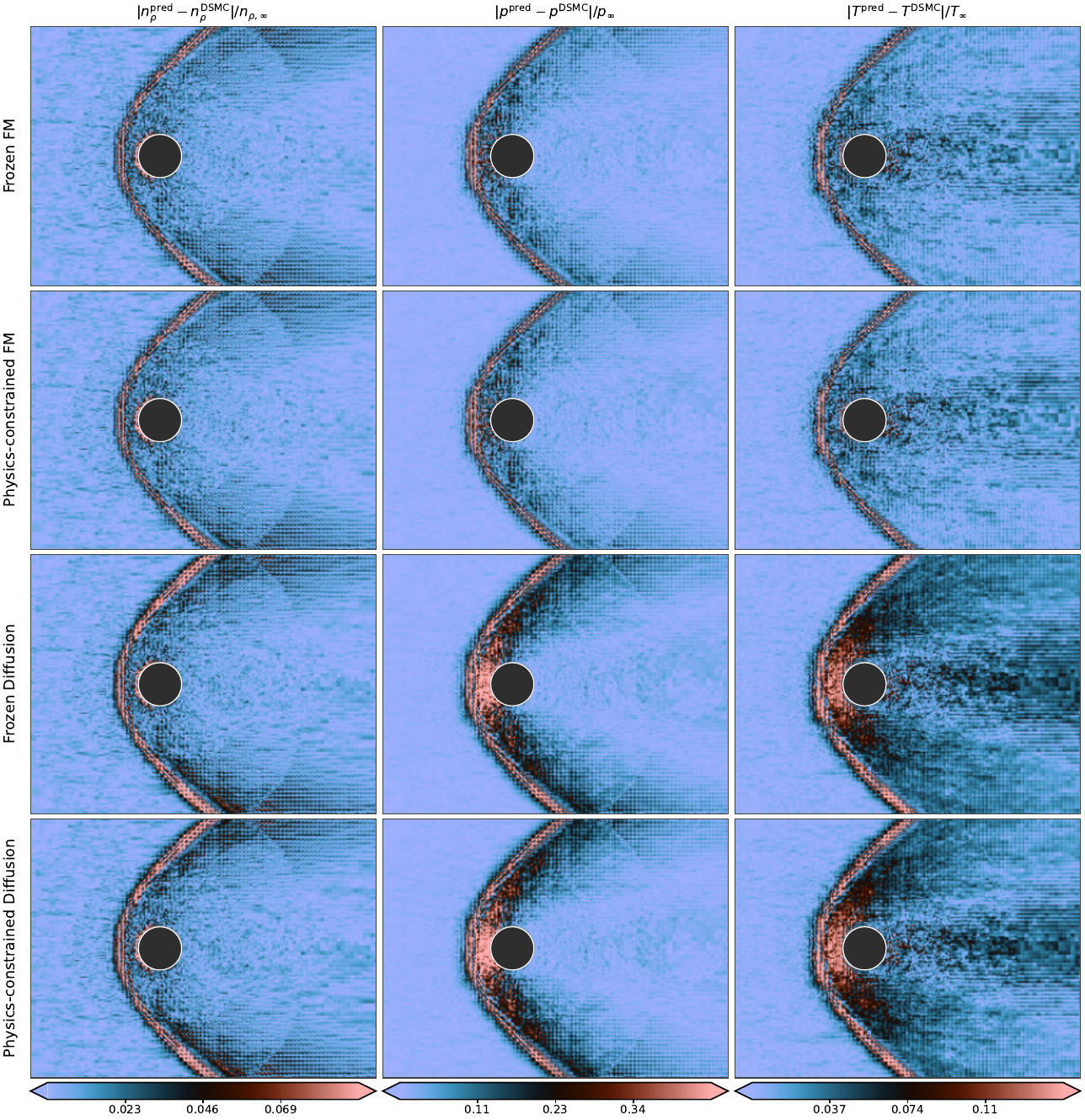}
 \caption{Direct dimensionless cylinder errors at $(\mathrm{Kn},\mathrm{Ma})
 =(0.38,3.7)$, using 99th-percentile upper clipping.}
 \label{fig:cylinder-kn038-ma37-errors}
\end{figure}

At the intermediate condition $(0.38,3.7)$,
\cref{fig:cylinder-kn038-ma37-fields,fig:cylinder-kn038-ma37-errors}, the
higher Mach number strengthens the compression and elongates the wake, while
$\mathrm{Kn}=0.38$ already thickens the bow shock into a broader compression
layer and enlarges the near-wall Knudsen layer relative to $(0.12,2.3)$.
Physics-constrained FM reduces $E_6$ from 0.02341 to 0.02319, while
Physics-constrained Diffusion is effectively unchanged (0.03389 to 0.03388).
The shock, wall, and wake structures are preserved after adaptation, consistent
with the trust-region and manifold anchors restricting the correction to a
small neighborhood of the frozen prediction.

\begin{figure}[htbp]
 \centering
 \includegraphics[width=0.96\textwidth,height=0.78\textheight,keepaspectratio]{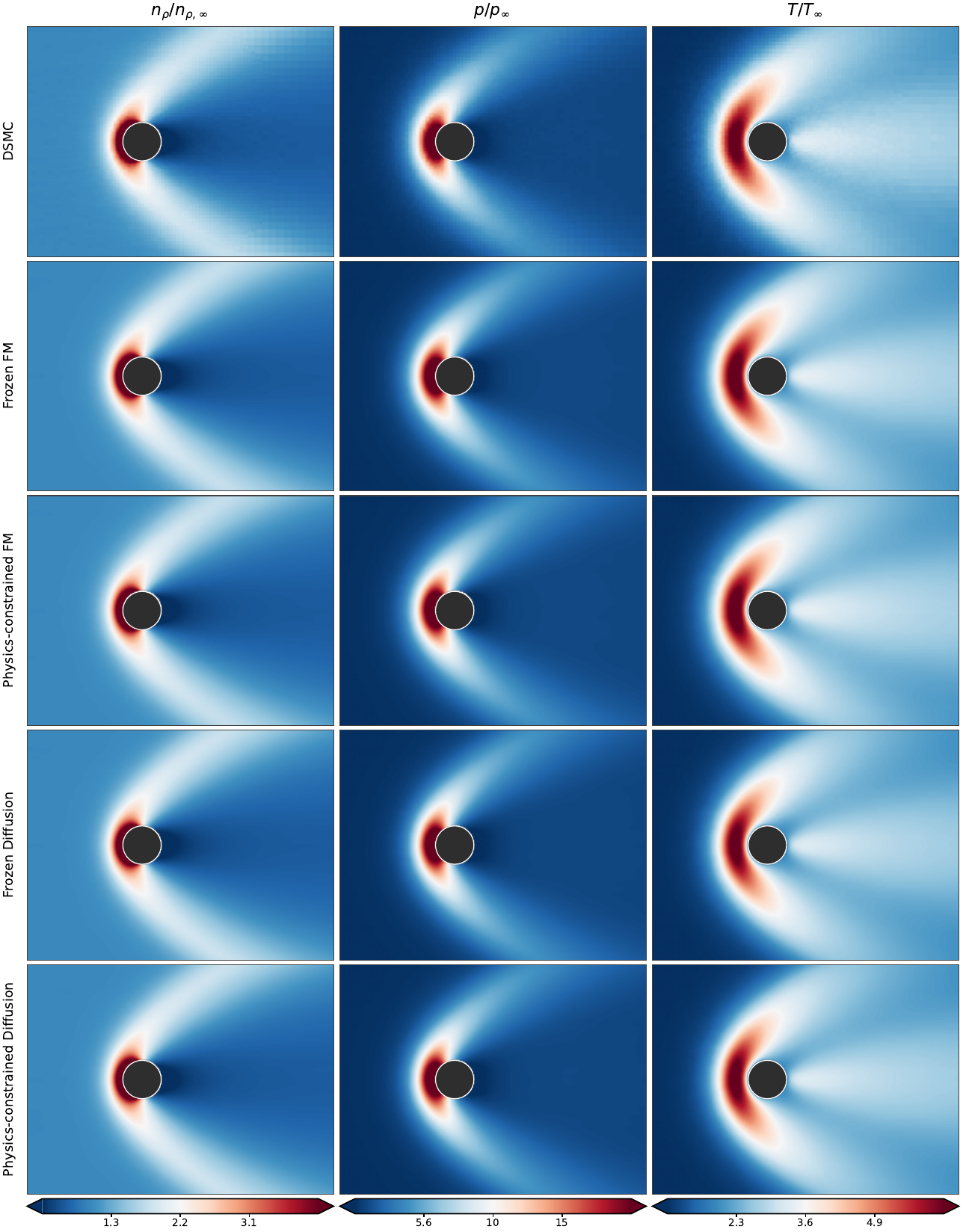}
 \caption{Dimensionless cylinder fields at
 $(\mathrm{Kn},\mathrm{Ma})=(0.76,4.5)$. Columnwise color limits are shared
 across rows and set independently for this condition.}
 \label{fig:cylinder-kn076-ma45-fields}
\end{figure}

\begin{figure}[htbp]
 \centering
 \includegraphics[width=0.96\textwidth,height=0.78\textheight,keepaspectratio]{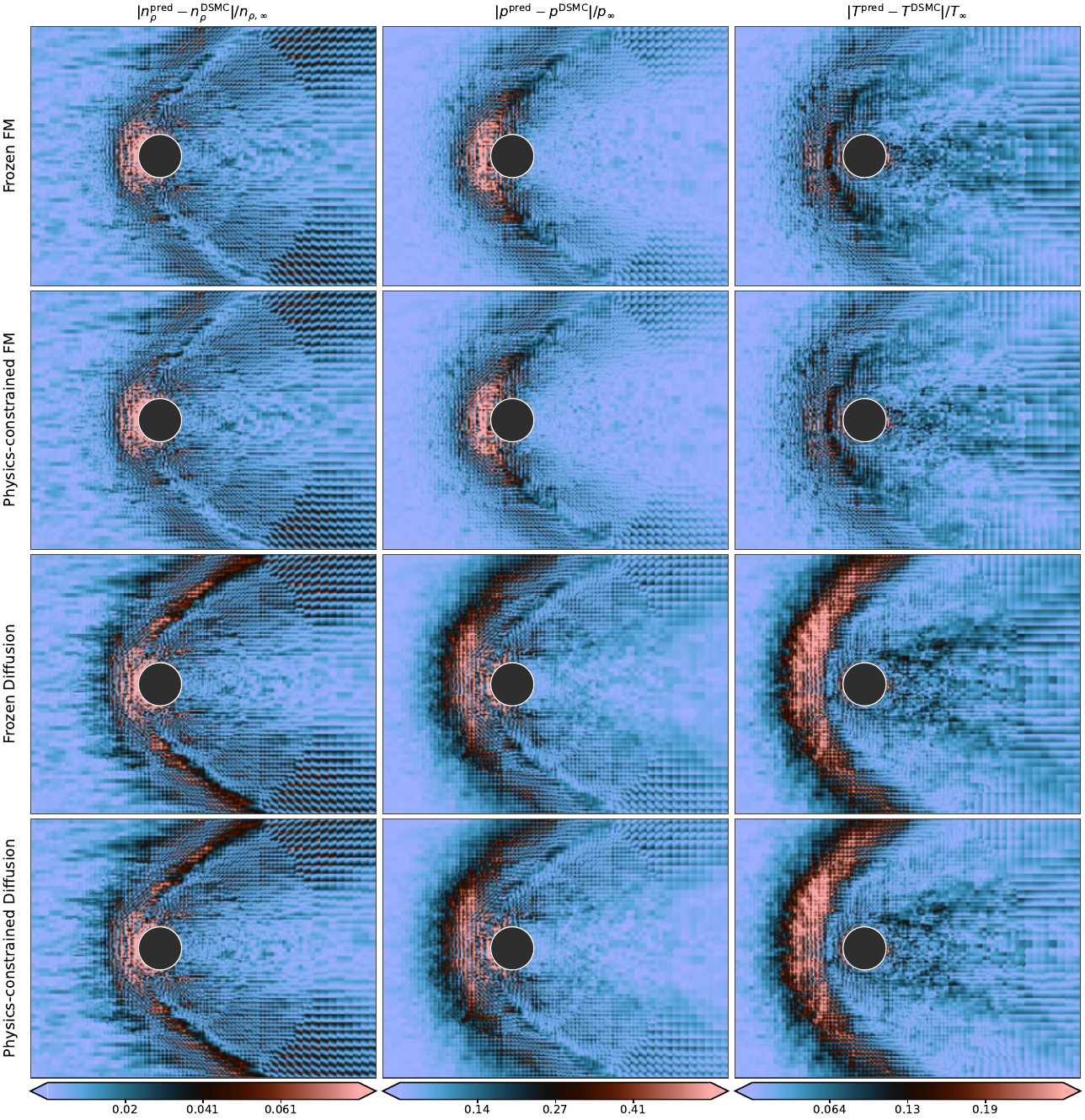}
 \caption{Direct dimensionless cylinder errors at $(\mathrm{Kn},\mathrm{Ma})
 =(0.76,4.5)$, using 99th-percentile upper clipping.}
 \label{fig:cylinder-kn076-ma45-errors}
\end{figure}

At the highest displayed Knudsen number, $(0.76,4.5)$,
\cref{fig:cylinder-kn076-ma45-fields,fig:cylinder-kn076-ma45-errors}, these
trends are most pronounced: the shock appears as a broad smeared compression
layer, the near-wall Knudsen layer is large, the wake is diffuse with weak
recirculation, and the higher Mach number yields the strongest stagnation
heating.  Both
adapters improve their own $E_6$: FM changes from 0.03756 to 0.03737 and
diffusion from 0.04812 to 0.04696.  The residuals remain concentrated along
the detached bow shock, in the near-wall layer, and through the downstream
wake.  Their location is physically consistent with the strongest
condition-dependent structures in rarefied cylinder flow
\citep{WangX2026}.

Across the three cases, the learned fields recover the principal shock and
wake morphology of the statistically sampled DSMC references.  Increasing
rarefaction generally broadens kinetic structures, but the simultaneous Mach
variation changes shock strength and prevents attribution of every visual
difference to $\mathrm{Kn}$ alone \citep{WangX2026}.  Fine-grained texture in
the direct-error maps partly reflects the finite-particle statistical
fluctuations intrinsic to DSMC averaging \citep{Bird1994}; the coherent
shock-aligned and wall-localized bands, by contrast, indicate systematic
condition-dependent surrogate error.  These two contributions cannot be
separated quantitatively from a single time-averaged DSMC realization.
Within each cylinder figure, color limits are shared across model rows but are
set independently for each condition; absolute colors should therefore not be
compared between different Knudsen--Mach pairs without reading their colorbars.

\subsection{Cross-benchmark discussion}
\label{sec:cross-benchmark-discussion}

The two benchmarks do not support a universal ranking of conditional
generators. Among the four displayed cavity conditions, diffusion has a clear
advantage at $\mathrm{Kn}=0.03$ and 1.00, while the two generators are closer
and exchange the macroscopic/kinetic ordering at the interpolation points.
In contrast, the decoder-aware curved FM has lower field error throughout the
reported two-parameter cylinder test. This
difference is consistent with the different conditional geometries: the
cavity generator transports from a simple reference along a one-dimensional
condition coordinate, while cylinder FM exploits local source cases and a
decoder-metric chart in the sparse $(\mathrm{Kn},\mathrm{Ma})$ plane.
Accordingly, the comparison concerns complete frozen pipelines, not an
architecture-only contest between FM and diffusion. This dependence on the
reference law, probability path, and conditioning construction is consistent
with the distinct transport formulations underlying flow matching and
diffusion \citep{Lipman2023,Tong2024,Ho2020,Song2021}.

Physical adaptation also has a narrower interpretation than field
reconstruction. For the cavity, access to $G$ and $B$ permits matched-grid
BGK, moment, and conservation audits. The adapters reduce the registered BGK
diagnostic in interpolation and upper extrapolation while approximately
preserving field accuracy, but do not do so in lower extrapolation. For the
cylinder, the available variables permit no-penetration, inlet, and weak-mass
constraints but not a closed kinetic or momentum-energy residual. The paired
frozen-versus-adapted audit therefore establishes improvements only in those
registered diagnostics, whereas the out-of-sample field evaluation shows that
such improvement does not guarantee a lower error at every additional
condition. More specifically, the formal Physics-constrained FM model improves
global boundary mass balance but not its local CV diagnostic. Its wall result
comes from the analytic projection, while the learned correction contributes
only as part of the jointly deployed adapter--projection system. The
Physics-constrained Diffusion model improves both mass diagnostics; neither
statement implies
closed momentum or energy conservation. The need to balance a physics residual
against fidelity to the learned data manifold is also observed in
physics-informed generative modeling more broadly
\citep{ShuDL2023,Jacobsen2024,Bastek2025}.

Finally, interpolation and extrapolation must be separated. At the displayed
$\mathrm{Kn}=0.23$ and 0.73 cases, cavity errors remain near $10^{-5}$, but
at the lower extrapolation point $\mathrm{Kn}=0.03$ they rise to the
$10^{-3}$ scale for both generator families. The similar behavior of frozen
and adapted predictions at this condition identifies the conditional latent
generator, rather than the local field adapter, as the main bottleneck. The
remedy is additional condition coverage or an extrapolation-aware transport
prior, not a larger decoder correction. Generated ensemble spread is likewise retained
as a numerical transport diagnostic: because each condition has one
deterministic steady reference, it is not evidence of calibrated physical
uncertainty.
The sampling counts should be interpreted with similar care. FM and diffusion
use different integration schemes, ensemble sizes, and, for the cylinder,
different reference constructions, so the number of ODE or denoising steps is
not by itself a fair measure of end-to-end inference cost.

Taken together, the comparisons support a component-wise interpretation of the
framework. The neural field supplies a common continuous representation for
heterogeneous DVM and DSMC outputs; FM and diffusion provide alternative
condition-dependent transports whose relative performance changes with the
parameter geometry and reference construction; and the frozen structured
adapter tests whether selected physical diagnostics can be improved without
retraining the representation or generator. The direct MLP comparison limits
the accuracy claim: on the present single-valued steady maps, generative
transport is not uniformly more accurate than deterministic regression. Its
additional value is instead the explicit transport construction, sample-based
inference, and compatibility with future conditional distributions containing
multiple admissible or stochastic states. The last capability is not validated
by the present one-reference-per-condition data and should be treated as a
direction for extension rather than a measured uncertainty result.

\section{Summary and conclusions}\label{Conclusions}

This work presented a conditional latent generative framework for two steady
rarefied-flow problems with different parameter dimensions and output
structures.  Neural-field auto-decoders compress the high-fidelity solutions,
and train-only principal-component charts organize the optimized lookup
latents.  FM or diffusion then generates a condition-dependent chart endpoint
directly from a simple reference distribution, without a deterministic
condition-to-latent regressor as the predictive backbone.  The frozen inverse
chart and decoder recover the physical fields.  This modular construction
accommodates both structured DVM fields and native-mesh DSMC fields while
separating representation, conditional-generation, and physical-correction
errors.

The findings establish four distinct points. First, one neural-field
representation can support structured DVM fields and native-mesh DSMC fields.
Second, FM and diffusion can predict the retained latent charts, but their
relative performance depends on condition-space geometry, latent metric,
reference construction, and numerical transport. Third, zero-initialized
structured adapters can improve selected, computable physical diagnostics while
the representation and generative backbone remain frozen. Fourth, the audit
answers the motivating question of where sampling-based transport adds value:
for the present single-valued steady problems, deterministic regression
remains a strong point predictor---five-seed MLPs at two capacity levels
match or exceed the generators on several metrics (cavity interpolation means
are statistically indistinguishable between the generators and lie above the
MLP band; cylinder MLPs outperform both generator five-seed means over all
seven additional conditions), whereas FM remains robustly above diffusion on
the cylinder. The distinctive assets of the generative formulation are
therefore its sampling dimension, the frozen physics-adaptation interface,
and extensibility to multivalued or stochastic solution families; the audit
delineates this regime of applicability. Matching the generator capacity does
not remove the deterministic advantage, so it is not a capacity artifact. The
principal empirical contribution is therefore a unified study of neural-field
representation, conditional latent transport, and frozen physical adaptation,
not evidence that sampling-based transport is universally preferable for a
single-valued steady surrogate.

The representation study showed that explained latent variance alone is not a
sufficient rank-selection criterion: weak latent directions can be accurately
reconstructed but remain difficult to generate from sparsely sampled physical
conditions.  Decoded-field screening selected a whitened rank-four chart for
the cavity and a rank-12 chart for the cylinder.  The retained cylinder FM
model uses a decoder-metric chart, whereas its diffusion counterpart uses
ordinary unwhitened PCA\@. Generator selection was likewise problem dependent:
plain-concatenation FM was effective for the one-parameter cavity, whereas
decoder-aware curved FM was the strongest frozen generator over the
two-parameter cylinder domain. These are rankings of complete frozen pipelines:
because the cylinder charts and reference constructions differ, they should not
be interpreted as an isolated algorithmic ranking of FM against diffusion.

On the displayed cavity interpolation conditions, $\mathrm{Kn}=0.23$ and
0.73, the macroscopic and reduced-distribution errors remain predominantly at
the $10^{-5}$ level.  At the upper extrapolation condition
$\mathrm{Kn}=1.00$, diffusion is more accurate than FM, whereas both generators
degrade to the $10^{-3}$ level at the lower extrapolation condition
$\mathrm{Kn}=0.03$.  The limited response of the structured adapters at the
latter condition indicates that the dominant error originates in extrapolation
of the latent endpoint rather than in local field decoding.  For the cylinder,
the equal-case six-field RMSE is 0.03655 for Frozen FM and 0.04748 for Frozen
Diffusion.  The corresponding contours recover the bow shock, near-wall layer,
and wake, although coherent shock-aligned discrepancies and DSMC sampling
texture remain visible.  These results emphasize that interpolation accuracy
does not by itself establish reliable continuation beyond the sampled
condition support.

Physics correction was introduced through zero-initialized structured low-rank
adapters while freezing the selected generator, latent transform, and principal
decoder. For the cavity, normalized manifold anchoring reduces the registered
matched-grid BGK diagnostic while approximately preserving macroscopic and
kinetic accuracy over the displayed interpolation cases and at
$\mathrm{Kn}=1.00$. The diffusion checkpoint improves its invariant
diagnostics, whereas the FM checkpoint has conservation and moment ratios of
1.0574 (+5.7\%); the FM result is not interpreted as improved conservation. For the cylinder,
the admissible constraints are limited
to wall no-penetration, inlet consistency, local control-volume mass balance,
and global boundary mass balance.  The formal Physics-constrained FM model
enforces no-penetration exactly (wall ratio $6.94\times10^{-11}$) with
essentially unchanged field error; the paired attribution shows that the wall
improvement is already obtained by the analytic hard map alone, while the
learned correction adds the inlet and global-balance improvements; the local control-volume
ratio remains 1.0218. The learned adapter by itself is not a feasible wall
model because it was trained upstream of the hard projection. The
Physics-constrained Diffusion model reduces
both local and global mass ratios but remains less accurate than FM in the
common field audit.  A lower
registered physics score consequently neither guarantees a smaller field error
nor establishes satisfaction of the complete Boltzmann or BGK system.

For the difficult lower-side cavity extrapolation, the principal limitation is
condition-space coverage rather than spatial query resolution.  Additional
low-Knudsen training conditions or an extrapolation-aware latent transport
prior are therefore more likely to help than increasing adapter capacity.  A
stronger cylinder audit would require pressure-tensor and heat-flux data to
close the momentum and energy balances, together with independent DSMC
realizations or averaging blocks to separate sampling variability from
surrogate error.  Because each present operating condition supplies only one
steady reference solution, generated ensemble spread measures numerical
transport sensitivity and is not calibrated physical uncertainty.  Replicated
stochastic data or genuinely non-unique unsteady states would be needed before
that spread could be interpreted probabilistically.  Overall, the results
support a modular strategy in which neural fields provide the common solution
representation, conditional transport supplies sampling-based cross-condition
prediction, and frozen structured adapters improve selected, explicitly
auditable physical diagnostics. For the present steady single-reference data,
this strategy is an extensible modeling framework rather than a demonstrated
replacement for the more economical deterministic MLP.

\appendix

This appendix provides the audits and additional field comparisons that
support the main text: endpoint-averaging noncommutativity
(\cref{app:endpoint-averaging}), leave-one-condition-out chart stability
(\cref{app:loco-chart-stability}), reduced-BGK numerical sensitivity
(\cref{sec:bgk-numerical-sensitivity}), a sampling-seed audit of the frozen
cavity pipelines (\cref{app:cavity-sampling-seeds}), computational cost
(\cref{app:computational-cost}), and a physics-adapter seed/weight
robustness audit (\cref{sec:adapter-robustness}).

\section{Endpoint-averaging noncommutativity}\label{app:endpoint-averaging}

The noncommutativity is quantified after generator selection using 128
endpoints, 256 fixed spatial probes, and 2,048 fixed velocity nodes at the
four cavity evaluation conditions $\{0.03,0.23,0.73,1.00\}$. The relative difference between
$D_{\theta_D}(M^{-1}\sum_m\widehat{\bm z}^{(m)})$ and
$M^{-1}\sum_mD_{\theta_D}(\widehat{\bm z}^{(m)})$ has FM macro/kinetic
mean values $1.75\times10^{-6}$/$4.69\times10^{-6}$ and maxima
$1.81\times10^{-6}$/$6.04\times10^{-6}$. For diffusion the corresponding
means are $7.09\times10^{-6}$/$1.84\times10^{-5}$ and the maxima are
$2.37\times10^{-5}$/$5.85\times10^{-5}$. The diffusion maximum occurs at
the lower extrapolation condition $\mathrm{Kn}=0.03$, where its endpoint RMS
spread is $1.89\times10^{-2}$; the three other diffusion kinetic biases are
$4.19\times10^{-6}$--$5.73\times10^{-6}$. Latent averaging is therefore an
accurate computational surrogate for decoded averaging in interpolation and
upper extrapolation, but contributes a non-negligible part of the already
larger low-Knudsen extrapolation error.

\section{Leave-one-condition-out chart stability}\label{app:loco-chart-stability}

Because this chart is estimated from only 16 lookup codes in a
24-dimensional ambient space, we additionally refit the rank-four PCA after
omitting each training condition and compare its span with the full-data span.
Across the 16 folds, the largest principal angle has a mean of
$2.10^{\circ}$ and a maximum of $11.12^{\circ}$; the worst fold omits the
lower endpoint $\mathrm{Kn}=0.05$. Across all 120 pairs of LOCO folds, the
corresponding mean and maximum are $3.38^{\circ}$ and $12.68^{\circ}$.
Thus, the dominant four-dimensional span is reasonably stable for most
omissions but is not statistically invariant at the boundary of the training
support. This finite-sample chart uncertainty is separate from the decoded
LOCO error in the main-text PCA ablation table and limits claims about an
intrinsic four-dimensional physical manifold.

\section{Reduced-BGK numerical sensitivity}
\label{sec:bgk-numerical-sensitivity}

The reductions in the main-text cavity adapter table are evaluated with the
near-wall-biased probe measure used for adapter selection. To separate that
measure from differentiation and spatial-quadrature effects, an independent
frozen-checkpoint audit pairs each adapted prediction with its frozen baseline
using the same generated latent, the same spatial points, and all 6,400
velocity nodes. Four training-support conditions,
$\mathrm{Kn}\in\{0.15,0.35,0.65,0.90\}$, are given equal weight. Spatial
integration uses cell-centered composite midpoint quadrature on
$[0.01,0.99]^2$.

\begin{table}[htbp]
 \centering
 \caption{Numerical sensitivity of the cavity reduced-BGK reduction. Values
 are equal-Kn reductions relative to each paired frozen generator under
 uniform-area quadrature.}
 \label{tab:bgk-numerical-sensitivity}
 \footnotesize
 \begin{tabular}{lcc}
  \toprule
  Audit setting & FM reduction & Diffusion reduction\\
  \midrule
  $h=1\times10^{-3}$, $24^2$ grid & 16.289\% & 11.274\%\\
  \textbf{$h=2\times10^{-3}$, $24^2$ grid} & \textbf{16.286\%} & \textbf{11.378\%}\\
  $h=5\times10^{-3}$, $24^2$ grid & 16.509\% & 11.366\%\\
  $h=2\times10^{-3}$, $36^2$ grid & 19.382\% & 13.076\%\\
  $h=2\times10^{-3}$, $48^2$ grid & 17.806\% & 12.011\%\\
  \bottomrule
 \end{tabular}
\end{table}

Across the three difference steps, the reduction spans only 0.223 percentage
points for FM and 0.104 percentage points for diffusion. The sign is also
unchanged for every audited condition on the $24^2$, $36^2$, and $48^2$
grids. A $12^2$ control is under-resolved and reverses the local ordering at
the lowest conditions, so it is excluded from
\cref{tab:bgk-numerical-sensitivity}. Between $24^2$ and $48^2$, the aggregate
reduction spans 3.095 percentage points for FM and 1.698 points for diffusion.
Thus, the conclusion that adaptation lowers the registered BGK diagnostic is
stable to the tested difference step and adequate query-grid refinement, but
the reduction magnitude is not independent of the spatial sampling measure.
In particular, the near-wall-biased values 28.65\% and 23.59\% in
the main-text cavity adapter table should not be interpreted as uniform-area
residual reductions.

\section{Sampling-seed audit of the frozen cavity pipelines}\label{app:cavity-sampling-seeds}

The primary checkpoints are single training fits. To separate training
replication from sampling variability, we additionally repeat only the frozen
cavity inference with 20 sampling seeds and 256 endpoints per seed. Over the
two displayed interpolation cases, the kinetic-error mean and population
standard deviation are $(5.4965\pm0.0073)\times10^{-5}$ for FM and
$(5.9123\pm0.0501)\times10^{-5}$ for diffusion; over the two extrapolation
cases they are $(3.09335\pm0.00062)\times10^{-3}$ and
$(1.6221\pm0.0138)\times10^{-3}$, respectively. These small sampling-seed variations do not quantify retraining uncertainty;
independent five-seed generator refits addressing that question are reported
in the main text.

\section{Computational cost}\label{app:computational-cost}

Training wall times are reported from the original run logs and therefore
retain their original, nonuniform hardware context. The cavity representation
fit required 200,000 updates and $20.38$ logged GPU hours (sum of per-update
times). The cavity FM and diffusion generator fits required 1,018 and 128 s,
respectively; their physics adapters required 5,755 and 5,920 s. The retained
cylinder representation fit required 131,239 s, the retained FM generator
screen 522 s, and the formal FM and diffusion adapters 1,486 and 4,779 s.
Because accelerator identifiers were not stored by all historical jobs, these
figures document cost but are not used for cross-model speed ranking.

For a controlled inference measurement, all frozen pipelines were rerun in
FP32 on one NVIDIA RTX 5070 Ti Laptop GPU after warm-up, using five repeats.
For one cavity query with 256 endpoints, 2,500 spatial points, and 6,400
velocity nodes, median latent-generation/full-decoding times were
0.071/0.444 s for FM and 0.049/0.658 s for diffusion. For one cylinder query
on a $256\times192$ output grid, the corresponding times were 0.134/0.090 s
for the 21-sample local-anchor FM and 0.046/0.090 s for 40-sample diffusion.
These are end-to-end neural inference components, excluding file I/O and
plotting. The distributed TransportBench fields do not include the original
DVM or DSMC job wall times and hardware records; consequently, a defensible
solver-to-surrogate acceleration ratio cannot be inferred from the available
artifacts and is not reported.

\section{Physics-adapter seed and weight robustness (cavity FM)}
\label{sec:adapter-robustness}

The retained cavity FM adapter is a single fit. To test whether its
conclusion depends on the training seed or on the manually set physics
weight, the adapter is retrained with two additional seeds and with the
physics total weight scaled by $0.5\times$ and $2\times$. Because these
runs were executed on a 12-GB accelerator, all five rows (including a
default-seed control) share a reduced collocation profile (16 conditions and
256 collocation points per update); the retained full-profile run is listed
for reference. Selection folds, steps, and all other weights are unchanged.

\begin{table}[htbp]
 \centering
 \caption{Cavity FM adapter robustness to training seed and physics weight.
 Reductions and ratios are evaluated against the paired frozen generator on
 independent probes; ``gates'' denotes the feasibility criteria of the main
 text.}
 \label{tab:adapter-robustness}
 \footnotesize
 \begin{tabular}{lcccc}
  \toprule
  Variant & BGK reduction & Invariant ratio & Adapted kinetic & Gates\\
  \midrule
  Retained (full profile) & 28.65\% & 1.0574 & $5.652\times10^{-5}$ & feasible\\
  Default seed, reduced profile & 24.54\% & 1.0432 & $5.650\times10^{-5}$ & all pass\\
  Seed +1 & 32.56\% & 1.0270 & $5.680\times10^{-5}$ & all pass\\
  Seed +2 & 25.47\% & 1.0689 & $5.696\times10^{-5}$ & all pass\\
  Weight $\times0.5$ & 10.24\% & 1.0140 & $5.460\times10^{-5}$ & all pass\\
  Weight $\times2$ & 14.19\% & 1.0454 & $5.709\times10^{-5}$ & all pass\\
  \bottomrule
 \end{tabular}
\end{table}

The BGK reduction remains positive and of the same magnitude for all seeds,
the invariant ratios stay within 1.027--1.069, and the adapted kinetic error
never exceeds the paired frozen value ($5.68\times10^{-5}$ under the reduced
profile) by more than $0.5\%$. Halving or doubling the
physics weight changes the reduction magnitude but not the feasibility or the
field accuracy. The adapter conclusion is therefore robust to seed and weight
perturbations; the weight scan additionally shows that larger physics weights
do not purchase larger reductions, consistent with the gate-based checkpoint
selection of the main text.

\begin{CJK*}{UTF8}{gbsn}
\section*{Acknowledgments}
Y. Qi would like to thank Dr. Tianyi Li (李天一), 
  Miss Hui Jin (金慧), and Mr. Kuilong Chen (陈奎龙)
  for their support during the research process.
This work was supported by the
  Science Foundation for Young Scientists of the State Key Laboratory of High
  Temperature Gas Dynamics (2025QN16), 
  Chinese Academy of Sciences Project for Young Scientists in Basic Research
  (YSBR107), 
  Strategic Priority Research Program of the Chinese Academy of Sciences
  (XDB0620403), 
  National Natural Science Foundation of China (12302381 and 12572340), 
  and Beijing Natural Science Foundation (L252039). 
\end{CJK*}

\section*{DATA AVAILABILITY}
The source code required to reproduce the reported models and evaluations will
be made publicly available upon publication together with a SHA-256 manifest
for the ten frozen checkpoints used in the reported pipeline comparisons. The cavity and cylinder data are
from TransportBench and are publicly available at
\href{https://huggingface.co/datasets/CFDML/TransportBench}{https://huggingface.co/datasets/CFDML/TransportBench}.

\bibliography{main.bib}
\newpage

\end{document}